%% file: paper.tex
\documentclass[twocolumn,aps,prd,superscriptaddress,nofootinbib,floatfix,preprintnumbers,a4]{revtex4-1}

\usepackage{bm,amsmath,amsfonts,amssymb,graphicx,xspace,placeins,multirow}

\usepackage{hyperref}
\usepackage{breakurl}
\usepackage{array}
\usepackage{slashed}
\usepackage{xspace}
\usepackage[italic]{hepnames}
\usepackage{booktabs}
\usepackage{hhline}
\usepackage{multirow}
\usepackage{header3}
\usepackage[]{lineno}
\usepackage{placeins}

\allowdisplaybreaks
\newcommand{\nn}{\nonumber}

\newcommand{\Bbar}{\,\overline{\!B}{}}
\newcommand{\Dbar}{\,\overline{\!D}{}}

\newcommand{\Vub}{V_{ub}}

\newcommand{\Eg}{\ensuremath{E_{\gamma}}\xspace}

\newcommand{\bfResa}{\ensuremath{\Delta \mathcal{B}(B \to X_u \ell^+ \, \nu_\ell) = \left( 1.09 \pm 0.05  \pm 0.08 \right) \times 10^{-3}}\xspace}
\newcommand{\bfResb}{\ensuremath{\Delta \mathcal{B}(B \to X_u \ell^+ \, \nu_\ell) = \left( 0.67 \pm 0.07  \pm 0.10 \right) \times 10^{-3}}\xspace}
\newcommand{\bfRescp}{\ensuremath{\Delta \mathcal{B}(B \to X_u \ell^+ \, \nu_\ell) = \left( 1.11 \pm 0.06  \pm 0.14 \right) \times 10^{-3}}\xspace}
\newcommand{\bfResc}{\ensuremath{\Delta \mathcal{B}(B \to X_u \ell^+ \, \nu_\ell) = \left( 1.69 \pm 0.09  \pm 0.26 \right) \times 10^{-3}}\xspace}
\newcommand{\bfResd}{\ensuremath{\Delta \mathcal{B}(B \to X_u \ell^+ \, \nu_\ell) = \left( 1.59 \pm 0.07  \pm 0.16 \right) \times 10^{-3}}\xspace}

\newcommand{\bfResdEl}{\ensuremath{\Delta \mathcal{B}(B \to X_u e^+ \, \nu_e) = \left( 1.57 \pm 0.10  \pm 0.16 \right) \times 10^{-3}}\xspace}
\newcommand{\bfResdMu}{\ensuremath{\Delta \mathcal{B}(B \to X_u \mu^+ \, \nu_\mu) = \left( 1.62 \pm 0.10  \pm 0.18 \right) \times 10^{-3}}\xspace}
\newcommand{\rElMu}{\ensuremath{ R_{e\mu} = \frac{\Delta \mathcal{B}(B \to X_u e^+ \, \nu_e)}{\Delta \mathcal{B}(B \to X_u \mu^+ \, \nu_\mu)} = 0.97 \pm 0.09 \pm 0.04 }\xspace}

\newcommand{\bfResdBp}{\ensuremath{\Delta \mathcal{B}(B^+ \to X_u \ell^+ \, \nu_\ell) = \left( 1.65 \pm 0.10  \pm 0.18 \right) \times 10^{-3}}\xspace}
\newcommand{\bfResdBn}{\ensuremath{\Delta \mathcal{B}(B^0 \to X_u \ell^+ \, \nu_\ell) = \left( 1.51 \pm 0.10  \pm 0.16 \right) \times 10^{-3}}\xspace}
\newcommand{\rBpBn}{\ensuremath{  R_{\mathrm{iso}}  = 1.01 \pm 0.09 \pm 0.06 }\xspace}

\newcommand{\bfResAverageVuba}{\ensuremath{\left| V_{ub} \right| = \left( 4.10 \pm 0.09  \pm 0.22 \pm 0.15  \right) \times 10^{-3}}\xspace}

\newcommand{\bfResBLNPVuba}{\ensuremath{ \left( 3.90 \pm 0.09  \pm 0.15  \pm 0.21 \right) \times 10^{-3}}\xspace}
\newcommand{\bfResBLNPVubb}{\ensuremath{\left( 4.24  {}^{+0.22}_{-0.23}  {}^{+0.30}_{-0.32} {}^{+0.26}_{-0.28} \right) \times 10^{-3}}\xspace}

\newcommand{\bfResBLNPVubd}{\ensuremath{ \left( 4.05 \pm 0.09 {}^{+0.20}_{-0.21} {}^{+0.18}_{-0.20} \right) \times 10^{-3}}\xspace}

\newcommand{\bfResDGEVuba}{\ensuremath{ \left( 4.08 \pm 0.09  \pm 0.16 {}^{+0.20}_{-0.26} \right) \times 10^{-3}}\xspace}
\newcommand{\bfResDGEVubb}{\ensuremath{ \left( 4.16 {}^{+0.21}_{-0.23} {}^{+0.30}_{-0.32}{}^{+0.18}_{-0.21} \right) \times 10^{-3}}\xspace}

\newcommand{\bfResDGEVubd}{\ensuremath{ \left( 4.16  \pm 0.09  {}^{+0.21}_{-0.22}  {}^{+0.11}_{-0.12} \right) \times 10^{-3}}\xspace}

\newcommand{\bfResGGOUVuba}{\ensuremath{ \left( 3.97 \pm 0.09  {}^{+0.15}_{-0.16} {}^{+0.15}_{-0.16} \right) \times 10^{-3}}\xspace}
\newcommand{\bfResGGOUVubb}{\ensuremath{ \left( 4.25 {}^{+0.22}_{-0.23} {}^{+0.30}_{-0.33} {}^{+0.24}_{-0.26} \right) \times 10^{-3}}\xspace}

\newcommand{\bfResGGOUVubd}{\ensuremath{ \left( 4.15  \pm 0.09   {}^{+0.21}_{-0.22}  {}^{+0.08}_{-0.09} \right) \times 10^{-3}}\xspace}

\newcommand{\bfResADFRVuba}{\ensuremath{ \left( 3.63 \pm 0.08 \pm 0.14 \pm 0.17 \right) \times 10^{-3}}\xspace}
\newcommand{\bfResADFRVubb}{\ensuremath{\left( 3.68 {}^{+0.19}_{-0.20}  {}^{+0.26}_{-0.28} \pm 0.17 \right) \times 10^{-3}}\xspace}

\newcommand{\bfResADFRVubd}{\ensuremath{ \left( 4.05 \pm 0.09  {}^{+0.20}_{-0.21}\pm 0.18 \right) \times 10^{-3}}\xspace}

\newcommand{\GF}{\ensuremath{G_{F}}}

\newcommand{\bxlnu}{\ensuremath{B \to X \, \ell^+\, \nu_{\ell}}\xspace}
\newcommand{\bulnu}{\ensuremath{B \to X_u \, \ell^+\, \nu_{\ell}}\xspace}
\newcommand{\bclnu}{\ensuremath{B \to X_c \, \ell^+\, \nu_{\ell}}\xspace}

\newcommand{\bpilnu}{\ensuremath{B \to \pi \, \ell^+\,\nu_{\ell}}\xspace}
\newcommand{\bpichlnu}{\ensuremath{B^0 \to \pi^- \, \ell^+\,\nu_{\ell}}\xspace}
\newcommand{\lbplnu}{\ensuremath{\Lambda_b \to p \, \mu^+ \, \nu_\mu}\xspace}
\newcommand{\lblclnu}{\ensuremath{\Lambda_b \to \Lambda_c \, \mu^+ \, \nu_\mu}\xspace}
\newcommand{\brholnu}{\ensuremath{B \to \rho \, \ell^+\,\nu_{\ell}}\xspace}
\newcommand{\bomegalnu}{\ensuremath{B \to \omega \, \ell^+\,\nu_{\ell}}\xspace}
\newcommand{\betalnu}{\ensuremath{B \to \eta \, \ell^+\,\nu_{\ell}}\xspace}
\newcommand{\betaplnu}{\ensuremath{B \to \eta' \, \ell^+\,\nu_{\ell}}\xspace}
\newcommand{\bdlnu}{\ensuremath{B \to D \, \ell^+\,\nu_{\ell}}\xspace}
\newcommand{\bdslnu}{\ensuremath{B \to D^* \, \ell^+\,\nu_{\ell}}\xspace}
\newcommand{\bddslnu}{\ensuremath{B \to D^{**} \, \ell^+\,\nu_{\ell}}\xspace}

\renewcommand{\arraystretch}{1.25}
\makeatletter
\g@addto@macro\bfseries{\boldmath}
\makeatother

\definecolor{red}{rgb}{0.9, 0,0}

\begin{document}


\title{Measurements of Partial Branching Fractions of Inclusive \bulnu Decays with Hadronic Tagging}

\author{L. Cao}
\email{cao@physik.uni-bonn.de}
\affiliation{University of Bonn, 53115 Bonn}

\author{W. Sutcliffe}
\affiliation{University of Bonn, 53115 Bonn}

\author{R. Van Tonder}
\affiliation{University of Bonn, 53115 Bonn}

\author{F. U.\ Bernlochner}
\email{florian.bernlochner@uni-bonn.de}
\affiliation{University of Bonn, 53115 Bonn}

\input{pub569_2}

\begin{abstract}
We present measurements of partial branching fractions of inclusive semileptonic \bulnu decays using the full Belle data set of 711 fb$^{-1}$ of integrated luminosity at the $\Upsilon(4S)$ resonance and for $\ell = e, \mu$. Inclusive semileptonic  \bulnu decays are CKM suppressed and measurements are complicated by the large background from CKM-favored \bclnu transitions, which have a similar signature. Using machine learning techniques, we reduce this and other backgrounds effectively, whilst retaining access to a large fraction of the \bulnu phase space and high signal efficiency. We measure partial branching fractions in three phase-space regions covering about 31\% to 86\% of the accessible \bulnu phase space. The most inclusive measurement corresponds to the phase space with lepton energies of $E_\ell^B > 1 $ GeV, and we obtain \bfResd from a two-dimensional fit of the hadronic mass spectrum and the four-momentum-transfer squared distribution, with the uncertainties denoting the statistical and systematic error. We find \bfResAverageVuba from an average of four calculations for the partial decay rate with the third uncertainty denoting the average theory error. This value is higher but compatible with the determination from exclusive semileptonic decays within 1.3 standard deviations. In addition, we report charmless inclusive partial branching fractions separately for $B^+$ and $B^0$ mesons as well as for electron and muon final states. No isospin breaking or lepton flavor universality violating effects are observed.
\end{abstract}


\preprint{ Belle Preprint 2020-22, KEK Preprint 2020-39}

\maketitle

\section{introduction}

Precision measurements of the absolute value of the Cabibbo-Kobayashi-Maskawa (CKM) matrix element $V_{ub}$ are important to challenge the Standard Model of particle physics (SM)~\cite{PhysRevLett.10.531,km_paper}. In the SM, the CKM matrix is a $3 \times 3$ unitary matrix and responsible for the known charge-parity (CP) violating effects in the quark sector~\cite{pdg_ckm:2020}. There are indications of CP violation in the neutrino sector~\cite{Abe:2019vii}, but it remains unclear if both sources of CP violation are sufficient to explain the matter dominance of today's universe. This motivates the search for new sources of CP-violating phenomena. If such exist in the form of heavy exotic particles that couple to quarks in some form, their presence might alter the properties of measurements constraining the unitarity of the CKM matrix~\cite{Kou:2018nap}. Precise measurements of $|V_{ub}|$ and the CKM angle $\gamma = \phi_3$ are imperative to isolate such effects, as their measurements involve tree-level processes, which are expected to remain unaffected by new physics and thus provide an unbiased measure for the amount of CPV due to the Kobayashi-Maskawa (KM) mechanism~\cite{km_paper} alone. 

\begin{figure}[t!]
  \includegraphics[width=0.48\textwidth]{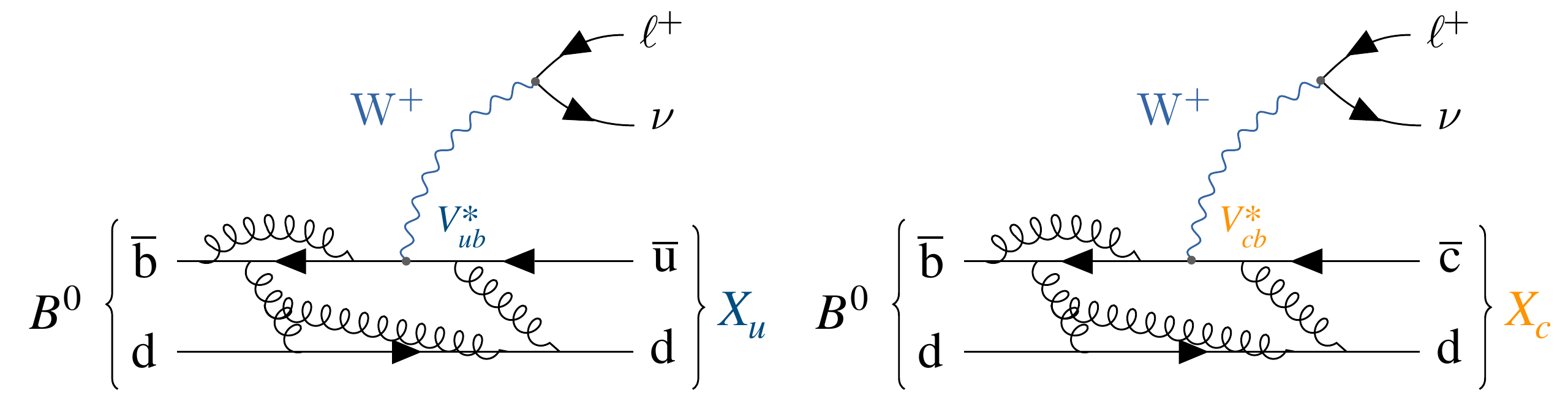} 
\caption{
The CKM suppressed and favored inclusive semileptonic processes \bulnu (left) and \bclnu (right) for a $B^0$ meson decay.
 }
\label{fig:bulnu_bclnu}
\end{figure}

Charmless semileptonic decays of \PB mesons provide a clean avenue to measure $|V_{ub}|$, as their decay rate is theoretically better understood than purely hadronic transitions and their decay signature is more accessible than leptonic \PB meson decays. The existing measurements either focus on exclusive final states, with \bpilnu~\cite{Amhis:2019ckw} and the ratio of \lbplnu\ and \lblclnu~\cite{Aaij:2015bfa} providing the most precise measurements to date, and measurements reconstructing the \bulnu decay fully inclusively\footnote{Charge conjugation is implied throughout this paper. In addition, \bulnu is defined as the average branching fraction of charged and neutral \PB meson decays and $\ell = e$ or $\mu$.}. Central for both approaches are reliable predictions of the (partial) decay rates $\Delta \Gamma(\bulnu)$ (omitting the CKM factor) from theory to convert measured (partial or full) branching fractions, $\Delta \mathcal{B}(\bulnu)$, into measurements of $|V_{ub}|$ via
\begin{align}
 |V_{ub}| = \sqrt{ \frac{ \Delta \mathcal{B}(\bulnu) }{ \tau_B \, \Delta \Gamma(\bulnu) } } \, , \nonumber
\end{align}
with $\tau_B$ denoting the $B$ meson lifetime. For exclusive measurements, the non-perturbative parts of the decay rates can be reliably predicted by lattice QCD~\cite{Aoki:2019cca} or light-cone sum rules~\cite{Bharucha:2012wy} and constrained by the measurements of the decay dynamics. The determination of $|V_{ub}|$ using inclusive decays is very challenging due to the large background from the CKM-favored \bclnu process. Both processes have a very similar decay signature in the form of a high momentum lepton, a hadronic system, and missing energy from the neutrino that escapes detection. Figure~\ref{fig:bulnu_bclnu} shows an illustration of both processes for a $B^0$-meson decay. A clear separation of the processes is only possible in kinematic regions where \bclnu is kinematically forbidden. In these regions, however, non-perturbative shape functions enter the description of the decay dynamics, making predictions for the decay rates dependent on the precise modeling. These functions parametrize at leading order the Fermi motion of the $b$ quark inside the $B$ meson. Properties of the leading-order $\Lambda_{\mathrm{QCD}}/m_b$ shape function can be determined using the photon energy spectrum of $B \to X_s \, \gamma$ decays and moments of the lepton energy or hadronic invariant mass in semileptonic $B$ decays~\cite{Gambino:2004qm,Bauer:2004ve,Benson:2004sg}, but the modeling of both the leading and subleading shape functions introduces large theory uncertainties on the decay rate. 
In the future, more model-independent approaches aim to directly measure the leading-order shape function~\cite{Bernlochner:2020jlt,Gambino:2016fdy}.

As such methods are not yet realized, it is  beneficial to extend the measurement region as much as possible into the \bclnu dominated phase space. This was done, e.g., by Refs.~\cite{Lees:2011fv,Urquijo:2009tp}. This reduces the theory uncertainties on the predicted partial rates~\cite{BLNP, GGOU, DGE1,DGE2,ADFR1,ADFR2}, although making the measurement more prone to systematic uncertainties. This strategy is also adopted in the measurement described in this paper. 

The corresponding world averages of $|V_{ub}|$ from both exclusive and inclusive determinations are~\cite{Amhis:2019ckw}:
\begin{align}
 |V_{ub}^{\mathrm{excl.}}| & =  \left(3.67 \pm 0.09 \pm 0.12 \right) \times 10^{-3} \, , \\
 |V_{ub}^{\mathrm{incl.}}| & = \left( 4.32 \pm 0.12 {}^{+0.12}_{-0.13} \right) \times 10^{-3} \, .
\end{align}
Here the uncertainties are experimental and from theory. Both world averages exhibit a disagreement of about 3 standard deviations between them. This disagreement is limiting the reach of present-day precision tests of the KM mechanism and searches for loop-level new physics, see e.g. Ref.~\cite{Bona:ICHEP2020} for a recent analysis. For a more complete review the interested reader is referred to Refs.~\cite{RevModPhys.88.035008,pdg_Vxb:2020}.

One important experimental method to extend the probed \bulnu phase space into regions dominated by \bclnu\ transitions is the full reconstruction of the second \PB meson of the $e^+ \, e^- \to \Upsilon(4S) \to B \bar B$ process. This process is referred to as ``tagging" and allows for the reconstruction of the hadronic $X$ system of the semileptonic process. In addition, the neutrino four-momentum can be reconstructed. Properties of both are instrumental to distinguish \bulnu and \bclnu processes. In this manuscript the reconstruction of the second \PB meson  and the separation of \bulnu from \bclnu processes were carried out using machine learning approaches. Several neural networks were trained to identify correctly reconstructed tag-side \PB mesons. The distinguishing variables of the classification algorithm were carefully selected in order not to introduce a bias in the measured partial branching fractions. In addition, the modeling of backgrounds was validated in \bclnu enriched selections. We report the measurement of three partial branching fractions, covering 31\% - 86\% of the accessible \bulnu phase space. The measurement of fully differential distributions, which allow one to determine the leading and subleading shape functions, is left for future work. 

The main improvement over the previous Belle result of Ref.~\cite{Urquijo:2009tp} lies in the adoption of a more efficient tagging algorithm for the reconstruction of the second $B$ meson and the improvements of the \bulnu signal and \bclnu background descriptions. In addition, the full Belle data set of 711 fb${}^{-1}$ is analyzed and we avoid the direct use of kinematic properties of the candidate semileptonic decay in the background suppression. After the final selection we retain a factor of approximatively 1.8 times more signal events than the previous analysis with a ca. 20\% improved purity.

The remainder of this manuscript is organized as follows: Section~\ref{sec:data_set_sim_samples} provides an overview of the data set and the simulated signal and background samples, that were used in the analysis. Section~\ref{sec:ana_strategy} details the analysis strategy and reconstruction of the hadronic $X$ system of the semileptonic decay.  Section~\ref{sec:fit} introduces the fit procedure used to separate \bulnu signal from background contributions. Section~\ref{sec:syst} lists the systematic uncertainties affecting the measurements and Section~\ref{sec:sidebands} summarizes sideband studies central to validate the modeling of the crucial \bclnu background processes. Finally, Section~\ref{sec:signal} shows the selected signal events and compares them with the expectation from simulation. In Section~\ref{sec:results}  the measured partial branching fractions and subsequent values of $|\Vub|$ are discussed. Section~\ref{sec:summary} presents our conclusions. 

\section{Data Set and Simulated Samples}\label{sec:data_set_sim_samples}

The analysis utilizes the full Belle data set of \mbox{$(772 \pm 10) \times 10^6$} \PB meson pairs, which were produced at the KEKB accelerator complex~\cite{KEKB} with a center-of-mass energy of $\sqrt{s} = \SI{10.58}{GeV}$ corresponding to the $\Upsilon(4S)$ resonance. In addition, $\SI{79}{fb^{-1}}$ of collision events recorded $\SI{60}{MeV}$ below the $\Upsilon(4S)$ resonance peak are used to derive corrections and for cross-checks.

The Belle detector is a large-solid-angle magnetic spectrometer that consists of a silicon vertex detector, a 50-layer central drift chamber (CDC), an array of aerogel threshold Cherenkov counters (ACC), a barrel-like arrangement of time-of-flight scintillation counters (TOF), and an electromagnetic calorimeter composed of CsI(Tl) crystals (ECL) located inside a superconducting solenoid coil that provides a \SI{1.5}{T} magnetic field. An iron flux return located outside of the coil is instrumented to detect $K^0_L$ mesons and to identify muons (KLM). A more detailed description of the detector and its layout and performance can be found in Ref.~\citep{Abashian:2000cg} and in references therein.

Charged tracks are identified as electron or muon candidates by combining the information of multiple subdetectors into a lepton identification likelihood ratio, $\mathcal{L}_\mathrm{LID}$. For electrons, the most important identifying features are the ratio of the energy deposition in the ECL with respect to the reconstructed track momentum, the energy loss in the CDC, the shower shape in the ECL, the quality of the geometrical matching of the track to the shower position in the ECL, and the photon yield in the ACC~\citep{HANAGAKI2002490}. Muon candidates can be identified from charged track trajectories extrapolated to the outer detector. The most important identifying features are the difference between expected and measured penetration depth as well as the transverse deviation of KLM hits from the extrapolated trajectory~\citep{ABASHIAN200269}. Charged tracks are identified as pions or kaons using a likelihood ratio $\mathcal{L}_\mathrm{K/\pi \, \mathrm{ID}} = \mathcal{L}_\mathrm{K\, \mathrm{ID}} / \left( \mathcal{L}_\mathrm{K \, \mathrm{ID}} + \mathcal{L}_\mathrm{\pi \, \mathrm{ID}} \right)$. The most important identifying features of the kaon ($\mathcal{L}_\mathrm{K \, \mathrm{ID}} $) and pion ($\mathcal{L}_\mathrm{\pi \, \mathrm{ID}} $) likelihoods for low momentum particles with transverse momentum below 1 GeV in the laboratory frame are the recorded energy loss by ionization, $\mathrm{d}E/\mathrm{d}x$, in the CDC, and the time of flight information from the TOF. Higher-momentum kaon and pion classification relies on the Cherenkov light recorded in the ACC. In order to avoid the difficulties in understanding the efficiencies of reconstructing $K^0_L$ mesons, they are not explicitly reconstructed or used in this analysis.

Photons are identified as energy depositions in the ECL, vetoing clusters to which an associated track can be assigned. Only photons with an energy deposition of \mbox{\Eg $ > \SI{100}{MeV}$, $\SI{150}{MeV}$, and $\SI{50}{MeV}$} in the forward endcap, backward endcap and barrel part of the calorimeter, respectively, are considered. We reconstruct $\pi^0$ candidates from photon candidates. The invariant mass is required to fall inside a window\footnote{We use natural units: $\hbar = c = 1$.}  of $m_{\gamma\gamma} \in [0.12, 0.15] \, \text{GeV}$, which corresponds to about 2.5 times the $\pi^0$ mass resolution. 

Monte Carlo (MC) samples of \PB meson decays and continuum processes ($e^+ e^- \to q \bar q$ with $q = u,d,s,c$) are simulated using the \texttt{EvtGen} generator~\citep{EvtGen}. These samples are used to evaluate reconstruction efficiencies and acceptance, and to estimate background contaminations. The sample sizes used correspond to approximately ten and five times, respectively, the Belle collision data for \PB meson and continuum decays. The interactions of particles traversing the detector are simulated using \texttt{Geant3}~\citep{Geant3}. Electromagnetic final-state radiation is simulated using the \texttt{PHOTOS}~\citep{Photos} package for all charged final-state particles. The efficiencies in the MC are corrected using data-driven methods to account for, e.g., differences in identification and reconstruction efficiencies. 

\begin{table}[t!]
\caption{
	Branching fractions for \bulnu\ and \bclnu background processes that were used are listed. More details on the applied corrections can be found in the text. We neglect the small contribution from $B^+ \to D_s^{(*)} \, K^+ \, \ell^+ \, \nu_\ell$ which has a branching fraction of similar size as $B \to D \pi \pi \, \ell^+ \, \nu_\ell$.
}
\label{tab:bfs}
\vspace{1ex}
\begin{tabular}{lcc}
\hline\hline
 $\mathcal{B}$ & Value $B^+$ & Value $B^0$  \\
 \hline
 \bulnu & \\
 \quad \bpilnu & $\left(7.8 \pm 0.3 \right) \times 10^{-5}$ & $\left(1.5 \pm 0.06 \right) \times 10^{-4}$ \\
 \quad \betalnu & $\left(3.9 \pm 0.5 \right)  \times 10^{-5}$ & - \\
 \quad \betaplnu & $\left(2.3 \pm 0.8 \right) \times 10^{-5}$ & - \\
 \quad  \bomegalnu & $\left(1.2 \pm 0.1 \right)  \times 10^{-4}$ & - \\
 \quad  \brholnu & $\left(1.6 \pm 0.1\right)  \times 10^{-4}$ & $\left(2.9 \pm 0.2\right)  \times 10^{-4}$ \\
 \quad  $B \to X_u \, \ell^+ \, \nu_\ell$ & $\left(2.2 \pm 0.3 \right)  \times 10^{-3}$ & $\left(2.0 \pm 0.3 \right)  \times 10^{-3}$ \\
 \hline 
 \bclnu & \\
 \quad $B \to D \, \ell^+ \, \nu_\ell$ & $\left(2.5 \pm 0.1\right) \times 10^{-2} $ & $\left(2.3 \pm 0.1\right)\times 10^{-2} $ \\
 \quad $B \to D^* \, \ell^+ \, \nu_\ell$ & $\left(5.4 \pm 0.1 \right)\times 10^{-2} $ &$\left( 5.1 \pm 0.1 \right)\times 10^{-2} $ \\
  \quad $B \to D_0^* \, \ell^+ \, \nu_\ell$  & $\left(4.2 \pm 0.8\right) \times 10^{-3}$ & $\left(3.9 \pm 0.7\right) \times 10^{-3}$ \\
  \quad\, $(\hookrightarrow D \pi)$ \\ 
  \quad $B \to D_1^* \, \ell^+ \, \nu_\ell$ & $\left(4.2 \pm 0.8\right) \times 10^{-3}$ & $\left(3.9 \pm 0.8\right) \times 10^{-3}$   \\
    \quad\, $(\hookrightarrow D^* \pi)$ \\
  \quad $B \to D_1 \, \ell^+ \, \nu_\ell$ &  $\left(4.2 \pm 0.3\right) \times 10^{-3}$ & $\left(3.9 \pm 0.3\right) \times 10^{-3}$  \\
  \quad\, $(\hookrightarrow D^* \pi)$ \\
  \quad $B \to D_2^* \, \ell^+ \, \nu_\ell$  & $\left(1.2 \pm 0.1\right) \times 10^{-3}$ & $\left(1.1 \pm 0.1\right) \times 10^{-3}$   \\
  \quad\, $(\hookrightarrow D^* \pi)$ \\ 
    \quad $B \to D_2^* \, \ell^+ \, \nu_\ell$  & $\left(1.8 \pm 0.2\right) \times 10^{-3}$ & $\left(1.7 \pm 0.2\right) \times 10^{-3}$   \\
  \quad\, $(\hookrightarrow D \pi)$ \\ 
    \quad $B \to D_1 \, \ell^+ \, \nu_\ell$ &  $\left(2.4 \pm 1.0\right) \times 10^{-3}$ & $\left(2.3 \pm 0.9\right) \times 10^{-3}$  \\
  \quad\, $(\hookrightarrow D \pi \pi)$ \\
   \quad $B \to D \pi \pi \, \ell^+ \, \nu_\ell$ & $\left(0.6 \pm 0.6 \right) \times 10^{-3}$ & $\left(0.6 \pm 0.6 \right) \times 10^{-3}$  \\
   \quad $B \to D^* \pi \pi \, \ell^+ \, \nu_\ell$ & $\left(2.2 \pm 1.0 \right) \times 10^{-3}$ & $\left(2.0 \pm 1.0 \right) \times 10^{-3}$  \\
   \quad $B \to D \eta \, \ell^+ \, \nu_\ell$ & $\left(4.0 \pm 4.0 \right) \times 10^{-3}$ & $\left(4.0 \pm 4.0 \right) \times 10^{-3}$  \\
   \quad $B \to D^{*} \eta \, \ell^+ \, \nu_\ell$ & $\left(4.0 \pm 4.0 \right) \times 10^{-3}$ & $\left(4.0 \pm 4.0 \right) \times 10^{-3}$  \\
     \hline
   \quad $B \to X_c \, \ell^+ \, \nu_\ell$ & $\left(10.8 \pm 0.4\right) \times 10^{-2} $ & $\left(10.1 \pm 0.4\right) \times 10^{-2} $ \\
 \hline\hline
\end{tabular}
\end{table}

The most important background processes are semileptonic \bclnu\ decays and continuum processes, which both can produce high-momentum leptons in a momentum range similar to the \bulnu\ process. The semileptonic background from \bclnu\ decays is dominated by \bdlnu\ and \bdslnu\ decays. The \bdlnu\ decays are modeled using the BGL parametrization~\cite{Boyd:1994tt} with form factor central values and uncertainties taken from the fit in Ref.~\cite{Glattauer:2015teq}. For \bdslnu\, we use the BGL implementation proposed by Refs.~\cite{Grinstein:2017nlq,Bigi:2017njr} with form factor central values and uncertainties from the fit to the measurement of Ref.~\cite{Waheed:2018djm}. Both backgrounds are normalized to the average branching fraction of Ref.~\cite{Amhis:2019ckw} assuming isospin symmetry. Semileptonic \bddslnu decays with $D^{**} = \{ D_0^*, D_1^*, D_1, D_2^* \}$ denoting the four orbitally excited charmed mesons are modeled using the heavy-quark-symmetry-based form factors proposed in Ref.~\cite{Bernlochner:2016bci}. We simulate all $D^{**}$ decays using masses and widths from Ref.~\cite{pdg:2020}. For the branching fractions we adopt the values of Ref.~\cite{Amhis:2019ckw} and correct them to account for missing isospin-conjugated and other established decay modes, following the prescription given in Ref.~\cite{Bernlochner:2016bci}. To correct for the fact that the measurements were carried out in the $D^{**\,0} \to D^{(*)+} \, \pi^-$ decay modes, we account for the missing isospin modes with a factor of
\begin{align}
 f_\pi = \frac{ \mathcal{B}( \overline D^{**\,0} \to D^{(*)\, -} \pi^+) }{  \mathcal{B}( \overline D^{**\,0} \to \overline D^{(*)} \pi) } =  \frac{2}{3} \, .
\end{align}
The measurements of the $B \to D_2^* \, \ell \bar \nu_\ell$ in Ref.~\cite{Amhis:2019ckw} are converted to only account for the $\overline D_2^{*\,0} \to D^{*\, -} \pi^+$ decay. To also account for $\overline D_2^{*\,0} \to D^{-} \pi^+$ contributions, we apply a factor of~\cite{pdg:2020}
\begin{align}
 f_{D_2^*}  = \frac{ \mathcal{B}(\overline D_2^{*\, 0} \to D^{-} \pi^+) }{ \mathcal{B}(\overline D_2^{*\, 0} \to D^{*\, -} \pi^+) } = 1.54 \pm 0.15  \, .
\end{align}
The world average of $B \to D_1^* \, \ell \bar \nu_\ell$ given in Ref.~\cite{Amhis:2019ckw} combines measurements, which show poor agreement, and the resulting probability of the combination is below 0.01\%. Notably, the measurement of Ref.~\cite{Liventsev:2007rb} is in conflict with the measured branching fractions of Refs.~\cite{Aubert:2008ea,Abdallah:2005cx} and with the expectation of $\mathcal{B}(B \to D_1^* \, \ell \bar \nu_\ell)$ being of similar size than $\mathcal{B}(B \to D_0 \, \ell \bar \nu_\ell)$~\cite{Leibovich:1997em,Bigi:2007qp}. We perform our own average excluding Ref.~\cite{Liventsev:2007rb} and use
\begin{align}
 \mathcal{B}(B^+ \to \overline D_1^{*\, 0}(\to D^{*\,-} \pi^+) \, \ell^+ \nu_\ell) & = \left(0.28 \pm 0.06 \right) \times 10^{-2} \, .
\end{align}
The world average of $B \to D_1 \, \ell \bar \nu_\ell$ does not include contributions from prompt three-body decays of $D_1 \to D \pi \pi$. We account for these using a factor~\cite{Aaij:2011rj}
\begin{align}
 f_{D_1} = \frac{ \mathcal{B}(\overline D_1^0 \to D^{* \, -} \pi^+ )}{\mathcal{B}( \overline D_1^0 \to \overline D^{0} \pi^+ \pi^- )} = 2.32 \pm 0.54 \, .
\end{align}
We subtract the contribution of $D_1 \to D \pi \pi$ from the measured non-resonant plus resonant $B \to D \pi \pi \ell \bar \nu_\ell$ branching fraction of Ref.~\cite{Lees:2015eya}. To account for missing isospin-conjugated modes of the three-hadron final states we adopt the prescription from Ref.~\cite{Lees:2015eya}, which calculates an average isospin correction factor of 
\begin{align}
 f_{\pi\pi} &= \frac{ \mathcal{B}( \overline D^{**\,0} \to \overline D^{(*)\, 0} \pi^+ \pi^-) }{ \mathcal{B}( \overline D^{**\,0} \to \overline D^{(*)} \pi \pi) } = \frac{1}{2} \pm \frac{1}{6} \, .
\end{align}
The uncertainty takes into account the full spread of final states ($f_0(500) \to \pi \pi$ or $\rho \to \pi \pi$ result in $f_{\pi \pi} = 2/3$ and $1/3$, respectively) and the non-resonant three-body decays ($f_{\pi\pi} = 3/7$). We further assume that
\begin{align}
 \mathcal{B}(\overline D_2^{*} \to \overline D \pi) + \mathcal{B}(\overline D_2^{*} \to \overline D^* \pi) = 1 \, , \nonumber \\
 \mathcal{B}(\overline D_1 \to  \overline D^{*} \pi) + \mathcal{B}(\overline D_1 \to \overline D \pi\pi)  = 1 \, , \nonumber \\
  \mathcal{B}(\overline D_1^{*} \to \overline D^{*} \pi) = 1 \, , \quad \text{and} \quad
  \mathcal{B}(\overline D_0 \to  \overline D \pi)  = 1 \,  \, .
\end{align}
For the remaining $B \to D^{(*)} \, \pi \, \pi \, \ell^+ \, \nu_\ell$ contributions we use the measured value of Ref.~\cite{Lees:2015eya}. The remaining ``gap" between the sum of all considered exclusive modes and the inclusive \bclnu branching fraction ($\approx 0.8 \times 10^{-2}$ or 7-8\% of the total \bclnu branching fraction) is filled in equal parts with $B \to D \, \eta \, \ell^+ \, \nu_\ell$ and $B \to D^{*} \, \eta \, \ell^+ \, \nu_\ell$ and we assume a 100\% uncertainty on this contribution. We simulate \mbox{$B \to D^{(*)} \, \pi \, \pi \, \ell^+  \nu_\ell$} and \mbox{$B \to D^{(*)} \, \eta  \, \ell^+  \nu_\ell$} final states assuming that they are produced by the decay of two broad resonant states $D^{**}_{\mathrm{gap}}$ with masses and widths identical to $D_1^{*}$ and $D_0$. Although there is currently no experimental evidence for decays of charm $1P$ states into these final states or the existence of such an additional broad state (e.g. a $2S$) in semileptonic transitions, this description provides a better kinematic description of the initial three-body decay, $B \to D^{**}_{\mathrm{gap}} \, \ell \bar \nu_\ell$, than e.g. a model based on the equidistribution of all final-state particles in phase space. For the form factors we adapt Ref.~\cite{Bernlochner:2016bci}.

Semileptonic \bulnu decays are modeled as a mixture of specific exclusive modes and non-resonant contributions. We normalize their corresponding branching fractions to the world averages from Ref.~\cite{pdg:2020}: semileptonic \bpilnu\ decays are simulated using the BCL parametrization~\citep{Bourrely:2008za} with form factor central values and uncertainties from the global fit carried out by Ref.~\citep{Lattice:2015tia}. The processes of \brholnu\ and \bomegalnu\ are modeled using the BCL form factor parametrization. We use the fit of Ref.~\cite{Bernlochner:2021rel}, that combines the measurements of Refs.~\cite{Sibidanov:2013rkk,Lees:2012mq,delAmoSanchez:2010af} with the light-cone sum rule predictions of Ref.~\cite{Bharucha:2012wy} to determine a set of form factor central values and uncertainties. The processes of \betalnu and \betaplnu are modeled using the LCSR calculation of Ref.~\cite{Duplancic:2015zna}. For the uncertainties we assume for these states that the pole-parameters $\alpha^{+/0}$ and the form factor normalization $f_{B\eta}^+(0)$ at maximum recoil can be treated as uncorrelated. In addition to these narrow resonances, we simulate non-resonant \bulnu\ decays with at least two pions in the final state following the DFN model~\cite{DeFazio:1999ptt}. The triple differential rate of this model is a function of the four-momentum-transfer squared ($q^2$), the lepton energy ($E_\ell^B$) in the \PB rest-frame, and the hadronic invariant mass squared ($M_X^2$) of the $X_u$ system at next-to-leading order precision in the strong coupling constant $\alpha_s$. This triple differential rate is convolved with a non-perturbative shape function using an ad-hoc exponential model. The free parameters of the model are the $b$ quark mass in the Kagan-Neubert scheme~\cite{Kagan:1998ym}, $m_{b}^{\text{KN}} = (4.66 \pm 0.04)\,\mathrm{GeV}$ and a non-perturbative parameter $a^{\text{KN}} = 1.3 \pm 0.5$. The values of these parameters were determined in Ref.~\cite{Buchmuller:2005zv} from a fit to $\bclnu$ and $B \to X_s \gamma$ decay properties. At leading order, the non-perturbative parameter $a^{\text{KN}}$ is related to the average momentum squared of the $b$ quark inside the $B$ meson and determines the second moment of the shape function. It is defined as $a^{\text{KN}} = - 3 \overline \Lambda^2 / \lambda_1 -1$ with the binding energy $\overline \Lambda = m_B - m_b^{\text{KN}}$ and the kinetic energy parameter $\lambda_1$. The hadronization of the parton-level \bulnu DFN simulation is carried out using the JETSET algorithm~\cite{SJOSTRAND199474}, producing final states with two or more mesons. The inclusive and exclusive \bulnu\ predictions are combined using a so-called `hybrid' approach, which is a method originally suggested by Ref.~\cite{hybrid}, and our implementation closely follows Ref.~\cite{Prim:2019gtj} and uses the library of Ref.~\cite{markus_prim_2020_3965699}. To this end, we combine both predictions such that the partial branching fractions in the triple differential rate of the inclusive ($ \Delta \mathcal{B}_{ijk}^{\rm incl}$) and combined exclusive ($ \Delta \mathcal{B}_{ijk}^{\rm excl}$) predictions reproduce the inclusive values. This is achieved by assigning weights to the inclusive contributions $w_{ijk}$ such that
\begin{equation}
\begin{aligned}
 \Delta \mathcal{B}_{ijk}^{\rm incl} = & \,   \Delta \mathcal{B}_{ijk}^{\rm excl} + w_{ijk} \times  \Delta \mathcal{B}_{ijk}^{\rm incl} \, ,
\end{aligned}
\end{equation}
with $i,j,k$ denoting the corresponding bin in the three dimensions of $q^2$, $E_\ell^B$, and $M_X$:
\begin{equation}
\begin{aligned}
 q^2 & = [0,2.5,5,7.5,10,12.5,15,20,25] \, \text{GeV}^2 \, , \nn \\
 E_\ell^B & = [0,0.5,1,1.25,1.5,1.75,2,2.25,3] \, \text{GeV} \, , \nn \\
 M_X & = [0,1.4,1.6,1.8,2,2.5,3,3.5]  \, \text{GeV}  \, .
\end{aligned}
\end{equation}
To study the model dependence of the DFN shape function, we also determine weights using the BLNP model of Ref.~\cite{Lange:2005yw} and treat the difference later as a systematic uncertainty. For the $b$ quark mass in the shape-function scheme we use $m_{b}^{\mathrm{SF}} = 4.61\,\mathrm{GeV}$ and $\mu_{\pi}^{2\, \text{SF}} = 0.20 \,\mathrm{GeV}^2$. Figures detailing the hybrid model construction can be found in Appendix A. 

\begin{figure}[ht!]
  \includegraphics[width=0.48\textwidth]{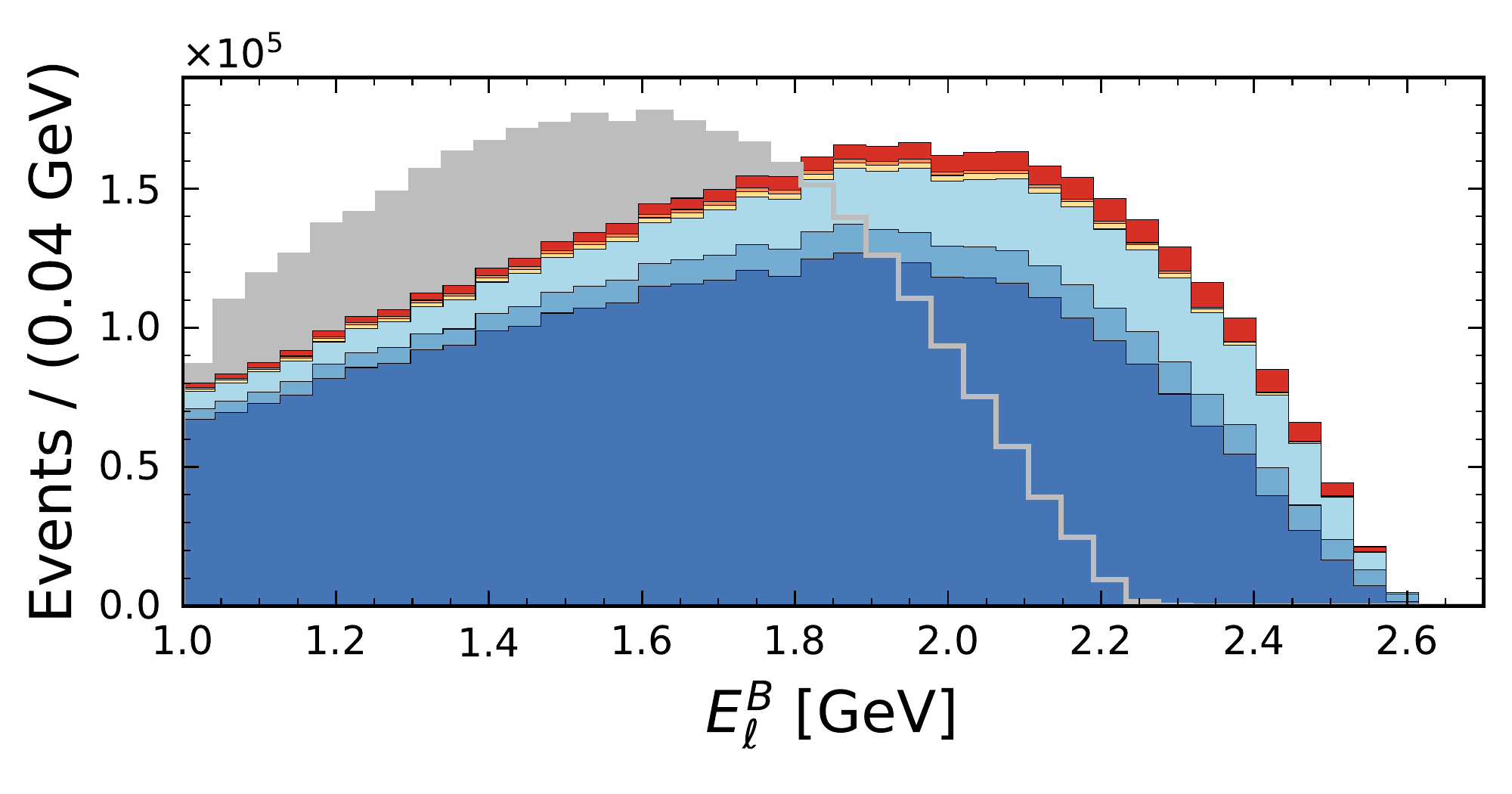} 
  \includegraphics[width=0.48\textwidth]{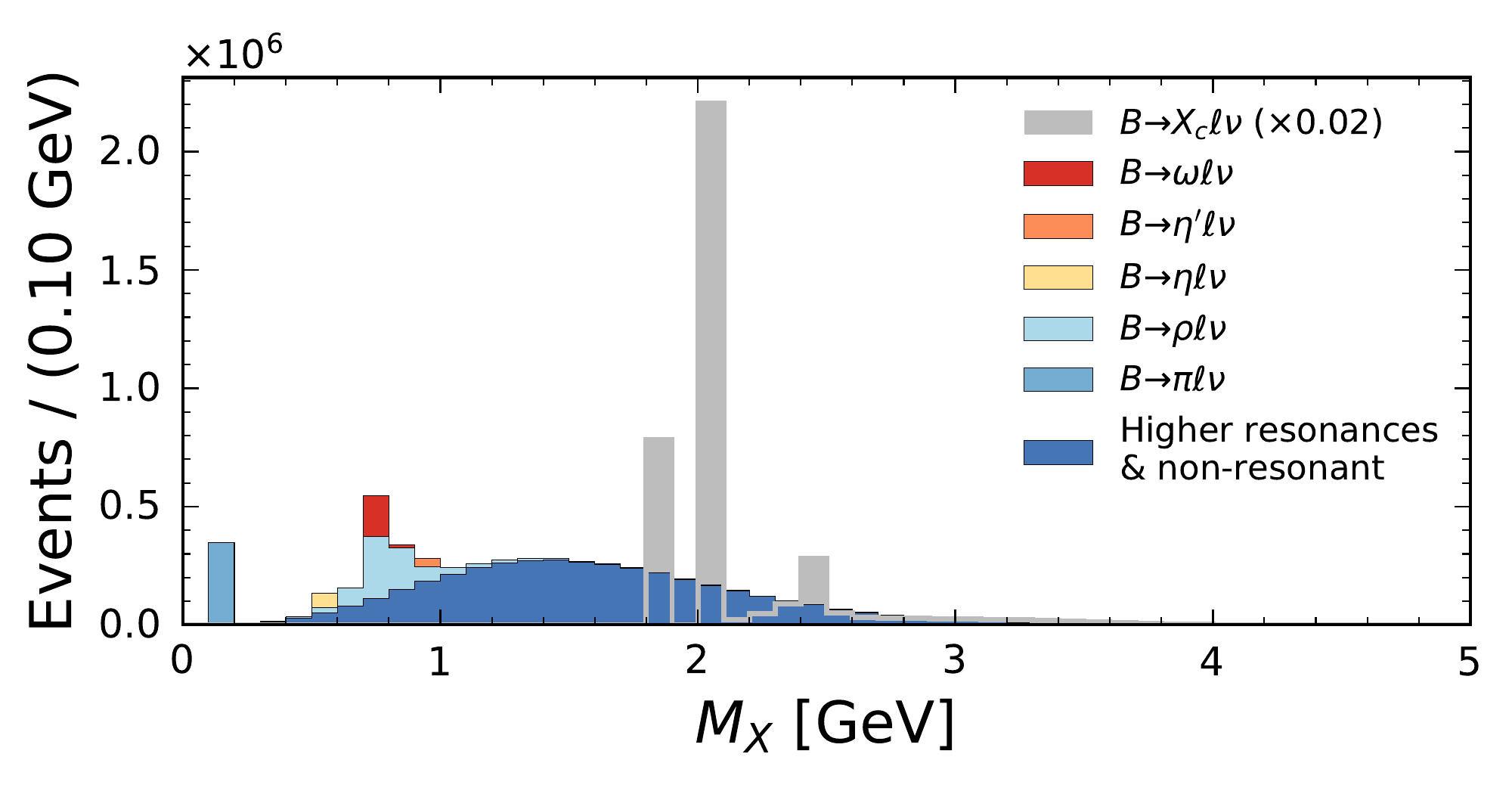} 
\caption{
The generator-level $E_\ell^B$ and $M_X$ distributions of the CKM suppressed and favored inclusive semileptonic processes, \bulnu (scaled up by a factor of 50) and \bclnu, respectively, are shown, using the models described in the text.
 }
\label{fig:bulnu_bclnu_mX_El}
\end{figure}

Table~\ref{tab:bfs} summarizes the branching fractions for the signal and the important \bclnu\ background processes that were used. Figure~\ref{fig:bulnu_bclnu_mX_El} shows the generator-level distributions and yields of \bclnu\ and \bulnu\ after the tag-side reconstruction (cf. Section~\ref{sec:ana_strategy}). The \bulnu\ yields were scaled up by a factor of 50 to make them visible. A clear separation can be obtained at low values of $M_X$ and high values of $E_\ell^B$.

\section{Analysis Strategy, Hadronic Tagging, and $X$ Reconstruction}\label{sec:ana_strategy}

\subsection{Neural Network Based Tag Side Reconstruction}

We reconstruct collision events using the hadronic full reconstruction algorithm of Ref.~\cite{Feindt:2011mr}. The algorithm reconstructs one of the \PB mesons produced in the collision event using hadronic decay channels. We label such \PB mesons in the following as $B_{\mathrm{tag}}$. Instead of attempting to reconstruct as many \PB meson decay cascades as possible, the algorithm employs a hierarchical reconstruction ansatz in four stages: at the first stage, neural networks are trained to identify charged tracks and neutral energy depositions as detector stable particles ($e^+, \mu^+, K^+, \pi^+, \gamma$), neutral $\pi^0$ candidates, or $K_S^0$ candidates. At the second stage, these candidate particles are combined into heavier meson candidates ($J/\psi, D^0, D^{+}, D_s$) and for each target final state a neural network is trained to identify probable candidates. In addition to the classifier output from the first stage, vertex fit probabilities of the candidate combinations, and the full four-momentum of the combination are passed to the input layer. At the third stage, candidates for $D^{*\, 0}, D^{*\, +}$, and $D_s^*$ mesons are formed and separate neural networks are trained to identify viable combinations. The input layer aggregates the output classifiers from all previous reconstruction stages. The final stage combines the information from all previous stages to form $B_{\mathrm{tag}}$ candidates. The viability of such combinations is again assessed by a neural network that was trained to distinguish correctly reconstructed candidates from wrong combinations and whose output classifier score we denote by $\mathcal{O}_{\mathrm{FR}}$. Over 1104 decay cascades are reconstructed in this manner, achieving an efficiency of 0.28\% and 0.18\% for charged and neutral \PB meson pairs~\cite{Bevan:2014iga}, respectively. Finally, the output of this classifier is used as an input and combined with a range of event shape variables to train a neural network to distinguish reconstructed \PB meson candidates from continuum processes. The output classifier score of this neural network is denoted as $\mathcal{O}_{\mathrm{Cont}}$. Both classifier scores are mapped to a range of $[0,1)$ signifying the reconstruction quality of poor to excellent candidates. We retain $B_{\mathrm{tag}}$ candidates that show at least moderate agreement based on these two outputs and require that $\mathcal{O}_{\mathrm{FR}} > 10^{-4}$ and  $\mathcal{O}_{\mathrm{Cont}} > 10^{-4}$. Despite these relatively low values, knowledge of the charge and momentum of the decay constituents in combination with the known beam-energy allows one to infer the flavor and four-momentum of the $B_{\mathrm{tag}}$ candidate. We require the $B_{\mathrm{tag}}$ candidates to have at least a beam-constrained mass of
\begin{align}
 M_{\mathrm{bc}} = \sqrt{ E_{\mathrm{beam}}^2 - | \bold{p}_{\mathrm{tag}}|^2 }  > 5.27 \, \text{GeV} \, ,
\end{align}
with $ \bold{p}_{\mathrm{tag}}$ denoting the momentum of the $B_{\mathrm{tag}}$ candidate in the center-of-mass frame of the colliding $e^+e^-$-pair. Furthermore, $E_{\mathrm{beam}} = \sqrt{s}/2$ denotes half the center-of-mass energy of the colliding $e^+e^-$-pair. The energy difference
\begin{align}
  \Delta E = E_{\mathrm{tag}} - E_{\mathrm{beam}} \, ,
\end{align}
is already used in the input layer of the neural network trained in the final stage of the reconstruction. Here $E_{\mathrm{tag}}$ denotes the energy of the $B_{\mathrm{tag}}$ candidate in the center-of-mass frame of the colliding $e^+e^-$-pair. In each event a single $B_{\mathrm{tag}}$ candidate is then selected according to the highest $\mathcal{O}_{\mathrm{FR}}$ score of the hierarchical full reconstruction algorithm. All tracks and clusters not used in the reconstruction of the $B_{\mathrm{tag}}$ candidate are used to define the signal side. 

\subsection{Signal Side Reconstruction}

\begin{figure}[th!]
  \includegraphics[width=0.45\textwidth]{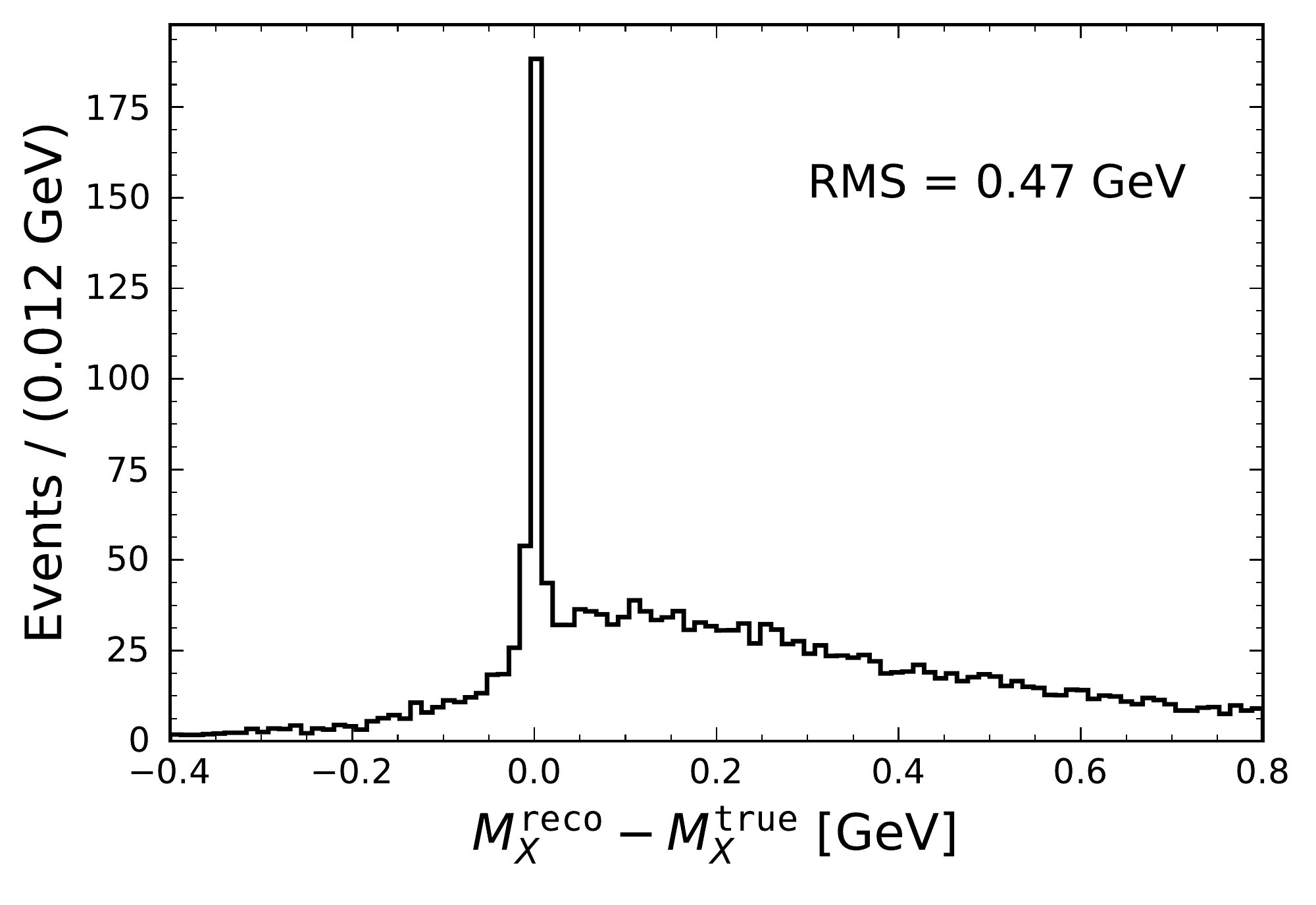} 
  \includegraphics[width=0.45\textwidth]{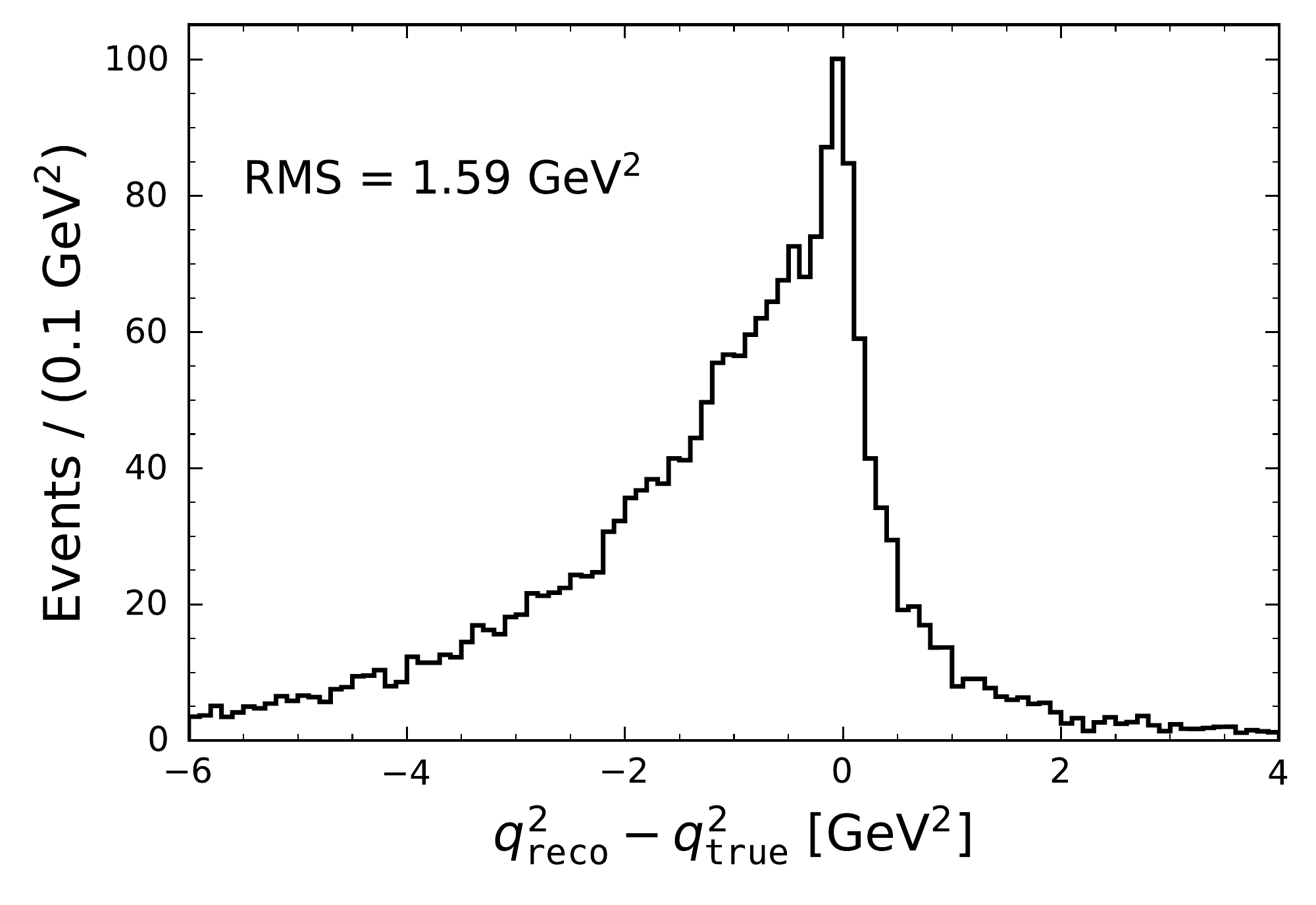} 
\caption{
The resolution of the reconstructed $M_X$ and $q^2$ values for \bulnu\ signal is shown as a residual with respect to the generated values. 
 }
\label{fig:mX_q2_resolution}
\end{figure}

The signal side of the event is reconstructed by identifying a well-reconstructed lepton with \mbox{$E_\ell^B = |\bold{p}_{\ell}^B| > 1 \, \mathrm{GeV}$} in the signal $B$ rest frame\footnote{We neglect the small correction of the lepton mass term to the energy of the lepton.} using the likelihood mentioned in Section~\ref{sec:data_set_sim_samples}. The signal $B$ rest frame is calculated using the momentum of the $B_{\mathrm{tag}}$ candidate via
\begin{align}
p_{\mathrm{sig}} =  p_{e^+ \, e^-}  - \left( \sqrt{m_B^2 + |\bold{p_{\mathrm{tag}}}|^2 } , \bold{p_{\mathrm{tag}}} \right) \, ,
\end{align}
with $p_{e^+ e^-}$ denoting the four-momentum of the colliding electron-positron pair. Leptons from $J/\psi$ and photon conversions in detector material are rejected by combining the lepton candidate with oppositely charged tracks ($t$) on the signal side and demanding that $m_{\ell t} > 0.14 \, \text{GeV}$ and \mbox{$m_{e t} \notin [3.05,3.15] \, \text{GeV}$} or \mbox{$m_{\mu t} \notin [3.06,3.12] \, \text{GeV}$}. If multiple lepton candidates are present on the signal side, the event is discarded as multiple leptons are likely to originate from a double semileptonic $b \to c \to s$ cascade. For charged  $B_{\mathrm{tag}}$ candidates, we demand that the charge assignment of the signal-side lepton be opposite that of the $B_{\mathrm{tag}}$ charge. The hadronic $X$ system is reconstructed from the remaining unassigned charged particles and neutral energy depositions. Its four momentum is calculated as
\begin{align}
  p_X = \sum_i \left( \sqrt{m_\pi^2 + |\bold{p}_i|^2 } , \bold{p}_i \right) + \sum_j \left( E_j , \bold{k}_j \right)  \, ,
\end{align}
with $E_i = |\bold{k_i}|$ the energy of the neutral energy depositions and all charged particles with momentum $\bold{p_i}$ are assumed to be pions. With the $X$ system reconstructed, we can also reconstruct the missing mass squared,
\begin{align}
 M_{\mathrm{miss}}^2 = \left(p_{\mathrm{sig}} - p_X - p_\ell  \right)^2 \, ,
\end{align}
which should peak at zero, $ M_{\mathrm{miss}}^2 \approx m_\nu^2 \approx 0\, \text{GeV}^2$, for correctly reconstructed semileptonic \bulnu\ and \bclnu\ decays. The hadronic mass of the $X$ system is later used to discriminate \bulnu\ signal decays from \bclnu and other remaining backgrounds. It is reconstructed using 
\begin{align}
 M_X = \sqrt{ \left( p_X \right)^\mu \left( p_X \right)_\mu } \, .
\end{align} 
In addition, we reconstruct the four-momentum-transfer squared, $q^2$, as
\begin{align}
  q^2 = \left(  p_{\mathrm{sig}} - p_X \right)^2 \, .
\end{align}
The resolution of both variables for \bulnu is shown in Figure~\ref{fig:mX_q2_resolution} as residuals with respect to the generated values of $q^2$ and $M_X$. The resolution for $M_X$ has a root-mean-square (RMS) deviation of $0.47 \, \text{GeV}$, but exhibits a large tail towards larger values. The distinct peak at 0 is from \bpichlnu and other low-multiplicity final states comprised of only charged pions. The four-momentum-transfer squared $q^2$ exhibits a large resolution, which is caused by a combination of the tag-side $B$ and the $X$ reconstruction. The RMS deviation for $q^2$ is $1.59 \, \text{GeV}^2$. The core resolution is dominated by the tagging resolution, whereas the large negative tail is dominated from the resolution of the reconstruction of the $X$ system.

\subsection{Background Suppression BDT}\label{sec:bkg_bdt}

At this point in the reconstruction, the \bclnu\ process completely dominates the selected events. To identify \bulnu, we combine several distinguishing features into a single discriminant. This is achieved by using a machine learning based classification with boosted decision trees (BDTs). Note that all momenta are in the center-of-mass frame of the colliding $e^+ e^-$-pair. These features are:
\begin{enumerate}
 \item $M_{\mathrm{miss}}^2$: The average \bclnu\ multiplicity is higher than \bulnu, broadening the missing mass squared distribution.
 \item $D^*$ veto: We search for low momentum neutral and charged pions in the $X$ system with \mbox{$| \bold{p}_\pi| < 220$ MeV}, compatible with a $D^* \to D \pi$ transition. The key idea of this is that due to the small available phase space from the small mass difference between the $D^*$ and $D$ mesons, the flight direction of the slow pion is strongly correlated with the $D^*$ momentum direction. The energy and momentum of a $D^*$ candidate can thus be approximated as
 \begin{align}
   E_{D^*} & = \frac{m_{D^*}}{m_{D^*} - m_D} \times E_{\pi} \, , \nonumber \\
    \bold{p}_{D^*} & = \bold{p_\pi} \times \frac{\sqrt{  E_{D^*}^2 - m_{D^*}^2}}{|\bold{p}_\pi|}  \, ,
 \end{align}
 with $m_{D^*}$ and $m_D$ denoting the $D^*$ and $D$ meson masses, respectively, and $E_{\pi} = \sqrt{  m_{\pi}^2 + |\bold{p}_\pi|^2 }$ is the energy of the slow pion. Using the $D^*$ candidate four momentum $p_{D^*} = (E_{D^*}, \bold{p}_{D^*} )$ we can calculate 
 \begin{align}
  M_{\mathrm{miss}, D^*}^2 & = \left(p_{\mathrm{sig}} - p_{D^*} - p_\ell  \right)^2 \, , \nonumber \\
  \cos \theta_{B, D^*\ell} & = \frac{ 2 E_{\mathrm{beam}} E_{D^*\ell} - m_B^2 - m_{D^*\ell}^2 }{ 2 |\bold{p}_{B}| | \bold{p}_{D^*\ell}| }\, , \nonumber \\
  \cos \theta^* & = \frac{ \bold{p}_\ell \cdot \bold{p}_{D^*} }{ |\bold{p}_\ell| |\bold{p}_{D^*}| } \, ,
 \end{align}
with $p_{D^* \ell} = p_{D^*} + p_\ell = (E_{D^*\ell}, \bold{p}_{D^*\ell})$ and \mbox{$ |\bold{p}_{B}| = \sqrt{E_B^2 - m_B^2 }$}. These three variables are used exclusively for events with charged and neutral slow pion candidates.
\item Kaons: We identify the number of $K^+$ candidates using the particle-identification likelihood, cf. Section~\ref{sec:data_set_sim_samples}. In addition, we reconstruct $K_S^0$ candidates from displaced tracks found in the $X$ system. 
\item $B_{\mathrm{sig}}$ vertex fit: The charmed mesons produced in \bclnu\ transitions exhibit a longer lifetime than their charmless counterparts produced in \bulnu\ decays. This can be exploited by carrying out a vertex fit using the lepton and all charged constituents, not identified as kaons, of the $X$ system and we use its $\chi^2$ value as a discriminator. 
\item $Q_{\mathrm{tot}}$: The total event charge as calculated from the $X$ system plus lepton on the signal and from the $B_{\mathrm{tag}}$ constituents. Due to the larger average multiplicity of \bclnu, the expected net zero event charge is more often violated in comparison to \bulnu candidate events. 
\end{enumerate}
We use the BDT implementation of Ref.~\cite{Chen:2016:XST:2939672.2939785} and train a classifier $\mathcal{O}_{\mathrm{BDT}}$ with simulated \bulnu\ and \bclnu\ events, which we discard in the later analysis. Ref.~\cite{Chen:2016:XST:2939672.2939785} uses optimized boosting and pruning procedures to maximize the classification performance. We choose a selection criteria on $\mathcal{O}_{\mathrm{BDT}}$ that rejects 98.7\% of \bclnu\ and retains 18.5\% of \bulnu\ signal. This working point was chosen by maximizing the significance of the most inclusive partial branching fraction, taking into account the full set of systematic uncertainties and the full analysis procedure. The stability of the result as a function of the BDT selection is further discussed in Section~\ref{sec:results}. 
 
 \begin{figure}[t!]
  \includegraphics[width=0.48\textwidth]{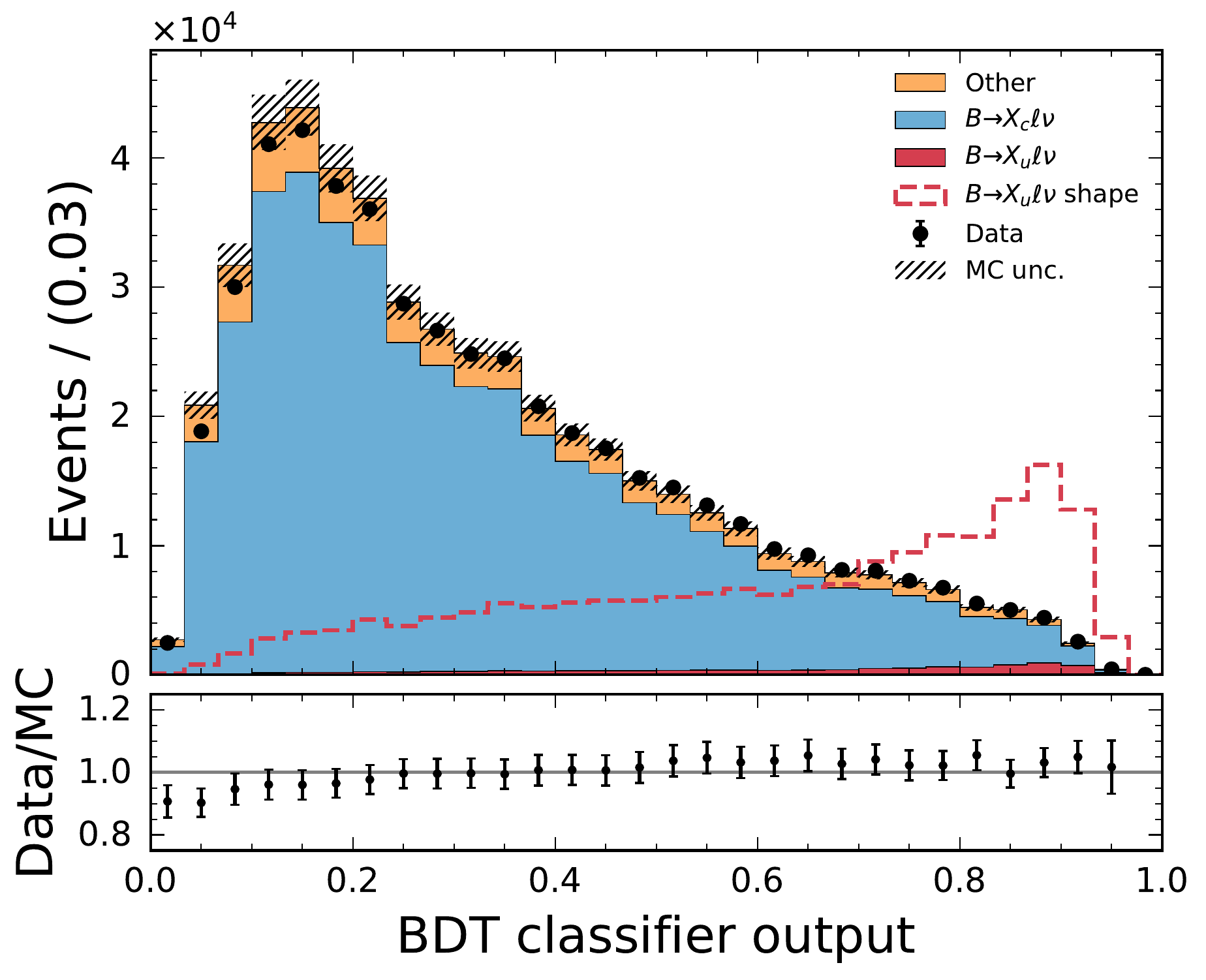} 
\caption{
The shape of the background suppression classifier $\mathcal{O}_{\mathrm{BDT}}$ is shown. MC is divided into \bulnu\ signal, the dominant \bclnu\ background, and all other contributions. To increase visibility, the \bulnu\ component is shown with a scaling factor (red dashed line). The uncertainties on the MC contain the full systematic errors and are further discussed in Section~\ref{sec:syst}.
 }
\label{fig:bdtoutput}
\end{figure}

Table~\ref{tab:sel} lists the efficiencies for signal and \bclnu background for the $M_{\mathrm{bc}}$ and the BDT selections. Figure~\ref{fig:bdtoutput} shows the output classifier of the background suppression BDT for MC and data. The classifier output shows good agreement between simulated and observed data, with the exception of the first two signal depleted bins. A comparison of the shape of all input variables for \bulnu\ and \bclnu, and further MC and data comparisons can be found in Appendix B.

\begin{table}[t!]
\caption{
	The selection efficiencies for \bulnu signal, \bclnu\, and for data are listed after the reconstruction of the $B_{\mathrm{tag}}$ and lepton candidate. The nominal selection requirement on the BDT classifier $\mathcal{O}_{\mathrm{BDT}}$ is 0.85. The other two requirements were introduced to test the stability of the result, cf. Section~\ref{sec:results}.
}
\label{tab:sel}
\vspace{1ex}
\begin{tabular}{lcccc}
\hline\hline
  Selection & \bulnu\ & \bclnu\ & Data \\
 \hline
  $M_{\mathrm{bc}} > 5.27 \, \text{GeV}$& 84.8\% & 83.8\% & 80.2\% \\
  $\mathcal{O}_{\mathrm{BDT}}> 0.85$ & 18.5\% & 1.3\% & 1.6\%  \\ \hline
  $\mathcal{O}_{\mathrm{BDT}}> 0.83$ & 21.9\% & 1.7\% & 2.1\% \\ 
  $\mathcal{O}_{\mathrm{BDT}}> 0.87$ & 14.5\% & 0.9\% & 1.1\% \\ 
 \hline\hline
\end{tabular}
\end{table}

\subsection{Tagging Efficiency Calibration}\label{sec:tagg_eff_corr}

The reconstruction efficiency of the hadronic full reconstruction algorithm of Ref.~\cite{Feindt:2011mr} differs between simulated samples and the reconstructed data. This difference mainly arises due to imperfections, e.g. in the simulation of detector responses, particle identification efficiencies, or incorrect branching fractions in the reconstructed decay cascades. To address this, the reconstruction efficiency is calibrated using a data-driven approach and we follow closely the procedure outlined in Ref.~\cite{Glattauer:2015teq}. We reconstruct full reconstruction events by requiring exactly one lepton on the signal side, and apply the same $B_{\mathrm{tag}}$ and lepton selection criteria outlined in the previous section. This \bxlnu\ enriched sample is divided into groups of subsamples according to the $B_{\mathrm{tag}}$ decay channel and the multivariate classifier output $\mathcal{O}_{\mathrm{FR}}$ used in the hierarchical reconstruction. Each of these groups of subsamples is studied individually to derive a calibration factor for the hadronic tagging efficiency: the calibration factor is obtained by comparing the number of inclusive semileptonic $B$-meson decays, $N(\bxlnu)$, in data with the expectation from the simulated samples, $N^{\mathrm{MC}}(\bxlnu)$. The semileptonic yield is determined via a binned maximum likelihood fit using the the lepton energy spectrum. To reduce the modeling dependence of the \bxlnu\ sample this is done in a coarse granularity of five bins. The calibration factor of each these groups of subsamples is given by 
\begin{align}
 C_{\mathrm{tag}}(B_{\mathrm{tag}} \, \text{mode}, \mathcal{O}_{\mathrm{FR}}) & = \frac{ N(\bxlnu) }{ N^{\mathrm{MC}}(\bxlnu)} \, .
\end{align}
The free parameters in the fit are the yield of the semileptonic \bxlnu decays, the yield of backgrounds from fake leptons and the yield of backgrounds from true leptons. Approximately 1200 calibration factors are determined this way. The leading uncertainty on the $C_{\mathrm{tag}}$ factors is from the assumed \bxlnu composition and the lepton PID performance, cf. Section~\ref{sec:syst}. We also apply corrections to the continuum efficiency. These are derived by using the off-resonance sample and comparing the number of reconstructed off-resonance events in data with the simulated on-resonance continuum events, correcting for differences in the selection. 

\section{Fitting Procedure}\label{sec:fit}

\begin{table}[b!]
\caption{
	The binning choices of the five fit scenarios are given. 
}
\label{tab:bins}
\vspace{1ex}
\begin{tabular}{lc}
\hline\hline
  Fit variable & Bins  \\
 \hline
 $M_X$ & $[0,1.5, 1.9, 2.5, 3.1, 4.0] \, \text{GeV}$ \\
 $q^2 $ & $[0, 2, 4, 6, 8, 10, 12, 14, 26] \, \text{GeV}^2$\\
$E_\ell^B$ & 15 equidist. bins in $[1, 2.5] \, \text{GeV}$ \& $[2.5, 2.7] \, \text{GeV}$ \\
 $M_X:q^2$ & $ [0,1.5]  \, \text{GeV}$ $\times [0, 2, 4, 6, 8, 10, 12, 14, 26]  \, \text{GeV}^2$ \\ 
 &$ [1.5,1.9]  \, \text{GeV}$ $\times [0, 2, 4, 6, 26]  \, \text{GeV}^2$\ \\
 &$ [1.9,2.5]  \, \text{GeV}$ $\times [0, 2, 4, 26]  \, \text{GeV}^2$\ \\
 &$ [2.5,4.0]  \, \text{GeV}$ $\times [0, 2, 26]  \, \text{GeV}^2$\ \\
 \hline\hline
\end{tabular}
\end{table}

After the selection, we retain 9875 events. In order to determine the \bulnu signal yield and constrain all backgrounds, we perform a binned likelihood fit of these events in several discriminating variables. To reduce the dependence on the precise modeling of the \bulnu signal, we use coarse bins over regions that are very sensitive to the admixture of resonant and non-resonant decays, cf.~Section~\ref{sec:data_set_sim_samples}, and explore different variables for the signal extraction. The total likelihood function is constructed as the product of individual Poisson distributions $\mathcal{P}$,
\begin{align}\label{eq:likelihood}
 \mathcal{L} =  \prod_i^{\rm bins} \, \mathcal{P}\left( n_i ; \nu_i \right)  \,  \times \prod_k \, \mathcal{G}_k \, ,
\end{align}
with $n_i$ denoting the number of observed data events and $\nu_i$ the total number of expected events in a given bin $i$. Here, $\mathcal{G}_k$ are nuisance-parameter (NP) constraints, whose role is to incorporate systematic uncertainties of a source $k$ into the fit. Their construction is further discussed in Section~\ref{sec:syst}. The number of expected events in a given bin, $\nu_i$, is estimated using simulated collision events and is given by
\begin{align} \label{eq:nui}
 \nu_i = \sum_k^{\mathrm{component}} \, f_{ik} \, \eta_k \, ,
\end{align}
with $\eta_k$ denoting the total number of events from a given fit component $k$, and $f_{ik}$ denoting the fraction of such events being reconstructed in bin $i$ as determined by the MC simulation. The three fit components we determine are:
\begin{itemize}
 \item[a)] Signal \bulnu events that fall inside a phase-space region for a partial branching fraction we wish to determine.
 \item[b)] Signal \bulnu events that fall outside said region if applicable. This component can have very similar shapes as other backgrounds. We thus constrain this component in all fits to its expectation using the world average of $\mathcal{B}(\bulnu) = \left(  2.13 \pm 0.30 \right) \times 10^{-3}$~\cite{pdg:2020}. We also investigated different approaches: for instance linking this component with the component of a). This leads to small shifts of $\mathcal{O}(0.3 - 1\%)$ of the reported partial branching fractions using this component.
 \item[c)] Background events; such are dominated by \bclnu and other decays that produce leptons in the final state (e.g. from $B \to h_1 h_2$ and $h_2 \to h_3 \ell^- \nu$  with $h_1, h_2$, and $h_3$ denoting hadronic final states). Other contributions are from misidentified lepton candidates and a small amount of continuum processes. A full description of all background processes is given in Section~\ref{sec:ana_strategy}. 
 \end{itemize}
We carry out five separate fits to measure three partial branching fractions, using different discriminating variables to determine the \bulnu\ yield. The fits and variables are:
\begin{enumerate}
  \item The hadronic mass, $M_X$: Signal is expected to pre-dominantly populate the low hadronic mass region, whereas remaining \bclnu\ background will produce a sharp peak at around $M_X \approx 2 \, \text{GeV}$. The sizeable resolution on the reconstruction of the $X$ system will result in a non-negligible amount of these backgrounds to also be present in the low and high $M_X$ region. The determined signal yields are used to measure the partial branching fraction of $M_X < 1.7 \, \mathrm{GeV}$ and $E_\ell^B > 1 \, \text{GeV}$. We thus use two signal templates and split events according to generator-level $M_X < 1.7 \, \mathrm{GeV}$ and $M_X > 1.7 \, \mathrm{GeV}$.
    \item The four-momentum-transfer squared, $q^2$: Signal will on average have a higher $q^2$ than \bclnu\ background, whose kinematic endpoint is $q^2 = \left( m_B - m_D \right)^2 \approx 11.6 \, \text{GeV}^2$. However, the reconstructed $q^2$ of \bclnu events is smeared over the entire kinematic range due to the sizeable resolution in the reconstruction of the inclusive $X$ system and the $B_{\mathrm{tag}}$ reconstruction. To reduce background from \bclnu events, we apply a cut on the reconstructed $M_X$ and require a value smaller than $1.7 \, \mathrm{GeV}$. The determined signal yields are used to measure the partial branching fraction of $M_X < 1.7 \, \mathrm{GeV}$, $q^2 > 8 \, \mathrm{GeV}^2$, and $E_\ell^B > 1 \, \text{GeV}$. We use two signal templates: Template a) is defined as signal events with generator-level values of $M_X < 1.7 \, \mathrm{GeV}$ and $q^2 > 8 \, \mathrm{GeV}^2$ and template b) contains all other signal events. 
  \item The lepton energy in the $B$ meson rest-frame, $E_\ell^B$: Signal and \bclnu\ can be separated beyond the kinematic endpoint of the \bclnu\ background, which is \mbox{$\frac{1}{2 m_B} \left( m_B^2 - m_D^2 + m_\ell^2 \right) \approx 2.3 \, \text{GeV}$}. The lepton energy is reconstructed using its momentum (\mbox{$E_\ell^B = |\bold{p}_\ell^B|$}), which has excellent resolution. This makes the measurement more sensitive to the exact composition of the \bclnu\ background and \bulnu\ signal. To minimize the dependence on the signal modeling, the endpoint of the lepton spectrum, ranging from $E_\ell^B \in [2.5, 2.7] \, \text{GeV}$, is treated as a single coarse bin in the fit. To reduce the dependence on the exact modeling of \bclnu\, we require $M_X < 1.7$ \, \text{GeV}. The determined signal yields are used to measure the partial branching fraction with $M_X < 1.7 \, \mathrm{GeV}$ and the signal templates are split accordingly into a matching generator-level template and all other signal events. 
  \item The next fit also analyzes $E_\ell^B$, but uses the determined signal yields to measure the partial branching fraction with $E_\ell^B > 1 \, \text{GeV}$. Thus no separation of signal events in different categories is used. 
 \item The final fit uses $M_X$ and $q^2$ simultaneously in a two dimensional fit ($M_X:q^2$). This fit also measure the partial branching fraction with $E_\ell^B > 1 \, \text{GeV}$ and no separation into different categories of signal events is used. 
 \end{enumerate}
A summary of the binning choices of the kinematic variables is provided in Table~\ref{tab:bins} and we further remove events with $M_X > 4 \, \mathrm{GeV}$ in all fits. Further, we also exclude events with negative $q^2$ values in the $M_X$, $M_X:q^2$, and $q^2$ fits. The likelihood Eq.~\ref{eq:likelihood} is numerically maximized to fit the value of the different components, $\eta_k$, from the observed events and by using the sequential least squares programming method implementation of Ref.~\cite{iminuit}. Confidence intervals are constructed using the profile likelihood ratio method. For a given component $\eta_k$ the ratio is
\begin{equation} \label{eq:test_stat}
  \Lambda(\eta_k) =  - 2 \ln \frac{ \mathcal{L}( \eta_k, \boldsymbol{\widehat \eta_{\eta_k}}, \boldsymbol{\widehat \theta_{\eta_k}}  ) }{  \mathcal{L}( \widehat  \eta_k,  \boldsymbol{ \widehat \eta}, \boldsymbol{ \widehat \theta}  )  } \, ,
\end{equation}
where $ \widehat  \eta_k$, $\boldsymbol{\widehat \eta}$, $\boldsymbol{ \widehat \theta} $ are the values of the component of interest, the remaining components, and a vector of nuisance parameters (NPs), respectively, that maximize the likelihood function, whereas the remaining components $\boldsymbol{\widehat \eta_{\eta_k}}$ and nuisance parameters $\boldsymbol{\widehat \theta_{\eta_k}} $ maximize the likelihood for the specific value $\eta_k$. In the asymptotic limit, the test statistic Eq.~\ref{eq:test_stat} can be used to construct approximate confidence intervals through
\begin{equation}
  1 - \text{CL} = \int_{ \Lambda(\eta_k) }^{\infty} \, f_{\chi^2}(x; 1 \, \text{dof}) \, \text{d} x \, ,
\end{equation}
with $f_{\chi^2}(x; 1 \, \text{dof})$ denoting the $\chi^2$ distribution of the variable $x$ with a single degree of freedom. Further, CL denotes the desired confidence level. The determined signal yields $ \widehat  \eta_{k} =  \widehat  \eta_{\mathrm{sig}}$ are translated into partial branching fractions via
\begin{align} \label{eq:deltaBF}
 \Delta\mathcal{B}(\bulnu; \mathrm{Reg.}) & = \frac{ \widehat  \eta_{\mathrm{sig}}  \cdot  \epsilon_{\Delta \mathcal{B}(\mathrm{Reg.})}  }{4 \left( \epsilon_{\mathrm{tag}} \cdot \epsilon_{\mathrm{sel}} \right) \cdot N_{BB}} \, .
\end{align} 
Here $\epsilon_{\mathrm{tag}}$ denotes the tagging efficiency, as determined after applying the calibration factor introduced in Section~\ref{sec:tagg_eff_corr}. Further, $\epsilon_{\mathrm{sel}}$ and $\epsilon_{\Delta \mathcal{B}(\text{Reg.})}$ denote the signal side selection efficiency and a correction to the efficiency to account for the fraction of \bulnu phase-space region that is measured. The factor of $4$ in the denominator is due to the factor $N_{BB} = \left(771.58 \pm 9.78 \right) \times 10^{6}$ $B$ meson pairs and our averaging over electron and muon final states. 

To validate the fit procedure we generated ensembles of pseudoexperiments for different input branching fractions for \bulnu signal and \bclnu\ background. Fits to these ensembles show no biases in central values and no under- or overcoverage of CI. Using the current world average of $\mathcal{B}(\bulnu) = \left( 2.13 \pm 0.30 \right) \times 10^{-3}$, we expect approximately between  930 \--\ 2070 \bulnu signal events with significances $s = \widehat  \eta_{\mathrm{sig}} / \epsilon$ ranging from about 9 to 15 standard deviations, depending on the signal region under study, and with $\epsilon$ being the expected fit error determined from Asimov data sets~\cite{Cowan:2010js}. 

\section{Systematic Uncertainties}\label{sec:syst}

Several systematic uncertainties affect the determination of the reported partial branching fractions.  The most important uncertainties arise from the modeling of the \bulnu signal component and from the tagging calibration correction. This is followed by uncertainties on particle identification of kaons and leptons, the uncertainty on the number of $B$-meson pairs, the statistical uncertainty on the used MC samples, and uncertainties related to the efficiency of the track reconstructions. Table~\ref{tab:systematic_uncertainties} summarizes the systematic uncertainties for the five measured partial branching fractions probing three phase-space regions. The table separates uncertainties that originate from the background subtraction ( `Additive uncertainties') and uncertainties related to the translation of the fitted signal yields into partial branching fractions ( `Multiplicative uncertainties'). 

The tagging calibration uncertainties are evaluated by producing different sets of calibration factors. These sets take into account the correlation structure from common systematic uncertainties (cf. Section ~\ref{sec:tagg_eff_corr}) and that individual channels and ranges of the output classifier are statistically independent. When applying the different sets of calibration factors, we notice only negligible shape changes on the signal and background template shapes, but the overall tagging efficiency is affected. The associated uncertainty on the calibration factors is found to be 3.6\% and is identical for the five measured partial branching fractions. The \bulnu and \bclnu\ modeling uncertainties do directly affect the shapes of $M_X$, $q^2$, and $E_\ell^B$ signal and background distributions. Further, the \bulnu modeling affects the overall reconstruction efficiencies and migrations of events inside and outside of the phase-space regions we measure. We evaluate the uncertainties on the composition of the hybrid \bulnu MC by variations of the \bpilnu, \brholnu, \bomegalnu, \betalnu, \betaplnu branching fractions and form factors. The uncertainty on non-resonant \bulnu contributions in the hybrid model is estimated by changing the underlying model from that of DFN~\cite{DeFazio:1999ptt} to that of BLNP~\cite{BLNP}. In addition, the uncertainty on the used DFN parameters $m_b^{1S}$ and $a$ (cf. Section~\ref{sec:data_set_sim_samples}) are incorporated. For each of these variations, new hybrid weights are calculated to propagate the uncertainties into shapes and efficiencies. We estimate the uncertainties of $X_u$ fragmentation into $s \bar s$ quark pairs by variations of the corresponding JETSET parameter $\gamma_s$ (cf. Ref.~\cite{SJOSTRAND199474}). As our BDT is trained to reject final states with kaon candidates, a change in this fraction will directly impact the signal efficiency. The $s \bar s$ production probability has been measured by Refs.~\cite{Althoff:1984iz,Bartel:1983qp} at center-of-mass energies of 12 and 36 GeV with values of $\gamma_s = 0.35 \pm 0.05$ and $\gamma_s = 0.27 \pm 0.06$, respectively. We adopt the value and error of $\gamma_s = 0.30 \pm 0.09$, which spans the range of both measurements including their uncertainties. The $X_u$ system of the non-resonant signal component is hadronized by JETSET into final states with two or more pions. We test the impact on the signal efficiency by changing the post-fit charged pion multiplicity of non-resonant \bulnu\ to the distribution observed in data in the signal enriched region of $M_X < 1.7$ GeV (cf. Section~\ref{subsec:pion_mult_res} and Appendix C). The \bclnu\ background after the BDT selection is dominated by \bdlnu and \bdslnu decays. We evaluate the uncertainties on the modeling of \bdlnu\, \bdslnu\, and \bddslnu\ by variations of the BGL parameters and heavy quark form factors within their uncertainties. In addition, we propagate the branching fraction uncertainties. The uncertainties on the \bclnu gap branching fractions are taken to be large enough to account for the difference between the sum of all exclusive branching fractions measured and the inclusive branching fraction measured. We also evaluate the impact on the efficiency of the lepton- and hadron-identification uncertainties, and the overall tracking efficiency uncertainty. The statistical uncertainty on all generated MC samples is also evaluated and propagated into the systematic errors.

We incorporate the effect of additive systematic uncertainties directly into the likelihood function. This can be done by introducing a vector of NPs, $\boldsymbol{\theta}_k$, for each fit template of a process $k$ (e.g. signal or background). Each element of this vector represents one bin of the fitted observables of interest (e.g. $M_X$,$q^2$, $E_\ell^B$ or a 2D bin of $M_X:q^2$). These NPs are constrained parameters in the likelihood Eq.~\ref{eq:likelihood} using multivariate Gaussian distributions,  $\mathcal{G}_k = \mathcal{G}_k( \boldsymbol{0}; \boldsymbol{\theta}_k, \Sigma_k ) $. Here $\Sigma_k$ denotes the systematic covariance matrix for a given template $k$ and $\boldsymbol{\theta}_k$ is a vector of NPs. The covariance $\Sigma_k$  is the sum over all possible uncertainty sources for a given template $k$,
\begin{equation}
 \Sigma_k = \sum_{s}^{\text{error sources}} \Sigma_{ks} \, ,
\end{equation}
with $\Sigma_{ks} $ denoting the covariance matrix of error source $s$. The covariance matrices $\Sigma_{ks} $ depend on uncertainty vectors $\boldsymbol{\sigma_{ks}}$, which represent the absolute error in bins of the fit variable of template $k$. Uncertainties from the same error source are either fully correlated, or for the case of MC or other statistical uncertainties, are treated as uncorrelated. Both cases can be expressed as $\Sigma_{ks}  = \boldsymbol{\sigma_{ks}} \otimes \boldsymbol{\sigma_{ks}}$ or $\Sigma_{ks} = \text{Diag}\left( \boldsymbol{\sigma_{ks}}^2 \right)$, respectively. For particle identification uncertainties, we estimate $\Sigma_{ks} $ using sets of correction tables, sampled according to their statistical and systematic uncertainties. The systematic NPs are incorporated in Eq.~\ref{eq:nui} by rewriting the fractions $f_{ik}$ for all templates as
\begin{equation}
 f_{ik} = \frac{ \eta_{ik}^{\rm MC} }{ \sum_j \eta_{jk}^{\rm MC} } \to  \frac{ \eta_{ik}^{\rm MC} \left( 1 + \theta_{ik} \right) }{ \sum_j \eta_{jk}^{\rm MC} \left( 1 + \theta_{jk} \right)  },
\end{equation}
to take into account changes in the signal or background shape. Here $\eta_{ik}^{\rm MC}$ denotes the predicted number of MC events of a given bin $i$ and a process $k$, and $\theta_{ik}$ is the associated  nuisance parameter constrained by $\mathcal{G}_k$. 

\section{$B \to X_c \ell \bar \nu_\ell$ Control Region}\label{sec:sidebands}

\begin{figure}
  \includegraphics[width=0.46\textwidth]{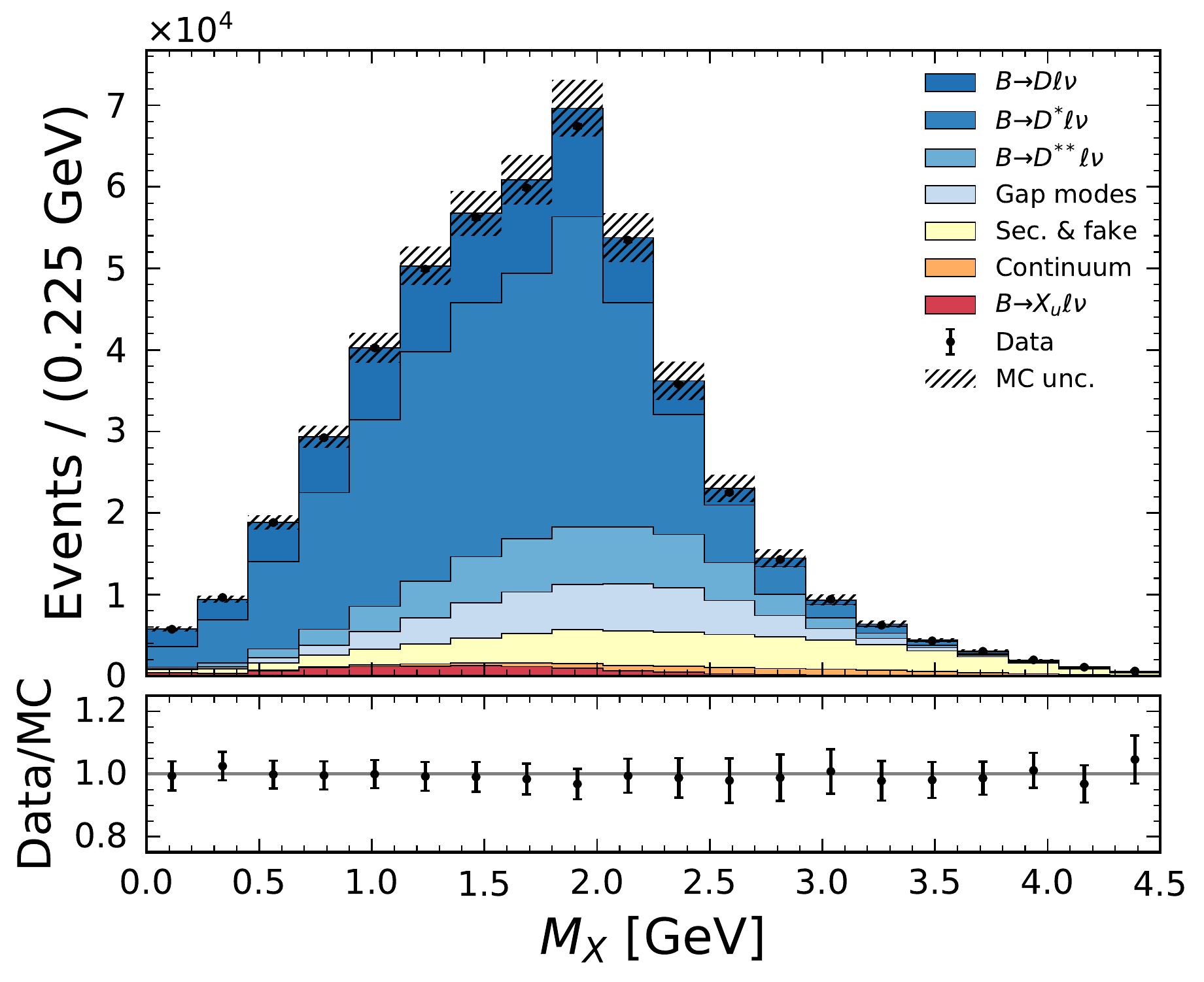} 
  \includegraphics[width=0.46\textwidth]{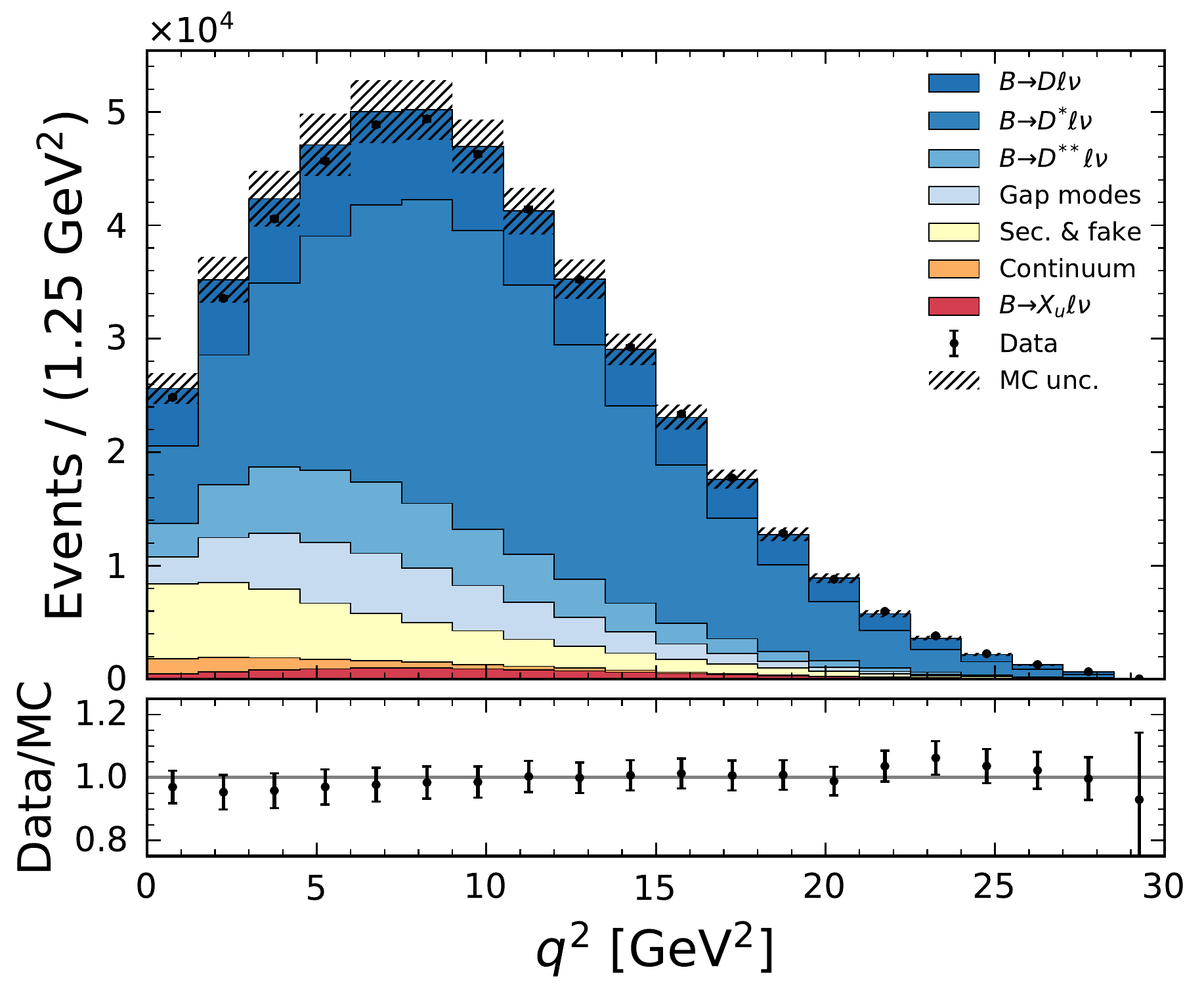} 
   \includegraphics[width=0.46\textwidth]{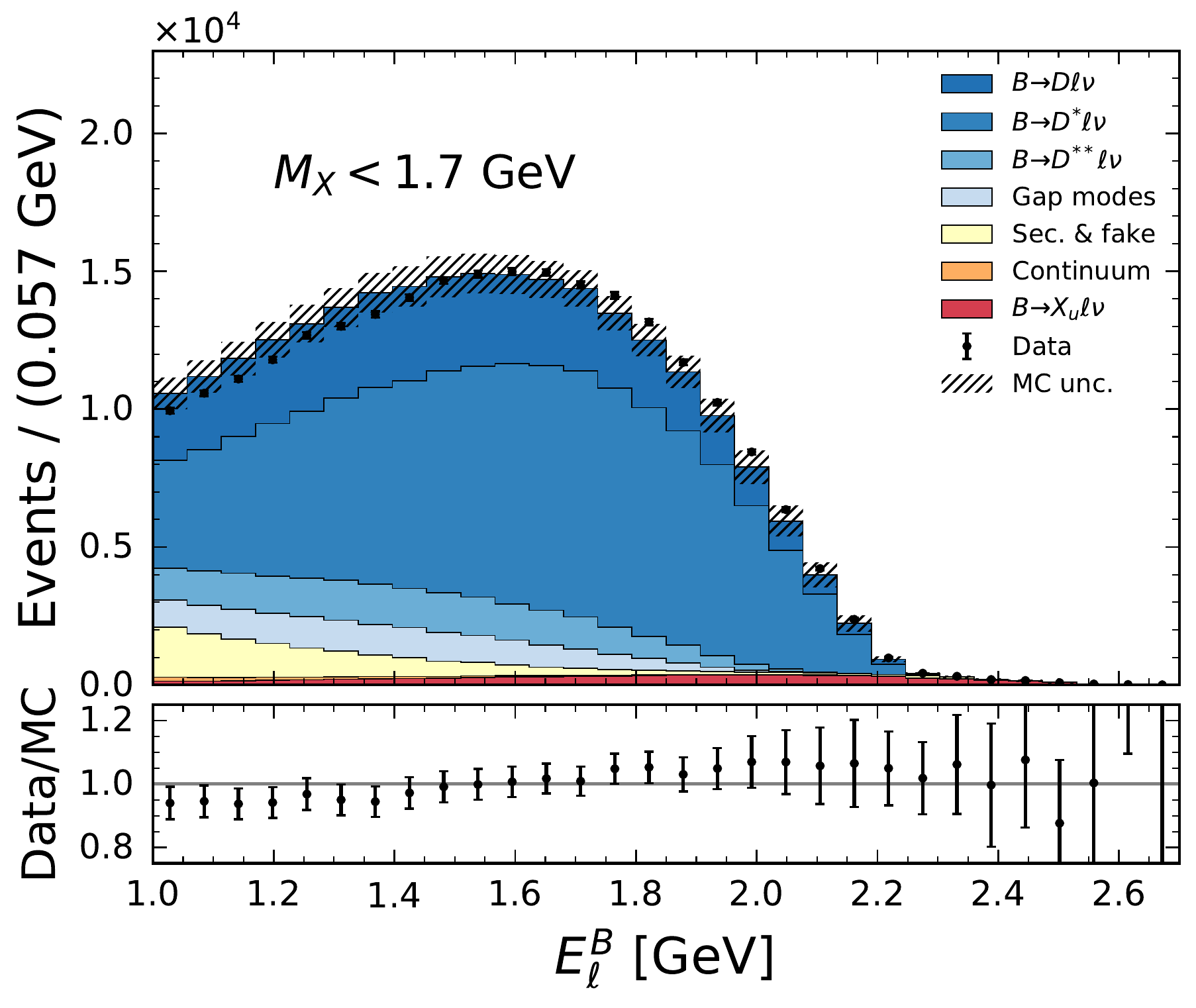} 
   \includegraphics[width=0.46\textwidth]{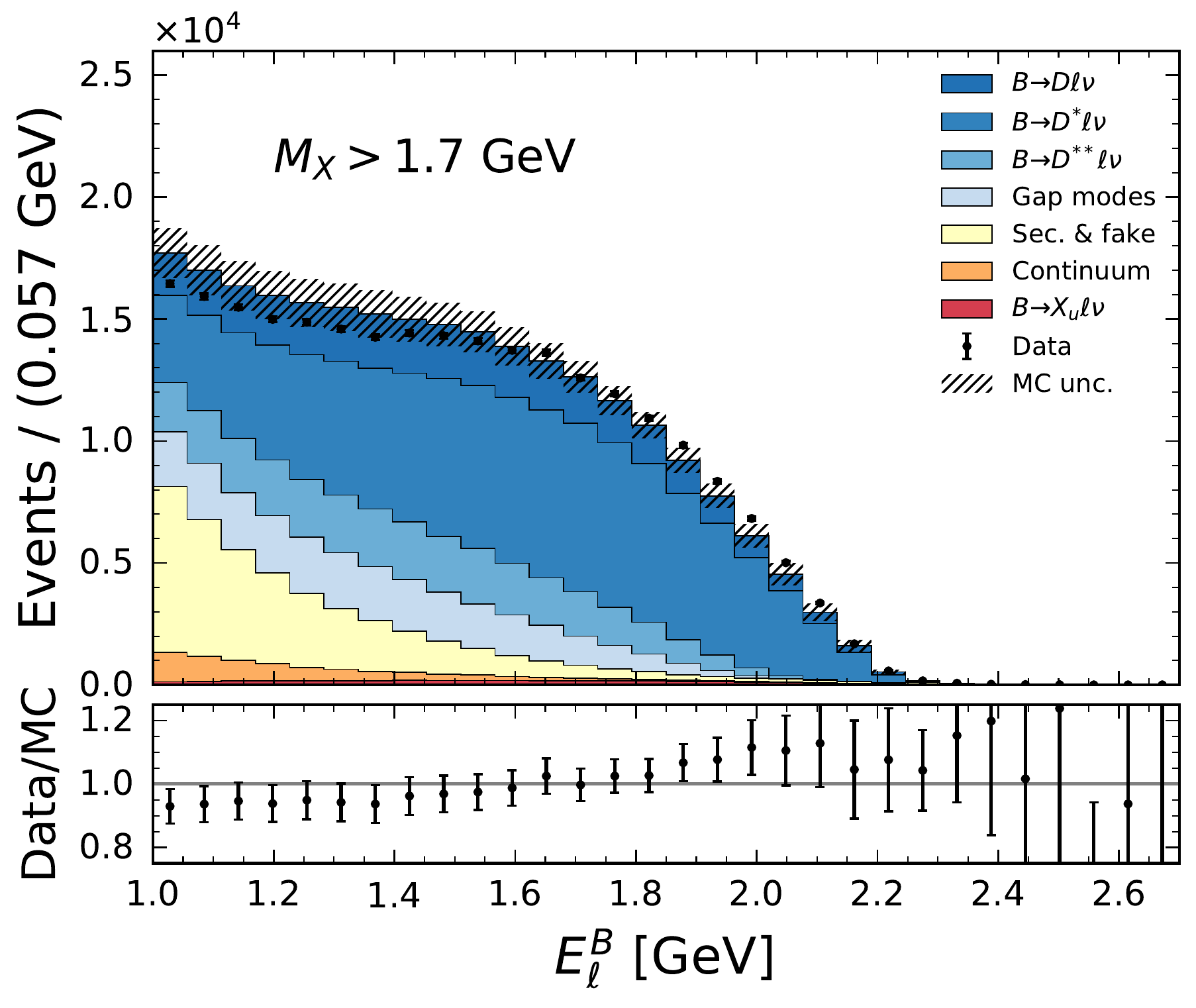}    
\caption{
 (Top) The $M_X$ and $q^2$ spectra of the selected candidates prior to applying the background BDT are shown. \\
 (Bottom)  The $E_\ell^B$ spectrum of the selected candidates prior to applying the background BDT are shown for events with $M_X < 1.7$ GeV and $M_X > 1.7$ GeV.
 }
\label{fig:incl_shapes}
\end{figure}

Figure~\ref{fig:incl_shapes} compares the reconstructed  $M_X$, $q^2$, and $E_\ell^B$ distributions with the expectation from MC before applying the background suppression BDT. All corrections are applied and the MC uncertainty contains all systematic uncertainties discussed in Section~\ref{sec:syst}. The agreement of $M_X$ and $q^2$ is excellent, but some differences in the shape of the lepton momentum spectrum are seen. This is likely due to imperfections of the modeling of the inclusive \bclnu background. The discrepancy reduces in the $M_X < 1.7$ GeV region. The main results of this paper will be produced by fitting $q^2$ and $M_X$ in two dimensions. We use the lepton spectrum to measure the same regions of phase space, to validate the obtained results.

\section{\bulnu Signal Region}\label{sec:signal}

\begin{figure}
  \includegraphics[width=0.46\textwidth]{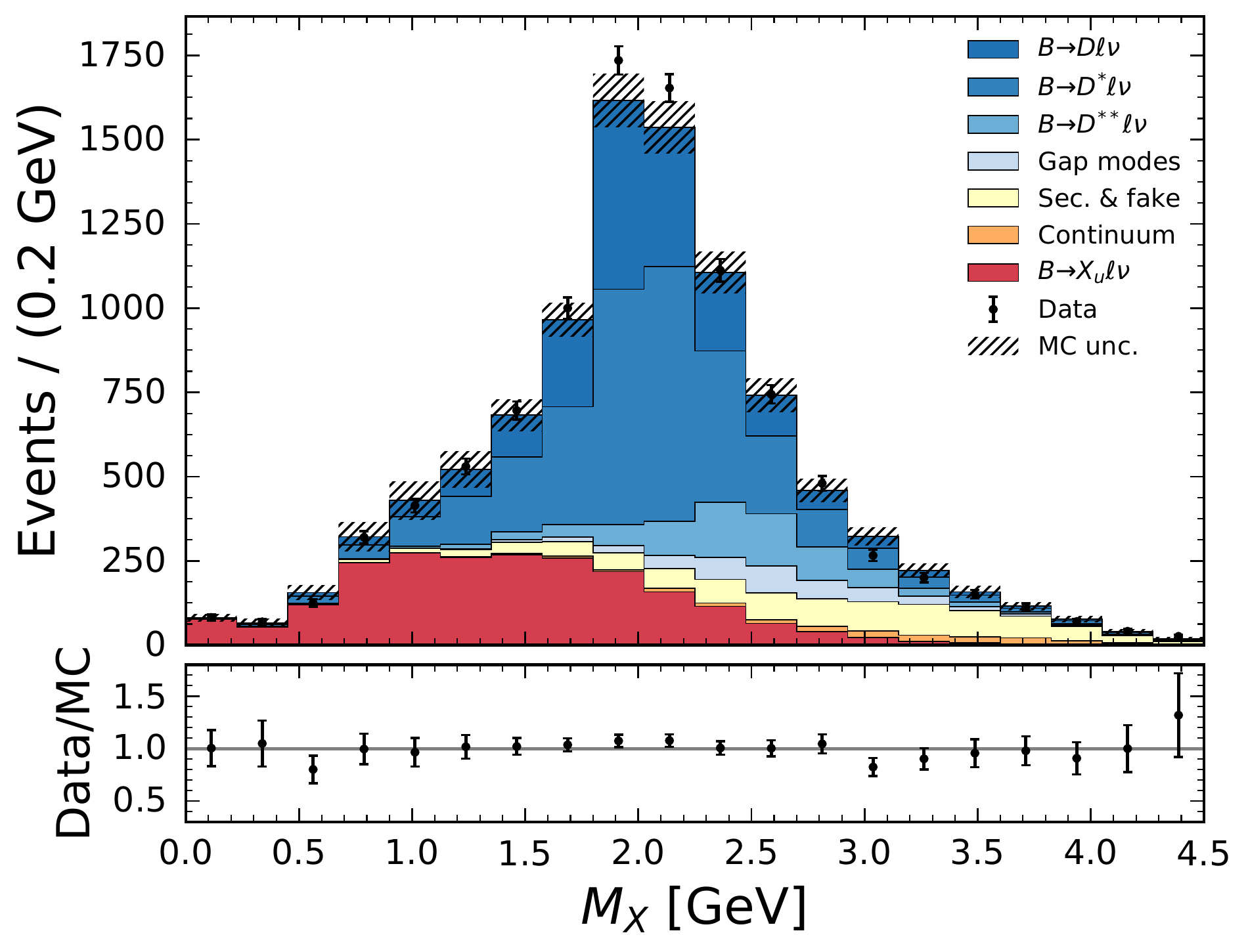} 
  \includegraphics[width=0.46\textwidth]{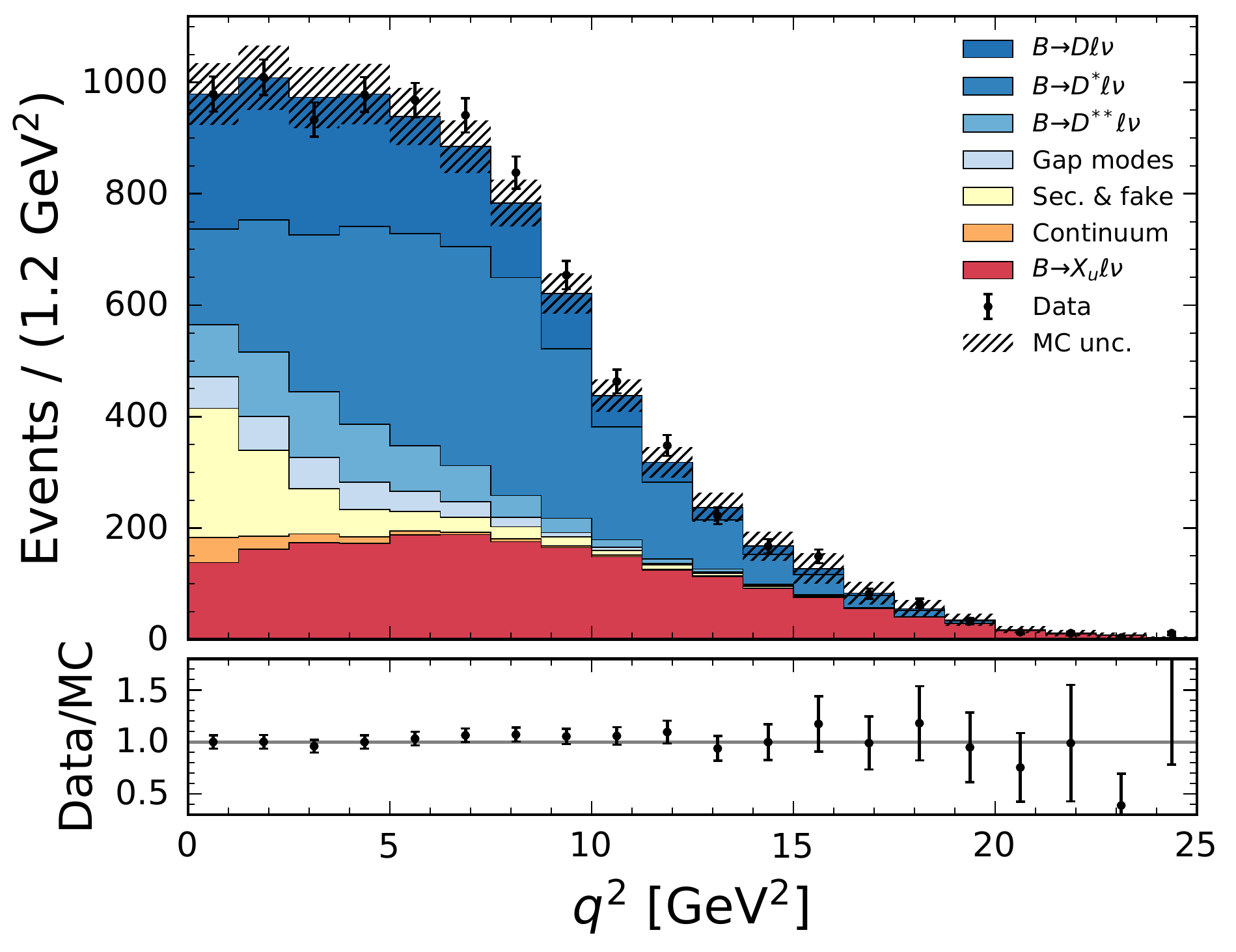} 
  \includegraphics[width=0.46\textwidth]{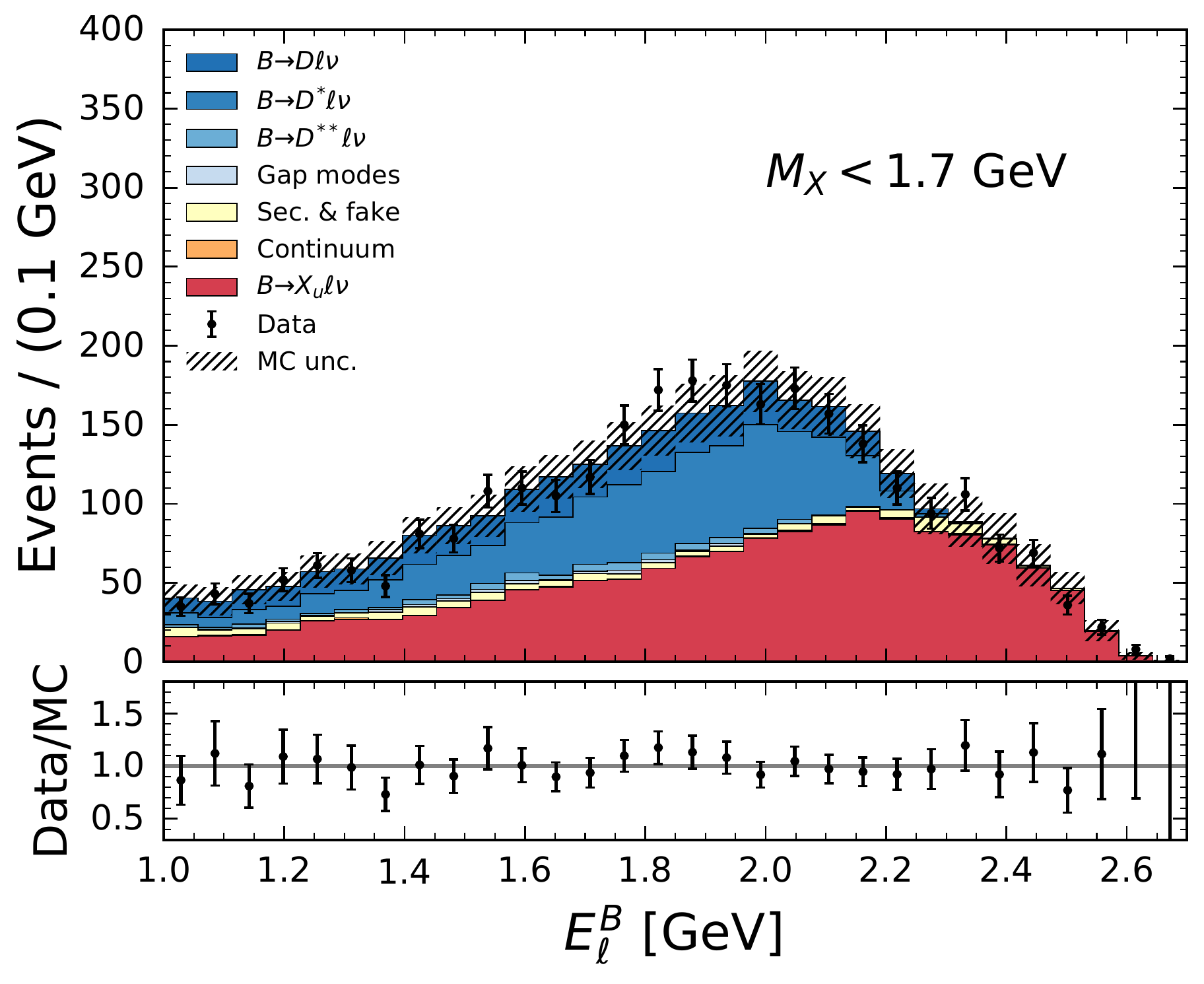} 
  \includegraphics[width=0.46\textwidth]{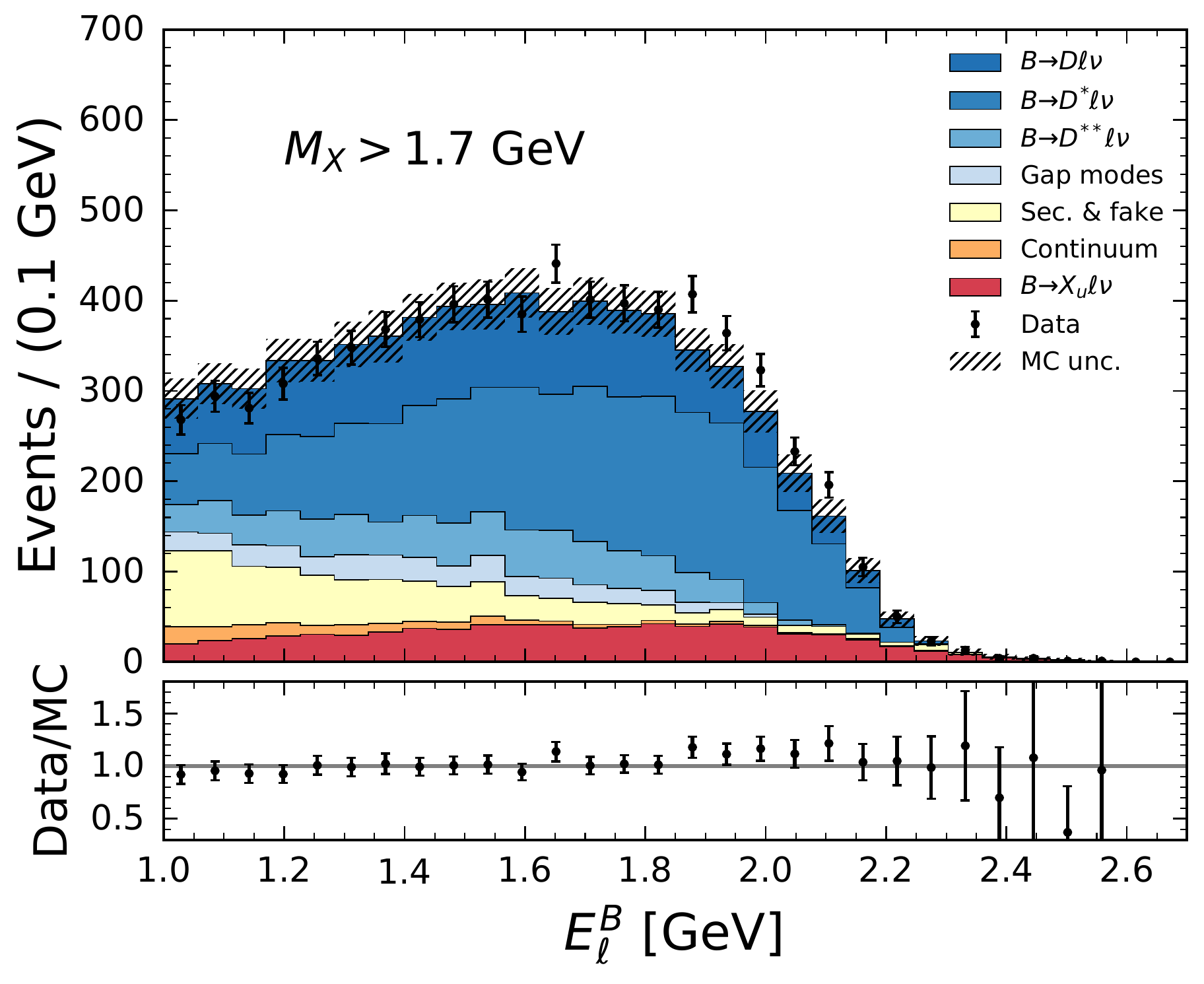} 
\caption{
 The $M_X$, $q^2$ and $E_\ell^B$ spectra after applying the background BDT but before the fit are shown. The \bulnu contribution is shown in red and scaled to the world average of $\mathcal{B}(\bulnu) = \left(  2.13 \pm 0.30 \right) \times 10^{-3}$. The data and MC agreement is reasonable in all variables. The $E_\ell^B$ spectra is shown with selections of $M_X < 1.7 \, \text{GeV}$ and  $M_X > 1.7 \, \text{GeV}$. The cut of  $M_X < 1.7 \, \text{GeV}$ is later used in the fit to reduce the dependence on the \bclnu modeling of higher charmed states. 
 }
\label{fig:ulnu_shapes}
\end{figure}

Figure~\ref{fig:ulnu_shapes} shows the reconstructed $M_X$, $q^2$, and $E_\ell^B$ distributions after the BDT selection is applied. The \bulnu\ contribution is now clearly visible at low $M_X$ and high $E_\ell^B$, while the reconstructed events and the MC expectation show good agreement. The \bclnu background is dominated by contributions from \bdlnu\ and \bdslnu\ decays, and the remaining background is predominantly from secondary leptons, and misidentified lepton candidates. 

\clearpage

\onecolumngrid

\twocolumngrid

\onecolumngrid

\begin{table}[t!]
 \renewcommand{\arraystretch}{0.9} 
\caption{
	The relative uncertainty on the extracted \bulnu partial branching fractions are shown. For definitions of additive and multiplicative errors, see text. 
}
\label{tab:systematic_uncertainties}
\vspace{1ex}
\begin{tabular}{lrrrrrr}
\hline\hline
  &  \multicolumn{5}{c}{ Relative uncertainties [\%]} \\ 
 Phase-space region & $M_X < 1.7 \, \mathrm{GeV}$, & $M_X < 1.7 \, \mathrm{GeV},$ & $M_X < 1.7 \, \mathrm{GeV},$ & $E_\ell^B > 1 \, \mathrm{GeV}$ & $ E_\ell^B > 1 \, \mathrm{GeV}$ \ \\ 
   &  $E_\ell^B  > 1 \, \mathrm{GeV}$ &  $E_\ell^B > 1 \, \mathrm{GeV}$ & $q^2 > 8 \, \text{GeV}^2,$     \\ 
  &   &  &   $E_\ell^B > 1 \, \mathrm{GeV}$  \vspace{1ex}  \\ 
Fit variable(s)  &  ($M_X$ fit) &  ($E_\ell^B$ fit)  &  ($q^2$ fit) & ($E_\ell^B $ fit) &  ($M_X:q^2$ fit)   \vspace{1ex}  \\  
\hline  \vspace{1ex} \\
 {\bf Additive uncertainties} \\
 \quad \bulnu modeling & \\
 \qquad \bpilnu\ FFs & 0.1 & 0.7 & 1.4& 0.6 & 0.4 \\
 \qquad \brholnu\ FFs & 0.2 & 1.9 & 4.3& 1.9 & 0.7 \\
 \qquad \bomegalnu\ FFs & 0.5 & 3.2 & 5.2& 3.1 & 0.8 \\
 \qquad \betalnu\ FFs  & 0.1 & 1.6 & 3.0 & 1.6 & 0.3 \\
 \qquad \betaplnu\ FFs & 0.1 & 1.6 & 3.0 & 1.6 & 1.6 \\
 \qquad $\mathcal{B}(\bpilnu)$ & 0.2 & 0.1 & 0.2& 0.1 & 0.2 \\
 \qquad $\mathcal{B}(\brholnu)$ & 0.3 & 0.7 & 0.8& 0.5 & 0.4 \\
 \qquad $\mathcal{B}(\bomegalnu)$ & $<$0.1 & 0.1 & 0.8& 0.1 & 0.1 \\
 \qquad $\mathcal{B}(\betalnu)$ & $<$0.1 & 0.1 & $<$0.1& 0.1 & $<$0.1 \\
 \qquad $\mathcal{B}(\betaplnu)$ & $<$0.1 & $<$0.1 & 0.1& 0.1 & $<$0.1 \\
 \qquad $\mathcal{B}(B \to X_u \, \ell^+ \, \nu)$  & 0.7 & 2.0 & 2.1& 2.1 & 2.1 \\
 \qquad DFN parameters& 2.3 & 3.5 & 1.1& 3.5 & 5.0 \\
 \qquad Hybrid model & 2.7 & 8.7 & 4.6& 8.7 & 3.1 \\
  \quad \bclnu modeling& \\
 \qquad \bdlnu\ FFs & 0.1 & 0.1 & 0.9& 0.1 & $<$0.1 \\
 \qquad \bdslnu\ FFs & 1.4 & 1.2 & 3.0& 1.3 & 1.1 \\
 \qquad \bddslnu\ FFs & 0.4 & 0.5 & 0.3& 0.5  & 0.4 \\
 \qquad $\mathcal{B}(\bdlnu)$ & 0.1 & $<$0.1 & 0.2& $<$0.1 & 0.2 \\
 \qquad $\mathcal{B}(\bdslnu)$ & $<$0.1 & $<$0.1 & 0.3& $<$0.1 & 0.2 \\
 \qquad $\mathcal{B}(\bddslnu)$ & 0.6 & 0.1 & 0.3& 0.1 & 0.5 \\
 \qquad Gap modeling & 1.1 & 0.1 & 0.3& 0.1 & 1.0 \\
 \quad MC statistics  & 1.3 & 1.6 & 3.8 & 1.7 & 1.6 \\ 
 \quad Tracking efficiency & 0.3 & - & 0.8 & - & 0.4   \\
 \quad $\mathcal{L}_{\rm \ell ID}$ shape & 1.0 & 0.5& 1.3 & 0.6 & 1.2 \\
 \quad $\mathcal{L}_{\rm K/\pi ID}$ shape & 1.2 & - & 1.3 & - & 1.0 \\
 \quad $D \to X \ell \, \nu_\ell$ &  0.1 & 0.1& 0.1 & 0.1 & 0.1 \\ 
 \quad $\pi_s$ efficiency & $<$0.1 & - & 0.1 & - & 0.1 \\
  \\
  {\bf Multiplicative uncertainties}  \\
 \quad \bulnu modeling & \\
 \qquad \bpilnu\ FFs & 0.2 & 0.2 & 1.9& 0.2 & 0.2 \\
 \qquad \brholnu\ FFs & 0.7 & 0.8 & 3.7 & 0.8 & 0.6 \\
 \qquad \bomegalnu\ FFs & 1.3 & 1.6 & 6.1 & 1.6 & 1.1 \\
 \qquad \betalnu\ FFs  & 0.3 & 0.3 & 1.8 & 0.3 & 0.2\\
 \qquad \betaplnu\ FFs & 0.2 & 0.3 & 1.8 & 0.3 & 0.2 \\
 \qquad $\mathcal{B}(\bpilnu)$ & 0.3 & 0.4 & 0.4 & 0.4 & 0.3\\
 \qquad $\mathcal{B}(\brholnu)$ & 0.4 & 0.6 & 0.6 & 0.6 & 0.4\\
 \qquad $\mathcal{B}(\bomegalnu)$ & $<$0.1 & $<$0.1 & 0.1 & $<$0.1 & $<$0.1\\
 \qquad $\mathcal{B}(\betalnu)$ & 0.1 & 0.1 & $<$0.1 & 0.1 & $<$0.1\\
 \qquad $\mathcal{B}(\betaplnu)$ & 0.1 & 0.1 & 0.1 & 0.1 & 0.1\\
 \qquad $\mathcal{B}(B \to X_u \, \ell^+ \, \nu)$  & 3.0& 3.2 & 2.9 & 4.8 & 3.8 \\
 \qquad DFN parameters & 2.5& 2.5 & 2.7 & 6.8& 3.6 \\
 \qquad Hybrid model & 0.2 & 0.8 & 1.4 & 4.7 & 2.8 \\
 \qquad $\pi^+$ multiplicity  & 1.7 & 2.5 & 2.3 & 3.1 & 1.7 \\
 \qquad $\gamma_s$ ($s \bar s$ fragmentation) & 0.5 & 0.8 & 1.1 & 1.1 & 0.8 \\
 \quad $\mathcal{L}_{\rm \ell ID}$ efficiency & 1.5 & 1.6 & 1.6& 1.6 & 1.5 \\
 \quad $\mathcal{L}_{\rm K/\pi \, ID}$ efficiency & 0.7 & 0.6 & 0.6 & 0.6 & 0.7 \\
 \quad $N_{B \bar B}$ & 1.3 & 1.3 & 1.3 & 1.3 & 1.3 \\
 \quad Tracking efficiency  & 0.8 & 0.3 & 0.8 & 0.3 & 0.9 \\
 \quad Tagging calibration & 3.6 & 3.6 & 3.6 & 3.6 & 3.6 \\ \\ \hline 
{\bf Total syst. uncertainty} & {\bf 7.8} & {\bf 12.6}   & {\bf 14.6}   & {\bf 15.4}  & {\bf 10.4} \vspace{1ex} \\
\hline\hline
\end{tabular}
\end{table}
\clearpage

\twocolumngrid

\begin{figure}[th!]
  \includegraphics[width=0.45\textwidth]{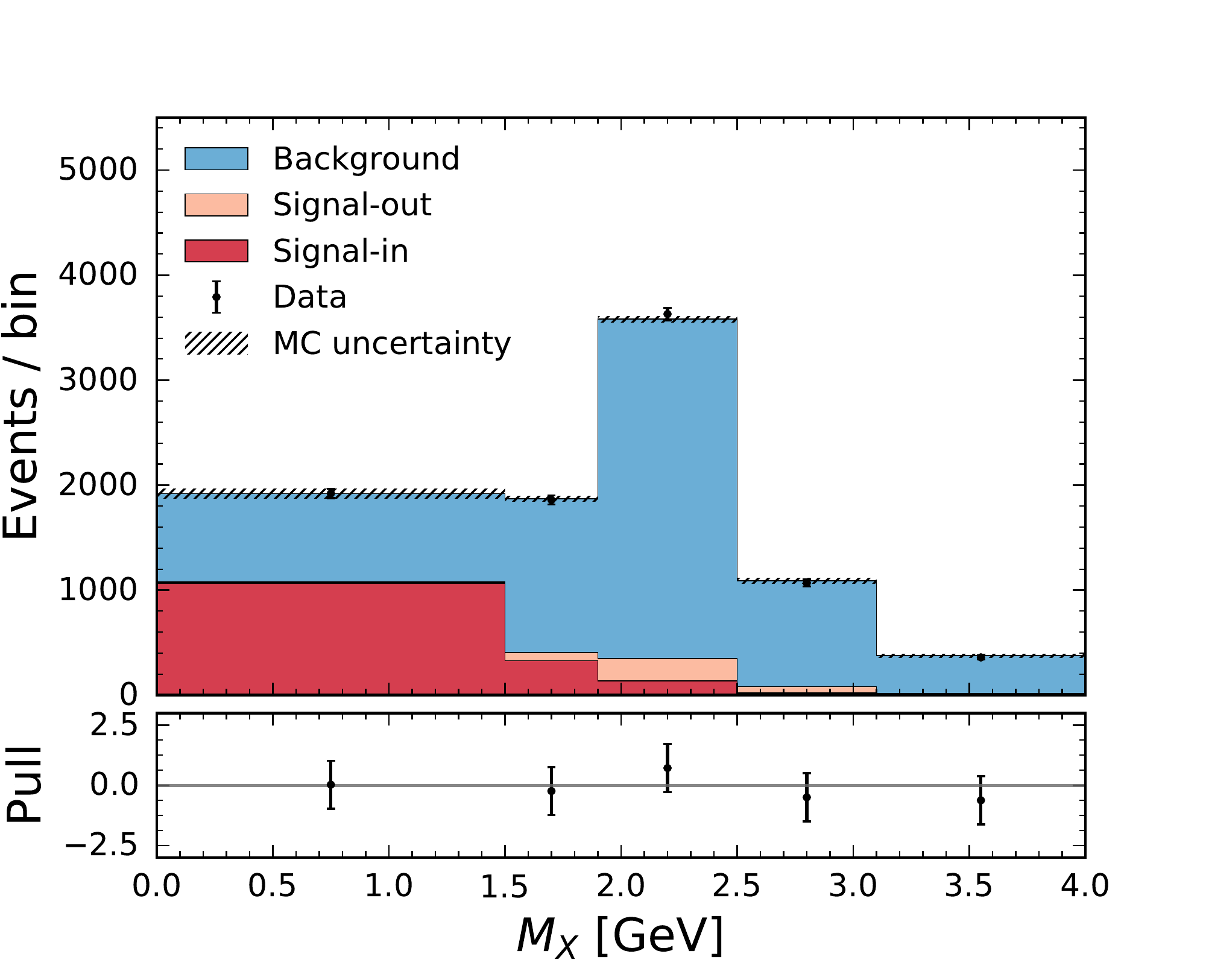} 
  \includegraphics[width=0.45\textwidth]{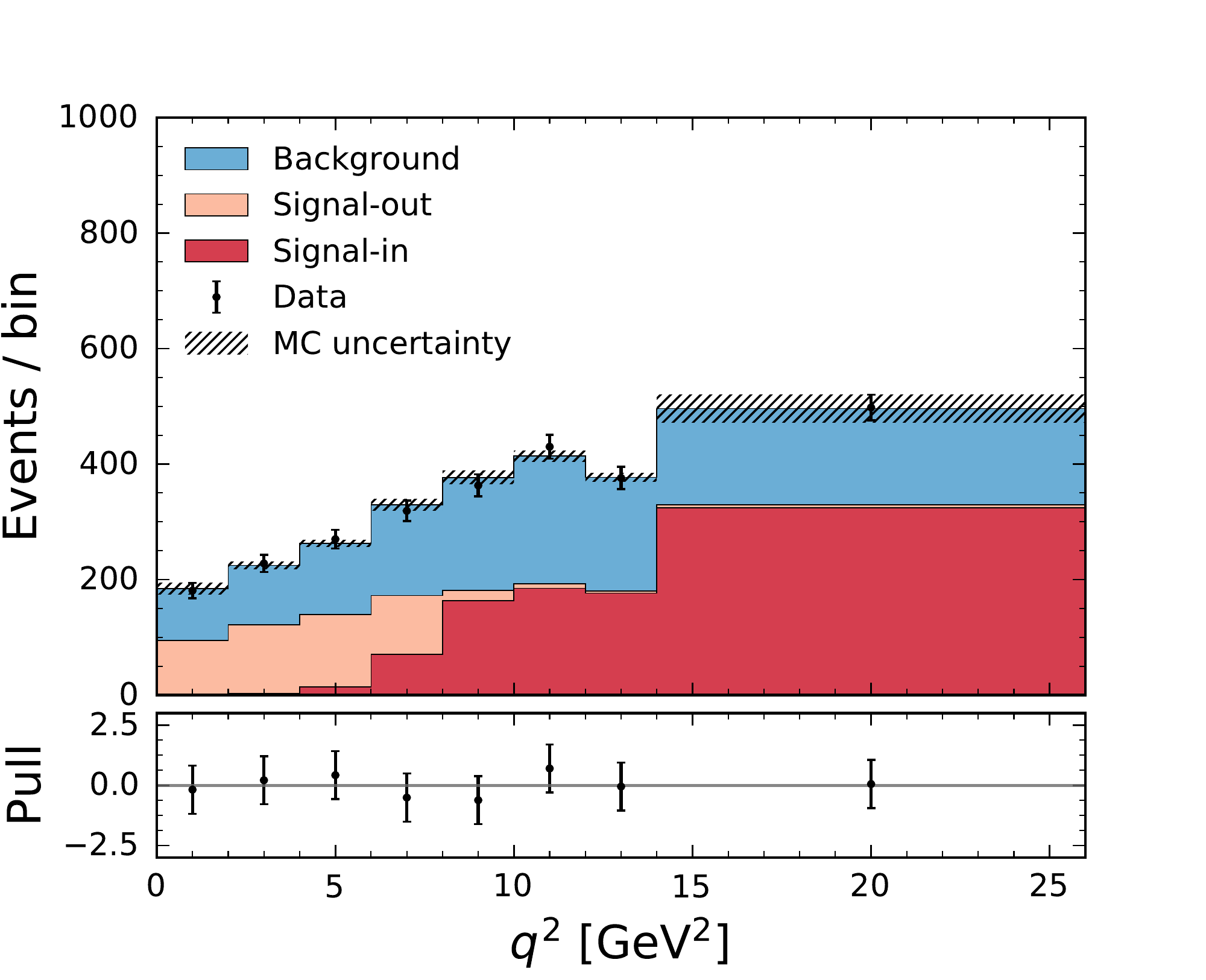} 
\caption{
  The post-fit distributions of the one-dimensional fits to $M_X$ and $q^2$ are shown, corresponding to the measured partial branching fractions for $E_\ell^B > 1 \, \text{GeV}$  with additional requirements of $M_X < 1.7 \, \text{GeV}$, and $M_X < 1.7 \, \text{GeV}$ and $q^2 > 8 \, \text{GeV}^2$, respectively.
 }
\label{fig:post_fit_1D_1}
\end{figure}

\section{Results}\label{sec:results}

We report partial branching fractions for three phase-space regions from five fits to the reconstructed variables introduced in Section~\ref{sec:fit}. All partial branching fractions correspond to a selection with $E_\ell^B > 1 \, \text{GeV}$, also reverting the effect of final state radiation photons, and possible additional phase-space restrictions. The resulting fit yields are listed in Table~\ref{tab:res_fit_yields}.

\begin{table*}[t]
\caption{The fitted signal yields in ($\widehat{\eta}_{\mathrm{sig}}$) and outside ($\widehat{\eta}_{\mathrm{sig-out}}$) the measured phase-space regions, the background yields ($\widehat{\eta}_{\mathrm{bkg}}$) and the product of tagging and selection efficiency are listed. The number of analyzed data events, $n_{\mathrm{data}}$, are also listed.
}
\renewcommand\arraystretch{1.2}
\centering
\begin{tabular}{lccccccc}
\hline
\hline
Phase-space region                                                                                                                    & \multicolumn{1}{l}{Additional Selection}   & \multicolumn{1}{l}{Fit variable(s)}   & $\widehat{\eta}_{\mathrm{sig}}$         & $\widehat{\eta}_{\mathrm{sig-out}}$      & $\widehat{\eta}_{\mathrm{bkg}}$      & $n_{\mathrm{data}}$       & $10^{3}\left(\epsilon_{\mathrm{tag}} \cdot \epsilon_{\mathrm{sel}}\right)$ \\ \hline
\multirow{2}{*}{\begin{tabular}[c]{@{}l@{}}$M_{X} < 1.7$ GeV, \\ $E_{\ell}^{B} >1$ GeV\end{tabular}}                     & -   & \multirow{2}{*}{$M_{X}$ fit}          & \multirow{2}{*}{$1558 \pm 69 \pm 71$}  & \multirow{2}{*}{$364 \pm 51 $}  & \multirow{2}{*}{$6912 \pm 138$}  &  \multirow{2}{*}{$8833 \pm 94$}  & \multirow{2}{*}{$0.26 \pm 0.07$}                                   \\ \vspace{0.3cm}
                                                                                                                                         &                                        &                                        &                                      &                                         &                                                                    \\ \hline \vspace{0.3cm}
\multirow{2}{*}{\begin{tabular}[c]{@{}l@{}}$M_{X} < 1.7$ GeV,\\ $E_{\ell}^{B} >1$ GeV\end{tabular}}                    &  \multirow{2}{*}{$M_X < 1.7 \, \mathrm{GeV}$}        & \multirow{2}{*}{$E_{\ell}^{B}$ fit}   & \multirow{2}{*}{$1285 \pm 68 \pm 139$} & \multirow{2}{*}{$22 \pm 3 $}   & \multirow{2}{*}{$1362 \pm 155$}  &\multirow{2}{*}{$2669 \pm 52$}   & \multirow{2}{*}{$0.21 \pm 0.07$}                                   \\
                                                                                                                                         &                                       &                                        &                                      &                                         &                                                                    \\  \hline \vspace{0.3cm}
\multirow{3}{*}{\begin{tabular}[c]{@{}l@{}}$M_{X} < 1.7$ GeV,\\ $q^{2} >8$ GeV$^{2}$,\\ $E_{\ell}^{B} >1$ GeV\end{tabular}} & \multirow{3}{*}{$M_X < 1.7 \, \mathrm{GeV}$}     & \multirow{3}{*}{$q^{2}$ fit}          & \multirow{3}{*}{$938 \pm 99 \pm 100$}  & \multirow{3}{*}{$474 \pm 57 $} & \multirow{3}{*}{$1253 \pm 192$} & \multirow{3}{*}{$2665 \pm 52$}   & \multirow{3}{*}{$0.14 \pm 0.07$}                                   \\
                                                                                                                                         &                                       &                                        &                                      &                                         &                                                                    \\
                                                                                                                                         &                                       &                                        &                                      &                                         &                                                                    \\  \hline \vspace{0.3cm}
$E_{\ell}^{B} >1$ GeV                                                                                                &  $M_X < 1.7 \, \mathrm{GeV}$                & $E_{\ell}^{B}$ fit                    & $1303 \pm 69 \pm 138$                  & -                                    & $1366 \pm 154$            &  $2669 \pm 52$      & $0.21 \pm 0.19$                                                    \\  \hline \vspace{0.3cm}
$E_{\ell}^{B} >1$ GeV                                                                                               &               & \multicolumn{1}{l}{$M_{X}:q^{2}$ fit} & $1801 \pm 81 \pm 123$                  & -                                    & $7031 \pm 164$             & $8833 \pm 94$     & $0.31 \pm 0.12$                                                    \\ \hline\hline
\end{tabular}
\label{tab:res_fit_yields}
\end{table*}

\subsection{Partial Branching Fraction Results}\label{sec:partial_BF_res}

\begin{figure}[h!]
 \includegraphics[width=0.45\textwidth]{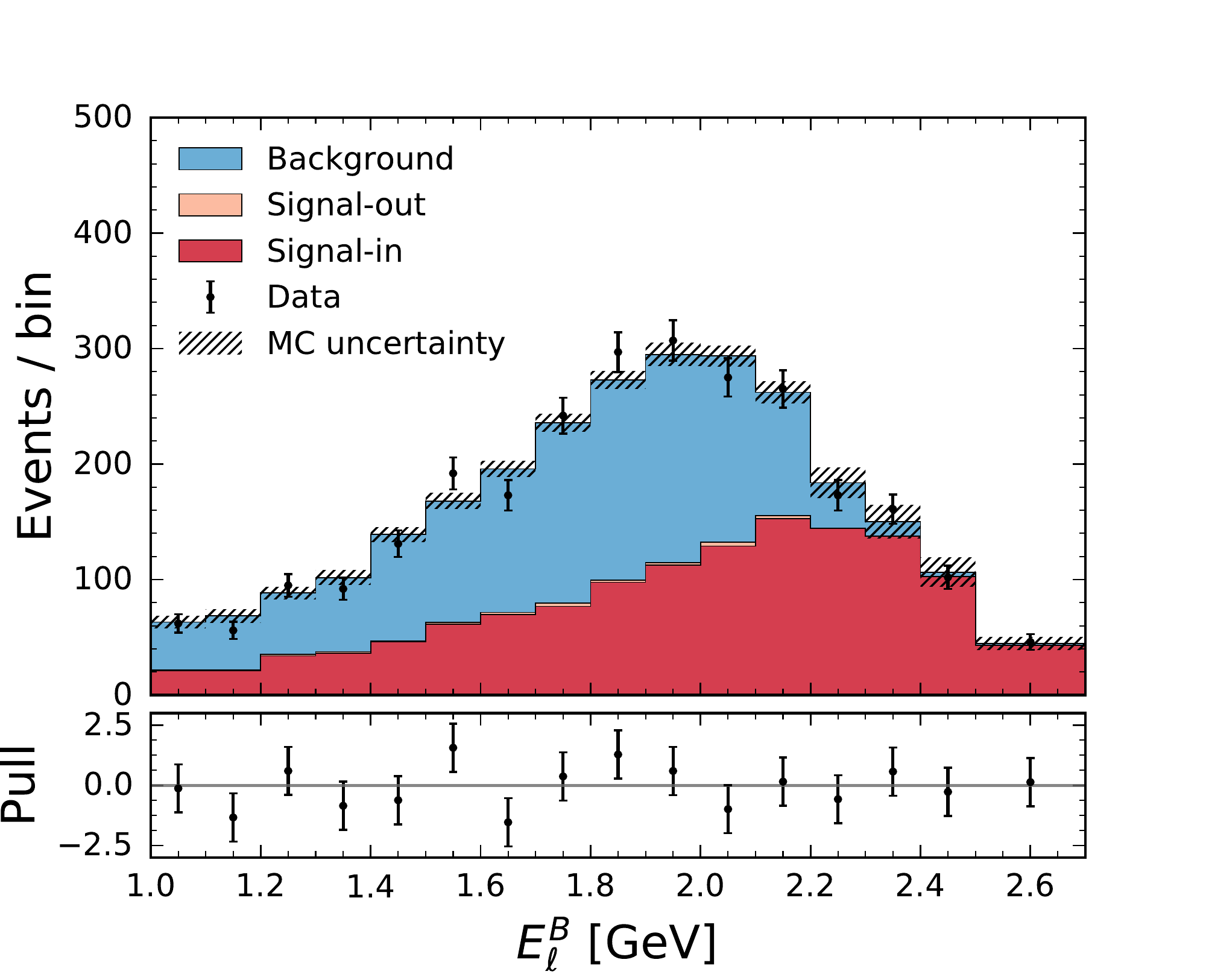} 
\caption{
  The post-fit distributions of the fit to $E_\ell^B$ with $M_X < 1.7 \, \text{GeV}$ is shown. The resulting yields were corrected to correspond to the partial branching fraction with $E_\ell^B > 1 \, \text{GeV}$ with and without an additional requirement of $M_X < 1.7 \, \text{GeV}$, respectively. 
 }
\label{fig:post_fit_1D_2}
\end{figure}

\begin{figure}[h!]
 \includegraphics[width=0.45\textwidth]{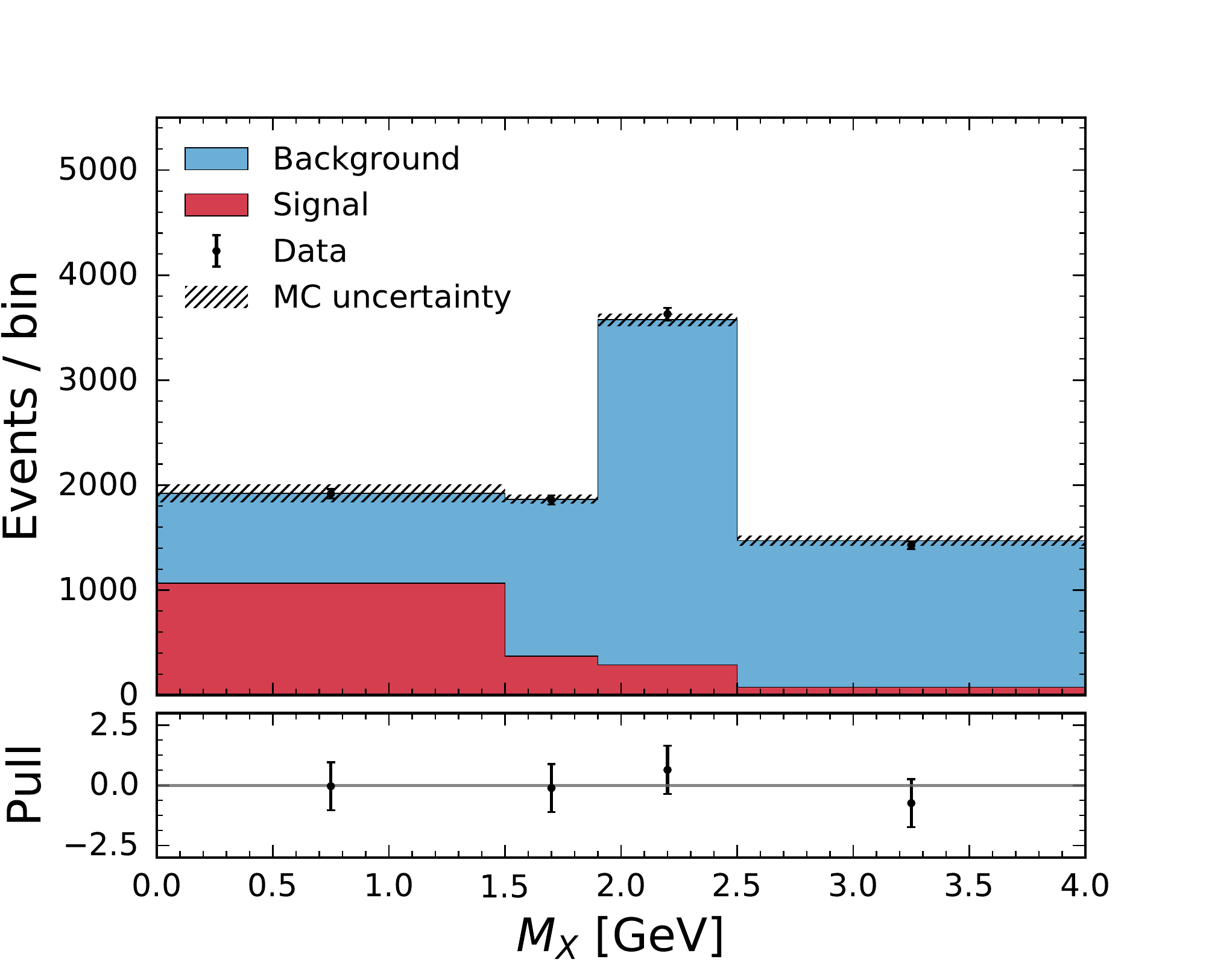} 
  \includegraphics[width=0.45\textwidth]{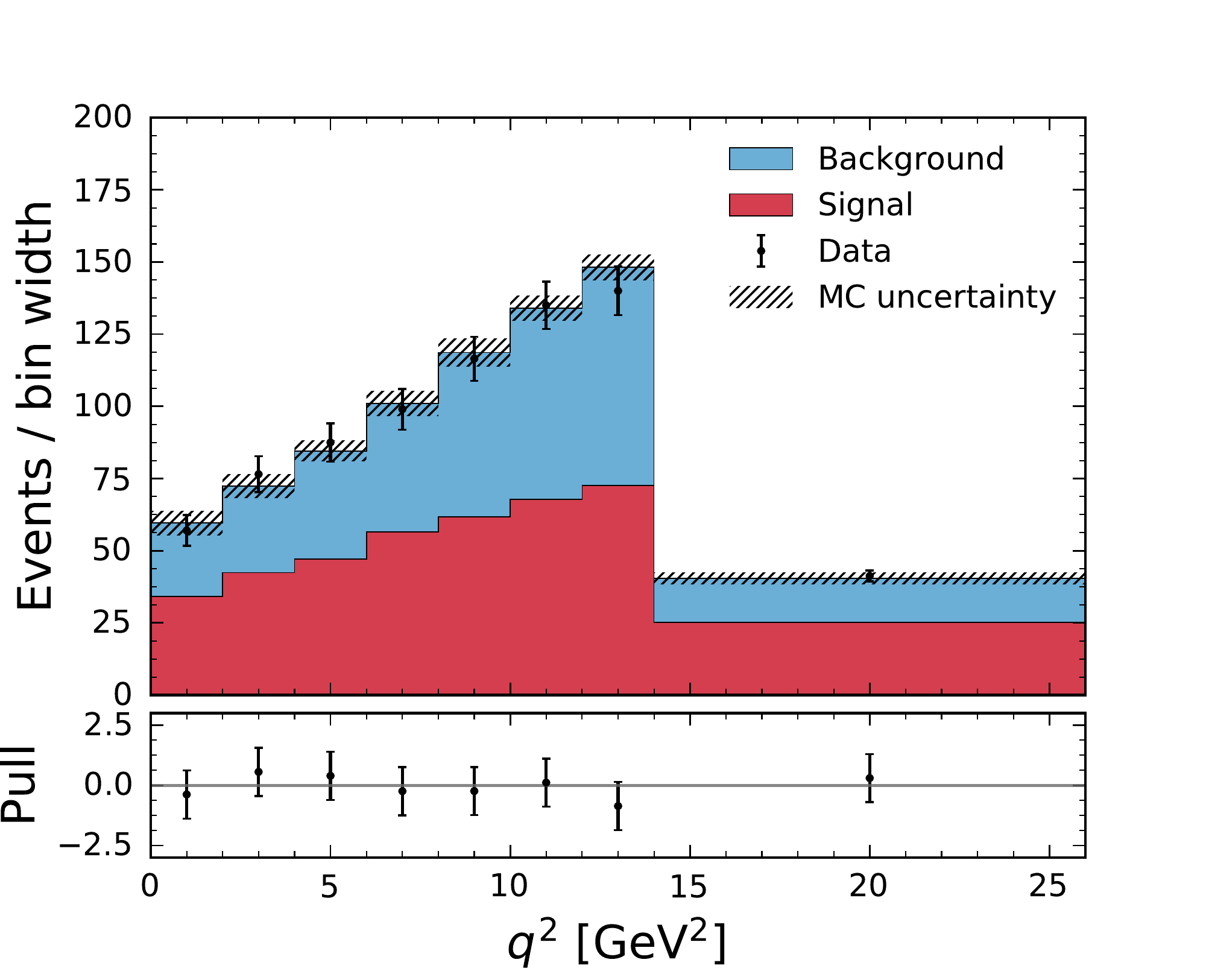}
\caption{
  The post-fit projection of $M_{X}$ of the two-dimensional fit to $M_X:q^2$ on $M_{X}$ and the $q^2$ distribution in the range of $M_X \in [0, 1.5] \, \mathrm{GeV}$ are shown. The resulting yields are corrected to correspond to a partial branching fraction with $E_\ell^B > 1 \, \text{GeV}$. The remaining $q^2$ distributions are given in Figure~\ref{fig:2Dfit_projections} (Appendix D).
 }
\label{fig:post_fit_2D}
\end{figure}

For the partial branching fraction with $M_X < 1.7 \, \text{GeV}$ from the fit to $M_X$ we find
\begin{align} \label{bf:resa}
\bfResa \, ,
\end{align}
with the first and second error denoting the statistical and systematic uncertainty, respectively. The resulting post-fit distribution is shown in the top panel of Figure~\ref{fig:post_fit_1D_1}. With this selection about 56\% of the available \bulnu\ phase space is probed. The partial branching fraction is in good agreement with the value obtained by fitting $E_\ell^B$ and corrected to the same phase space. The fit is shown in Figure~\ref{fig:post_fit_1D_2} and we measure
\begin{align}
\bfRescp \, ,
\end{align}
with a larger systematic and statistical uncertainty than Eq.~\ref{bf:resa}. 
To further probe the \bulnu\, enriched region, we carry out a measurement for $M_X < 1.7 \, \text{GeV}$ and $q^2 > 8 \, \text{GeV}^2$ from a fit to the $q^2$ spectrum. This selection only probes about 31\% of the available \bulnu phase space. We find
\begin{align}
\bfResb \, .
\end{align}
The corresponding post-fit distribution of $q^2$ is shown in the bottom panel of Figure~\ref{fig:post_fit_1D_1}.
The most precise determinations of \bulnu are obtained from a two-dimensional fit, exploiting the full combined discriminatory power of $M_X$ and $q^2$. The resulting partial branching fraction probes about 86\% of the available \bulnu phase space. We measure 
\begin{align} \label{eq:bfResD}
\bfResd \, .
\end{align}
The projection of the 2D fit onto $M_X$ and the $q^2$ distribution for the signal enriched region of $M_X < 1.5 \, \mathrm{GeV}$ are shown in Figure~\ref{fig:post_fit_2D}. The remaining $q^2$ distributions are given in Appendix D. The partial branching fraction is also in good agreement from the measurement obtained by fitting $E_\ell^B$, covering the same phase space (c.f. Figure~\ref{fig:post_fit_1D_2}):
\begin{align}
\bfResc \, .
\end{align}
The uncertainties are larger, but both results are compatible. The nuisance parameter pulls of all fits are provided in Appendix D. The result of Eq.~\ref{eq:bfResD} can be further compared with the most precise measurement to date of this region of Ref.~\cite{TheBABAR:2016lja}, where \mbox{$\Delta \mathcal{B}(B \to X_u \ell \, \nu_\ell) = \left( 1.55 \pm 0.12 \right) \times 10^{-3}$}, and shows good agreement. The measurement can also be compared to Ref.~\cite{Lees:2011fv} using a similar experimental approach. The measured partial branching fraction of \mbox{$E_\ell^B > 1 \, \text{GeV}$} is \mbox{$\Delta \mathcal{B}(B \to X_u \ell \, \nu_\ell) = \left( 1.82 \pm 0.19 \right) \times 10^{-3}$}, which is compatible with Eq.~\ref{eq:bfResD} within 0.9 standard deviations. Belle previously reported in Ref.~\cite{Urquijo:2009tp} using also a similar approach for the same phase space a higher value of \mbox{$\Delta \mathcal{B}(B \to X_u \ell \, \nu_\ell) = \left( 1.96 \pm 0.19 \right) \times 10^{-3}$}. We cannot quantify the statistical overlap between both results, but by comparing the number of determined signal events one can estimate it to be below 55\%. The dominant systematic uncertainties of Ref.~\cite{Urquijo:2009tp} were evaluated using different approaches, but fully correlating the dominant systematic uncertainties and assuming a statistical correlation of 55\% we obtain a compatibility of 1.7 standard deviations. The main difference of this analysis with Ref.~\cite{Urquijo:2009tp} lies in the modeling of signal and background processes: since its publication our understanding improved and more precise measurements of branching fractions and form factors were made available. Further, for the \bulnu signal process in this paper a hybrid approach was adopted (see Section~\ref{sec:data_set_sim_samples} and Appendix A), whereas Ref.~\cite{Urquijo:2009tp} used an alternative approach to model signal as a mix of inclusive and exclusive decay modes. Note that this work supersedes Ref.~\cite{Urquijo:2009tp}.

\subsection{$|\Vub|$ Determination}

We determine $|\Vub|$ from the measured partial branching fractions using a range of theoretical rate predictions. In principle, the total \bulnu decay rate can be calculated using the same approach as \bclnu using the heavy quark expansion (HQE) in inverse powers of $m_b$. Unfortunately, the measurement requirements necessary to separate \bulnu from the dominant \bclnu background spoil the convergence of this approach. In the predictions for the partial rates corresponding to our measurements, perturbative and non-perturbative uncertainties are largely enhanced and as outlined in the introduction the predictions are sensitive to the shape function modeling.

The relationship between measured partial branching fractions, predictions of the rate (omitting CKM factors) $\Delta \Gamma(\bulnu)$, and $|\Vub|$ is
\begin{align}\label{eq:vubdet}
|V_{ub}| = \sqrt{ \frac{ \Delta \mathcal{B}(\bulnu) }{ \tau_B \cdot \Delta \Gamma(\bulnu) } } \, .
\end{align}
with $\tau_B = \left(1.579 \pm 0.004\right) \, \text{ps}$ denoting the average of the charged and neutral \PB meson lifetime~\cite{pdg:2020}. We use four predictions for the theoretical partial rates. All predictions use the same input values as Ref.~\cite{Amhis:2019ckw} chooses for their world averages. 
%
%
The four predictions are:
\begin{itemize}
 \item[-] {\bf BLNP}: The prediction of Bosch, Lange, Neubert,  and Paz (short BLNP) of Ref.~\cite{BLNP} provides a prediction at next-to-leading-order accuracy in terms of the strong coupling constant $\alpha_s$ and incorporates all known corrections. Predictions are interpolated between the shape-function dominated region (endpoint of the lepton spectrum, small hadronic mass) to the region of phase space, that can be described via the operator product expansion (OPE). As input we use $m_b^{\mathrm{SF}} = 4.58 \pm 0.03 \, \text{GeV}$ and $\mu_\pi^{2\, \mathrm{SF}} =  0.20 {}^{+0.09}_{-0.10} \, \text{GeV}^2$. \vspace{1ex}
 \item[-] {\bf DGE}: The Dressed Gluon Approximation (short DGE) from Andersen and Gardi~\cite{DGE1,DGE2} makes predictions by avoiding the direct use of shape functions, but produces predictions for hadronic observables using the on-shell $b$-quark mass. The calculation is carried out in the $\overline{\mathrm{MS}}$ scheme and we use $m_b(\overline{\mathrm{MS}}) = 4.19 \pm 0.04 \, \text{GeV}$.
 \item[-] {\bf GGOU}: The prediction from Gambino, Giordano, Ossola, and Uraltsev~\cite{GGOU} (short GGOU) incorporates all known perturbative and non-perturbative effects up to the order $\mathcal{O}(\alpha_s^2 \, \beta_0)$ and $\mathcal{O}(1/m_{b}^3)$, respectively. The shape function dependence is incorporated by parametrizing its effects in each structure function with a single light-cone function. The calculation is carried out in the kinetic scheme and we use as inputs $m_b^{\mathrm{kin}} = 4.55 \pm 0.02 \,\, \text{GeV}$ and $\mu_\pi^{2\, \mathrm{kin}} =  0.46 \pm 0.08 \, \text{GeV}^2$. 
 \item[-] {\bf ADFR}: The calculation of Aglietti, Di Lodovico, Ferrera, and Ricciardi~\cite{ADFR1,ADFR2} makes use of the ratio of \bulnu to \bclnu rates and soft-gluon resummation at next-to-next-to-leading-order and an effective QCD coupling approach. The calculation uses the $\overline{\mathrm{MS}}$ scheme and we use $m_b(\overline{\mathrm{MS}}) = 4.19 \pm 0.04 \, \text{GeV}$.
\end{itemize}
Table~\ref{tab:theory} lists the decay rates and their associated uncertainties for the probed regions of phase space, which we use to extract $|V_{ub}|$ from the measured partial branching fractions with Eq.~\ref{eq:vubdet}. 

\begin{table*}
\caption{
	The theory rates $\Delta \Gamma(\bulnu)$ from various theory calculations are listed. The rates are given in units of $\text{ps}^{-1}$.
}
\label{tab:theory}
\begin{tabular}{lcccc}
\hline\hline
  Phase-space region& BLNP~\cite{BLNP}  & DGE~\cite{DGE1,DGE2} & GGOU~\cite{GGOU} & ADFR~\cite{ADFR1,ADFR2}  \\
 \hline
    $M_X < 1.7 \, \text{GeV}$   					      & $45.2^{+5.4}_{-4.6}$  & $42.3^{+5.8}_{-3.8}$ & $43.7^{+3.9}_{-3.2}$ & $52.3^{+5.4}_{-4.7}$\\
    $M_X < 1.7 \, \text{GeV}$, $q^2 > 8 \, \text{GeV}^2$   & $23.4^{+3.4}_{-2.6}$ & $24.3^{+2.6}_{-1.9}$ & $23.3^{+3.2}_{-2.4}$ & $31.1^{+3.0}_{-2.6}$\\
    $E_\ell^B > 1 \, \mathrm{GeV}$ 				     & $61.5^{+6.4}_{-5.1}$  & $58.2^{+3.6}_{-3.0}$ & $58.5^{+2.7}_{-2.3}$ & $61.5^{+5.8}_{-5.1}$\\
 \hline\hline
\end{tabular}
\end{table*}

\subsection{$|\Vub|$ Results}

From the partial branching fractions with $E_\ell^B > 1 \, \text{GeV}$ and $M_X < 1.7 \, \text{GeV}$ determined from fitting $M_X$ we find
\begin{align}
\left|\Vub\right| \, (\mathrm{BLNP}) & = \bfResBLNPVuba \, , \nn \\
\left|\Vub\right| \, (\mathrm{DGE}) & = \bfResDGEVuba \, , \nn \\
\left|\Vub\right| \, (\mathrm{GGOU}) & = \bfResGGOUVuba \, , \nn \\
\left|\Vub\right| \, (\mathrm{ADFR}) & = \bfResADFRVuba \, . 
\end{align}
The uncertainties denote the statistical uncertainty, the systematic uncertainty and the theory error from the partial rate prediction. 
For the partial branching fraction with  $E_\ell^B > 1 \, \text{GeV}$, $M_X < 1.7 \, \text{GeV}$, and $q^2 > 8 \, \text{GeV}^2$ we find
\begin{align}
\left|\Vub\right| \, (\mathrm{BLNP}) & = \bfResBLNPVubb \, , \nn \\
\left|\Vub\right| \, (\mathrm{DGE}) & = \bfResDGEVubb \, , \nn \\
\left|\Vub\right| \, (\mathrm{GGOU}) & = \bfResGGOUVubb \, , \nn \\
\left|\Vub\right| \, (\mathrm{ADFR}) & = \bfResADFRVubb \, . 
\end{align}
Finally, the most inclusive determination with $E_\ell^B > 1 \, \text{GeV}$ from the two-dimensional fit of $M_X$ and $q^2$ results in
\begin{align}
\left|\Vub\right| \, (\mathrm{BLNP}) & = \bfResBLNPVubd \, , \nn \\
\left|\Vub\right| \, (\mathrm{DGE}) & = \bfResDGEVubd \, , \nn \\
\left|\Vub\right| \, (\mathrm{GGOU}) & = \bfResGGOUVubd \, , \nn \\
\left|\Vub\right| \, (\mathrm{ADFR}) & = \bfResADFRVubd \, .  \label{eq:Vubprec}
\end{align}

In order to quote a single value for $|V_{ub}|$ we adapt the procedure of Ref.~\cite{pdg_Vxb:2020} and calculate a simple arithmetic average of the most precise determinations in Eq.~\ref{eq:Vubprec} to obtain
\begin{align}
 \bfResAverageVuba \, .
\end{align}
This value is larger, but compatible with the exclusive measurement of $|\Vub|$ from \bpilnu of \mbox{$\left|\Vub\right| = \left(3.67 \pm 0.09 \pm 0.12 \right) \times 10^{-3}$} within 1.3 standard deviations.

\subsection{Stability Checks}

\begin{figure}[b!]
 \includegraphics[width=0.48\textwidth]{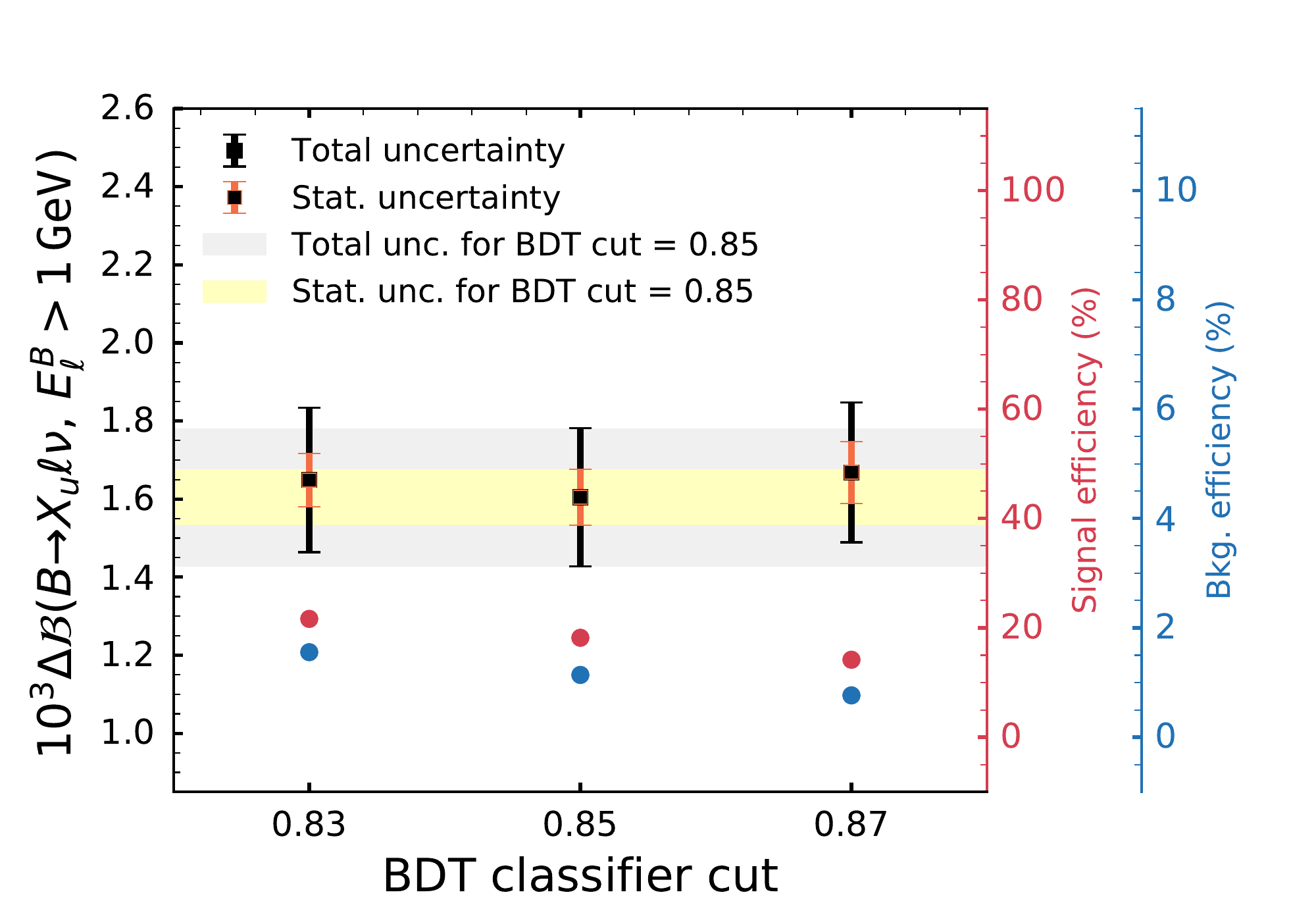} 
\caption{
   The stability of the determined partial branching fraction $\Delta \mathcal{B}(B \to X_u \ell \, \nu_\ell)$ using the $M_{X}:q^2$ fit is studied as a function of the BDT selection requirement. The classifier output selection of $0.83$ and $0.87$ correspond to signal efficiencies after the pre-selection of 22\% and 15\%, respectively. These selections increase, or decrease the background from \bclnu and other processes by 37\% and 33\%, respectively. The grey and yellow bands show the total and statistical error, respectively, with the nominal BDT working point of 0.85.
 }
\label{fig:stability}
\end{figure}

To check the stability of the result we redetermine the partial branching fractions using two additional working points. We change the BDT selection to increase and decrease the amount of \bclnu and other backgrounds, and repeat the full analysis procedure. The resulting values of $\Delta \mathcal{B}(B \to X_u \ell \, \nu_\ell)$ are determined using the two-dimensional fit of $M_X:q^2$ and are shown in Figure~\ref{fig:stability}. The background contamination changes by $+37\%$ and $-33\%$, respectively. The small shifts in central value are well contained within the quoted systematic uncertainties. To further estimate the compatibility of the result we determine the full statistical and systematic correlations of the results and recover that the partial branching fraction with looser and tighter BDT selection are in agreement with the nominal result within 1.1 and 1.4 standard deviations, respectively.

\subsection{\bulnu Charged Pion Multiplicity}\label{subsec:pion_mult_res}

\begin{figure}[b!]
 \includegraphics[width=0.48\textwidth]{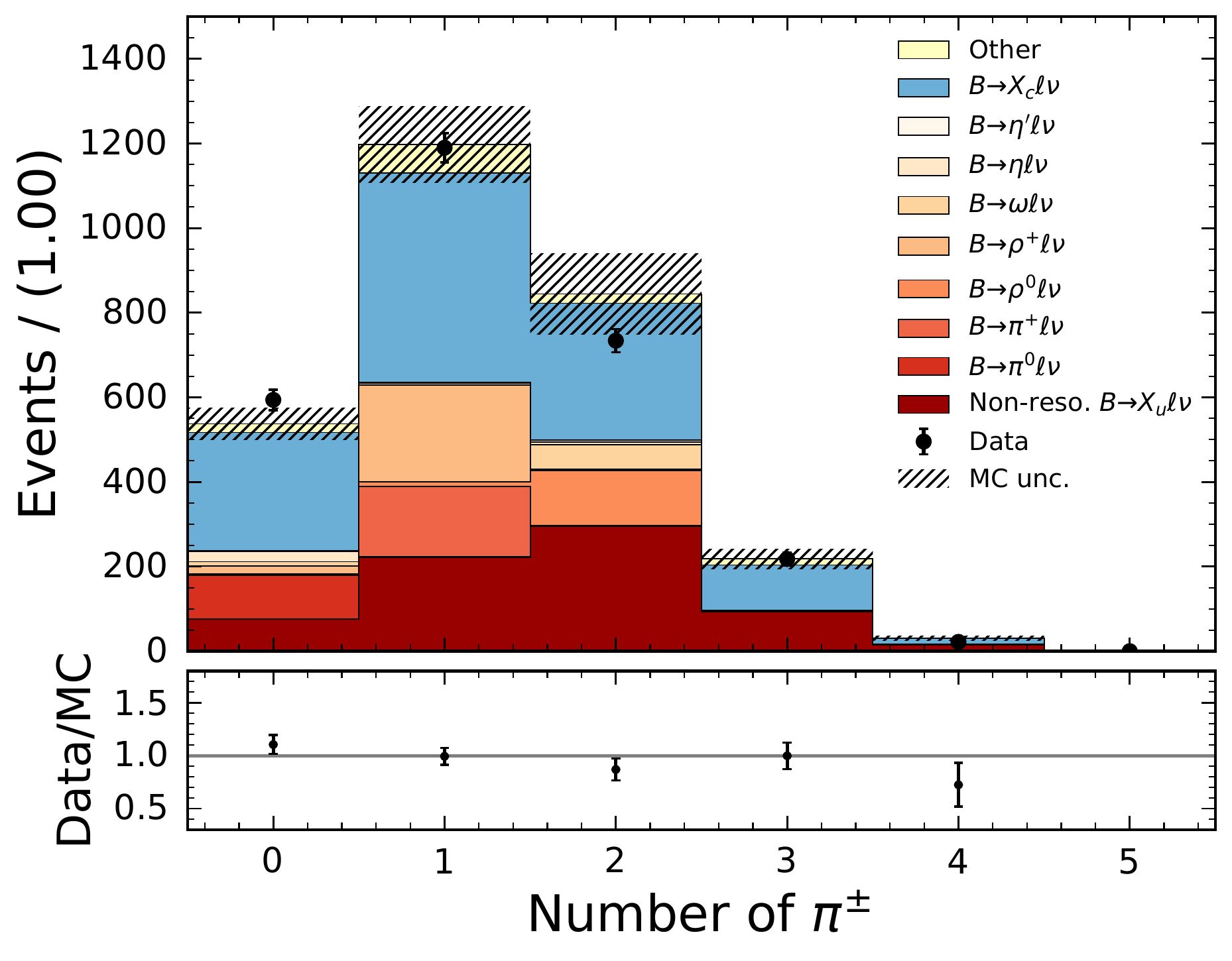} 
\caption{
  The post-fit charged pion multiplicity is shown for events with $M_{X} < 1.7$ GeV. The uncertainties on the MC stack include all systematic uncertainties. 
 }
\label{fig:np_chi_dist_post_fit}
\end{figure}

The modeling the \bulnu\ signal composition is crucial to all presented measurements. One aspect difficult to assess is the $X_u$ fragmentation simulation: the charmless $X_u$ state can decay via many different channels producing a number of charged or neutral pions or kaons. In Section~\ref{sec:syst} we discussed how we assess the uncertainty on the number of $s \bar s$ quark pairs produced in the $X_u$ fragmentation. Due to the BDT removing such events to suppress the dominant \bclnu\ background, no signal-enriched region can be easily obtained. The accuracy of the fragmentation into the number of charged pions can be tested in the signal enriched region of $M_X < 1.7 \, \mathrm{GeV}$. Figure~\ref{fig:np_chi_dist_post_fit} compares the charged pion multiplicity between simulated signal and background processes and data. The signal and background predictions are scaled to their respective normalizations obtained from the two-dimensional fit in $M_{X}:q^2$. The uncertainty band shown on the MC includes the full systematic uncertainties discussed in Section~\ref{sec:syst}. The agreement overall lies within the assigned uncertainties, with the data having more events in the zero multiplicity bin and less in the two charged pion multiplicity bin. We use this distribution to correct our simulation to assign an additional uncertainty from the charged pion fragmentation. More details can be found in Section~\ref{sec:syst} and Appendix C.

\subsection{Lepton Flavor Universality and Weak Annihilation Contributions}

To test the lepton flavor universality in \bulnu we also carry out fits to determine the partial branching fraction for electron and muon final states. For this we categorzie the selected events accordingly and carry out a fit to the $M_X:q^2$ distributions using the same granularity as the fit described in Section~\ref{sec:partial_BF_res}. We carry out a simultaneous analysis of both samples, such that shared NPs for the modeling of the signal or background components can be correctly correlated afterwards. The resulting yields are corrected to a partial branching fraction with $E_\ell^B > 1 \, \text{GeV}$ and we obtain 
\begin{align}
\bfResdEl \, , \\
\bfResdMu \, ,
\end{align}
with a total correlation of $\rho = 0.53$. The ratio of the electron to the muon final state is
\begin{align}
\rElMu \, ,
\end{align}
with the first error denoting the statistical uncertainty and the second the systematic uncertainty. We observe no significant deviation from lepton flavor universality. More details on the fit can be found in Appendix E.

Isospin breaking effects can be studied by separately measuring the partial branching fraction for charged and neutral $B$ meson final states. We determine the ratio 
\begin{align} \label{eq:isospin_ratio}
 R_{\mathrm{iso}} = \frac{ \tau_{B^0} }{ \tau_{B^+} } \times \frac{ \Delta \mathcal{B}(B^+ \to X_u \ell^+ \, \nu_\ell) }{ \Delta \mathcal{B}(B^0 \to X_u \ell^+ \, \nu_\ell) } \, ,
\end{align}
by using the information from the composition of the fully reconstructed tag-side $B$-meson decays to separate charged and neutral $B$ candidates. The partial branching fraction is then determined by a simultaneous fit of both samples in $M_X:q^2$ to correctly correlate common systematic uncertainties. To account for the small contamination of wrongly assigned $B$ tag flavors, we use the wrong-tag fractions from our simulation. The measured number of signal events in the reconstructed neutral and charged $B$ candidate categories (denoted in the following as $N_{\mathrm{reco}}^0 $ and $ N_{\mathrm{reco}}^+ $) are related to the number of neutral and charged $B$ mesons ($N_{\mathrm{true}}^0 $ and $ N_{\mathrm{true}}^+ $) via
\begin{align}
 N_{\mathrm{reco}}^0 = \mathcal{P}_{B^0_{\mathrm{true}} \to B^0_{\mathrm{reco}} } \, N_{\mathrm{true}}^0 + \mathcal{P}_{B^+_{\mathrm{true}} \to B^0_{\mathrm{reco}} } \, N_{\mathrm{true}}^+ \, , \\
 N_{\mathrm{reco}}^+ = \mathcal{P}_{B^0_{\mathrm{true}} \to B^+_{\mathrm{reco}} } \, N_{\mathrm{true}}^0 + \mathcal{P}_{B^+_{\mathrm{true}} \to B^+_{\mathrm{reco}} } \, N_{\mathrm{true}}^+ \, .
\end{align}
Here e.g. $\mathcal{P}_{B^0_{\mathrm{true}} \to B^+_{\mathrm{reco}} } $ denotes the probability to identify in the reconstruction of the tag-side $B$-meson a true $B^0$ as a $B^+$ candidate. In the simulation we find 
\begin{align}
  \mathcal{P}_{B^0_{\mathrm{true}} \to B^0_{\mathrm{reco}} }  = 0.985 \, \quad   \mathcal{P}_{B^0_{\mathrm{true}} \to B^+_{\mathrm{reco}} }  = 0.015 \, , \\
  \mathcal{P}_{B^+_{\mathrm{true}} \to B^+_{\mathrm{reco}} }  = 0.977 \, \quad   \mathcal{P}_{B^+_{\mathrm{true}} \to B^0_{\mathrm{reco}} }  = 0.023 \, .
\end{align}
Using this procedure we determine for the individual partial branching fractions with $E_\ell^B > 1 \, \text{GeV}$
\begin{align}
\bfResdBp \, , \\
\bfResdBn \, , 
\end{align}
with a total correlation of $\rho = 0.50$ and for the ratio Eq.~\ref{eq:isospin_ratio}
\begin{align} \label{res:isospin_ratio}
 \rBpBn \, ,
\end{align}
compatible with the expectation of equal semileptonic rates for both isospin states. Isospin breaking effects would for instance arise from weak annihilation contributions, which only can contribute to charged $B$ meson final states. Using Eq.~\ref{res:isospin_ratio} the relative contribution from weak annihilation processes to the total semileptonic $B \to X_u \ell^+ \, \nu_\ell$ rate can be constrained via
\begin{align}
 \frac{ \Gamma_{\mathrm{wa}} }{ \Gamma(B \to X_u \ell^+ \, \nu_\ell) } = \frac{f_u}{f_{\mathrm{wa}}} \times \left( R_{\mathrm{iso}} -1 \right) \, .
\end{align}
Here $f_u$ is a factor that corrects the measured partial branching fraction to the full inclusive phase space. We estimate it using the DFN model~\cite{DeFazio:1999ptt} (cf. Section~\ref{sec:data_set_sim_samples} for details) and find $f_u = 0.86$. We further assume that $f_{\mathrm{wa}} = 1$, as such processes would produce a high momentum lepton. We recover
\begin{align}
 \frac{ \Gamma_{\mathrm{wa}} }{ \Gamma(B \to X_u \ell^+ \, \nu_\ell) } = 0.01 \pm 0.09 \, , 
\end{align}
which translates into a limit of $ [-0.14,0.17]$ at 90\% CL. This result is more stringent than the limit of Ref.~\cite{Lees:2011fv}, but weaker than the result of Ref.~\cite{Rosner:2006zz}, that directly used the shape of the $q^2$ distribution to constrain weak annihilation processes. Our result is also weaker than the estimates of Refs.~\cite{, Gambino:2010jz, Ligeti:2010vd,Voloshin:2001xi,Bigi:1993bh} that constrain weak annihilation contributions to be of the order 2-3\%.

\clearpage

\section{Summary and Conclusions}\label{sec:summary}

\begin{figure}[h!]
 \includegraphics[width=0.48\textwidth]{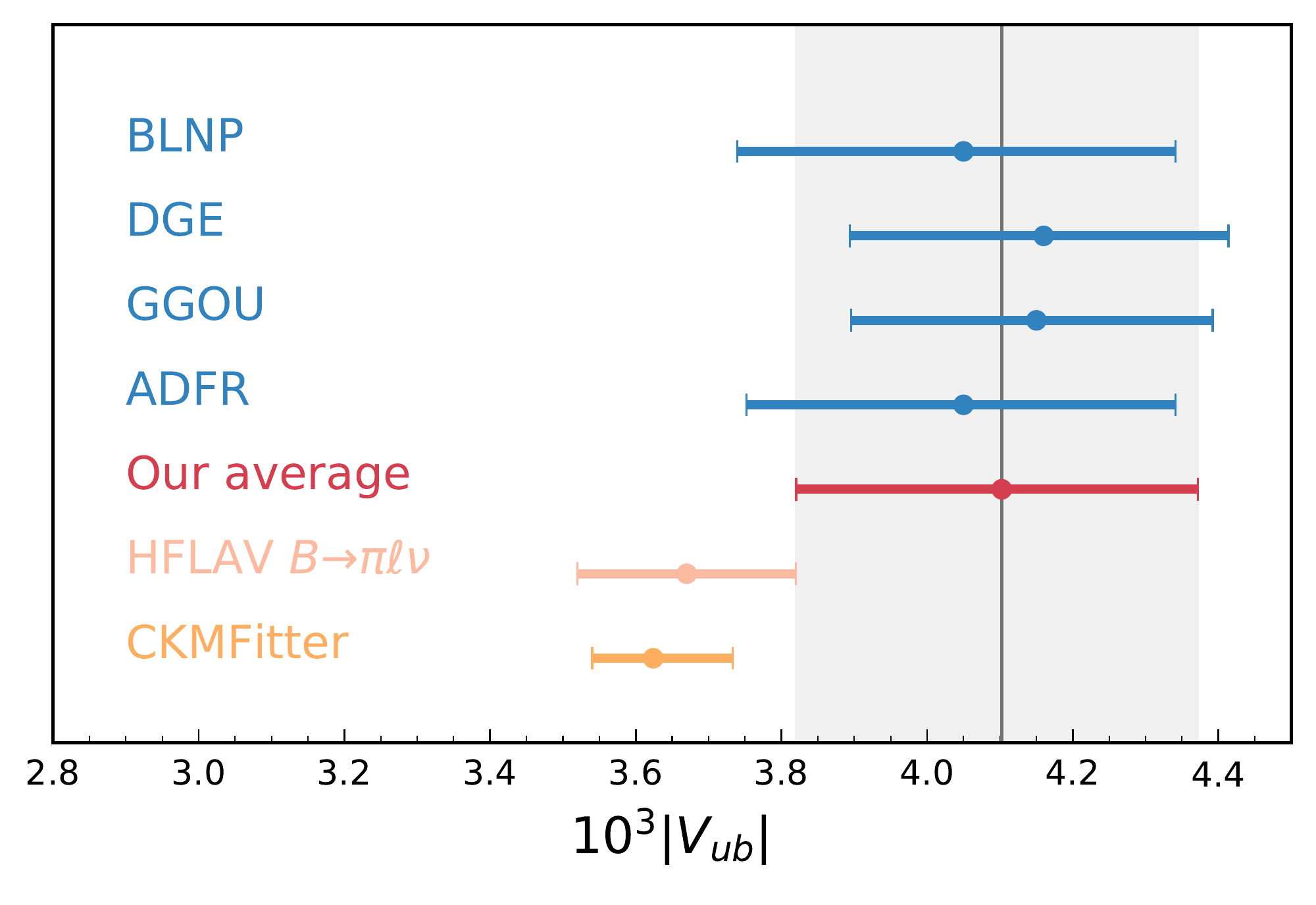} 
\caption{
   The obtained values of $|\Vub|$ from the four calculations and the arithmetic average is compared to the determination from exclusive \bpilnu and the expectation from CKM unitarity~\cite{Charles:2004jd} without the direct constraints from semileptonic and leptonic decays. 
 }
\label{fig:summary}
\end{figure}

We report measurements of partial branching fractions with different requirements on the properties of the hadronic system of the \bulnu decay and with a lepton energy of $E_\ell^B > 1\, \text{GeV}$ in the \PB rest-frame, covering 31-86\% of the available phase space. The sizeable background from semileptonic \bclnu decays is suppressed using multivariate methods in the form of a BDT. This approach allows us to reduce such backgrounds to an acceptable level, whilst retaining a high signal efficiency. Signal yields are obtained using a binned likelihood fit in either the reconstructed hadronic mass $M_X$, the four-momentum-transfer squared $q^2$, or the lepton energy $E_\ell^B$. The most precise result is obtained from a two-dimensional fit of $M_X$ and $q^2$. Translated to a partial branching fraction for $E_\ell^B > 1 \, \text{GeV}$ we obtain
\begin{align}
\bfResd \, ,
\end{align}
with the errors denoting statistical and systematic uncertainties. The partial branching fraction is compatible with the value obtained by a fit of the lepton energy spectrum $E_\ell^B$ and with the most precise determination of Ref.~\cite{TheBABAR:2016lja}. In addition, it is stable under variations of the background suppression BDT. From this partial branching fraction we obtain a value of 
\begin{align}
\bfResAverageVuba
\end{align}
from an average over four theoretical calculations. This value is higher than, but compatible with, the value of $|\Vub|$ from exclusive determinations by 1.3 standard deviations. The compatibility with the value expected from CKM unitarity from a fit of Ref.~\cite{Charles:2004jd} of $|V_{ub}| = \left( 3.62 {}^{+0.11}_{-0.08} \right) \times 10^{-3}$ is 1.6 standard deviations. Figure~\ref{fig:summary} summarizes the situation. The result presented here supersedes Ref.~\cite{Urquijo:2009tp}: this paper uses a more efficient tagging algorithm, incorporates improvements of the \bulnu signal and \bclnu background descriptions, and analyzes the full Belle data set of 711 fb${}^{-1}$. The measurement of kinematic differential shapes of $M_X$, $q^2$, and other properties are left for future work. These results will be crucial for future direct measurements with Belle II that will attempt to use data-driven methods to directly constrain the shape function using \bulnu information. 

\acknowledgments

We thank Kerstin Tackmann, Frank Tackmann, Zoltan Ligeti, Ian Stewart, Thomas Mannel, and Keri Voss for useful discussions about the subject matter of this manuscript. LC, WS, RVT, and FB were supported by the DFG Emmy-Noether Grant No.\ BE~6075/1-1. WS was supported by the Alexander von Humboldt Foundation. FB is dedicating this paper to his father Urs Bernlochner, who sadly passed away during the writing of this manuscript. We miss you so much. We thank the KEKB group for the excellent operation of the
accelerator; the KEK cryogenics group for the efficient
operation of the solenoid; and the KEK computer group, and the Pacific Northwest National
Laboratory (PNNL) Environmental Molecular Sciences Laboratory (EMSL)
computing group for strong computing support; and the National
Institute of Informatics, and Science Information NETwork 5 (SINET5) for
valuable network support.  We acknowledge support from
the Ministry of Education, Culture, Sports, Science, and
Technology (MEXT) of Japan, the Japan Society for the 
Promotion of Science (JSPS), and the Tau-Lepton Physics 
Research Center of Nagoya University; 
the Australian Research Council including grants
DP180102629, 
DP170102389, 
DP170102204, 
DP150103061, 
FT130100303; 
Austrian Federal Ministry of Education, Science and Research (FWF) and
FWF Austrian Science Fund No.~P~31361-N36;
the National Natural Science Foundation of China under Contracts
No.~11435013,  
No.~11475187,  
No.~11521505,  
No.~11575017,  
No.~11675166,  
No.~11705209;  
Key Research Program of Frontier Sciences, Chinese Academy of Sciences (CAS), Grant No.~QYZDJ-SSW-SLH011; 
the  CAS Center for Excellence in Particle Physics (CCEPP); 
the Shanghai Pujiang Program under Grant No.~18PJ1401000;  
the Shanghai Science and Technology Committee (STCSM) under Grant No.~19ZR1403000; 
the Ministry of Education, Youth and Sports of the Czech
Republic under Contract No.~LTT17020;
Horizon 2020 ERC Advanced Grant No.~884719 and ERC Starting Grant No.~947006 ``InterLeptons'' (European Union);
the Carl Zeiss Foundation, the Deutsche Forschungsgemeinschaft, the
Excellence Cluster Universe, and the VolkswagenStiftung;
the Department of Atomic Energy (Project Identification No. RTI 4002) and the Department of Science and Technology of India; 
the Istituto Nazionale di Fisica Nucleare of Italy; 
National Research Foundation (NRF) of Korea Grant
Nos.~2016R1\-D1A1B\-01010135, 2016R1\-D1A1B\-02012900, 2018R1\-A2B\-3003643,
2018R1\-A6A1A\-06024970, 2018R1\-D1A1B\-07047294, 2019K1\-A3A7A\-09033840,
2019R1\-I1A3A\-01058933;
Radiation Science Research Institute, Foreign Large-size Research Facility Application Supporting project, the Global Science Experimental Data Hub Center of the Korea Institute of Science and Technology Information and KREONET/GLORIAD;
the Polish Ministry of Science and Higher Education and 
the National Science Center;
the Ministry of Science and Higher Education of the Russian Federation, Agreement 14.W03.31.0026, 
and the HSE University Basic Research Program, Moscow; 
University of Tabuk research grants
S-1440-0321, S-0256-1438, and S-0280-1439 (Saudi Arabia);
the Slovenian Research Agency Grant Nos. J1-9124 and P1-0135;
Ikerbasque, Basque Foundation for Science, Spain;
the Swiss National Science Foundation; 
the Ministry of Education and the Ministry of Science and Technology of Taiwan;
and the United States Department of Energy and the National Science Foundation.

\bibliographystyle{apsrev4-1}
\bibliography{BtoXulnu}

\vspace{4ex}

\clearpage

\begin{appendix}

\onecolumngrid

\section*{A. \bulnu Hybrid MC Details}

\begin{figure}[h!]
  \includegraphics[width=0.4\textwidth]{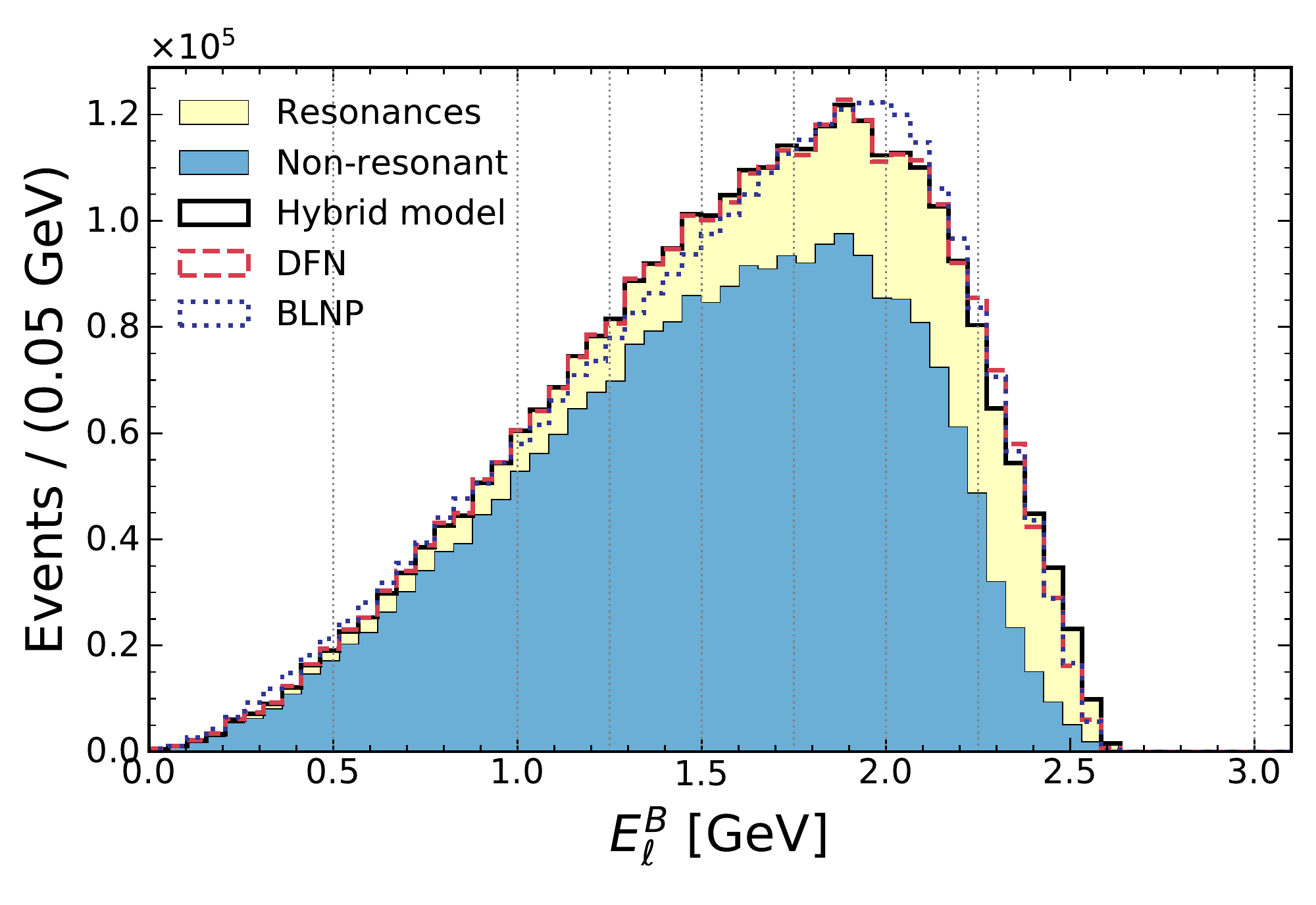} 
  \includegraphics[width=0.4\textwidth]{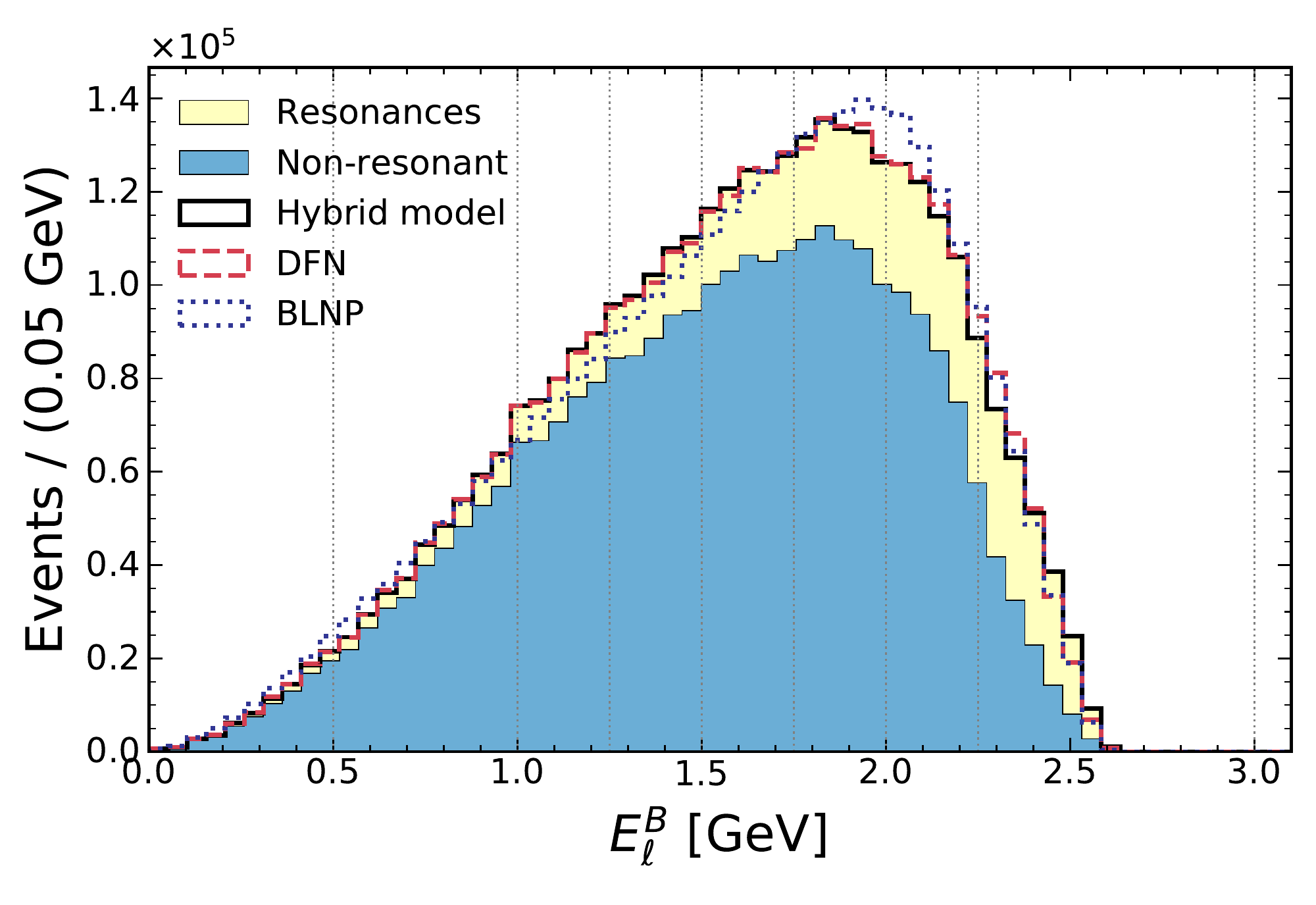}  \\
  \includegraphics[width=0.4\textwidth]{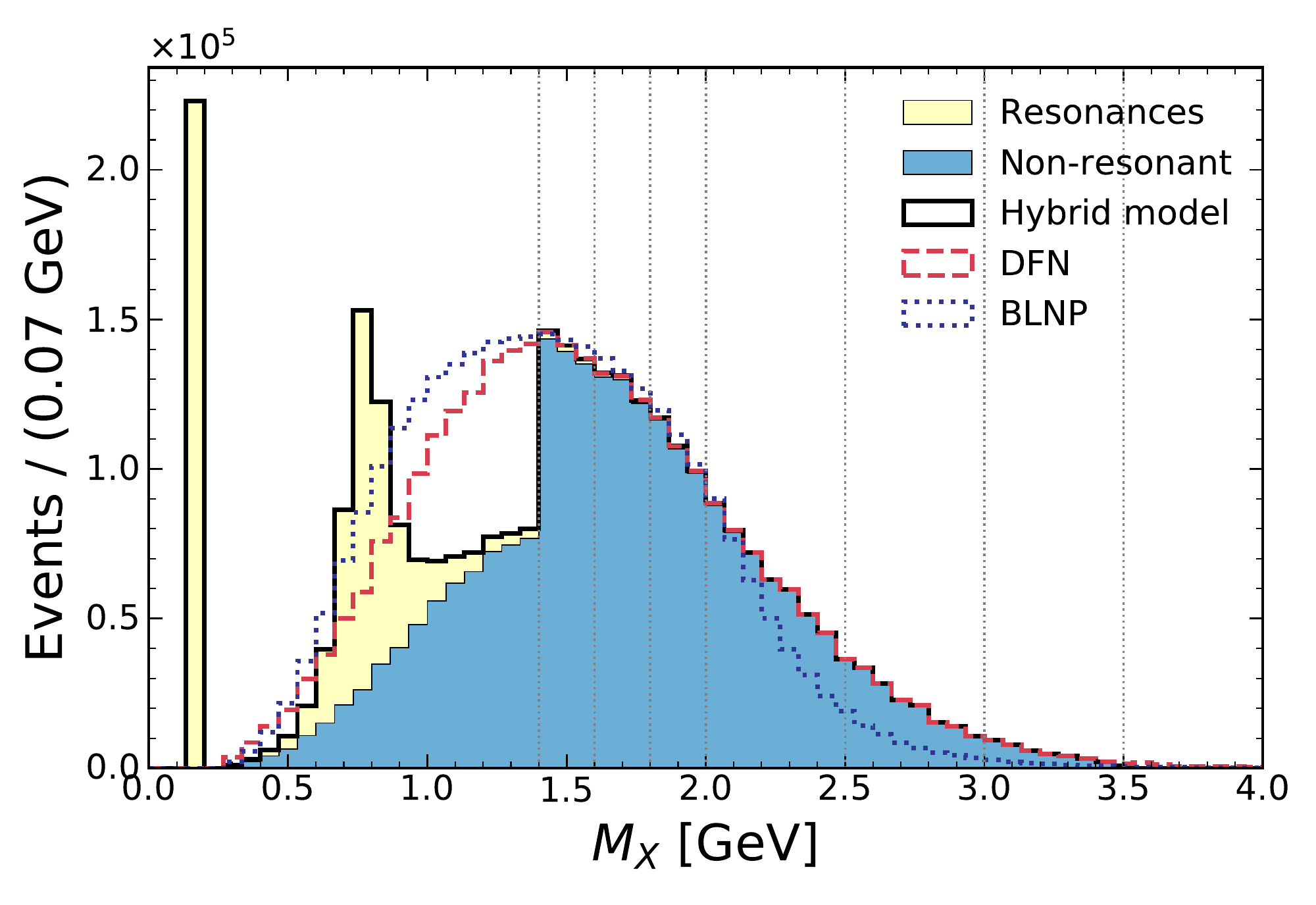} 
  \includegraphics[width=0.4\textwidth]{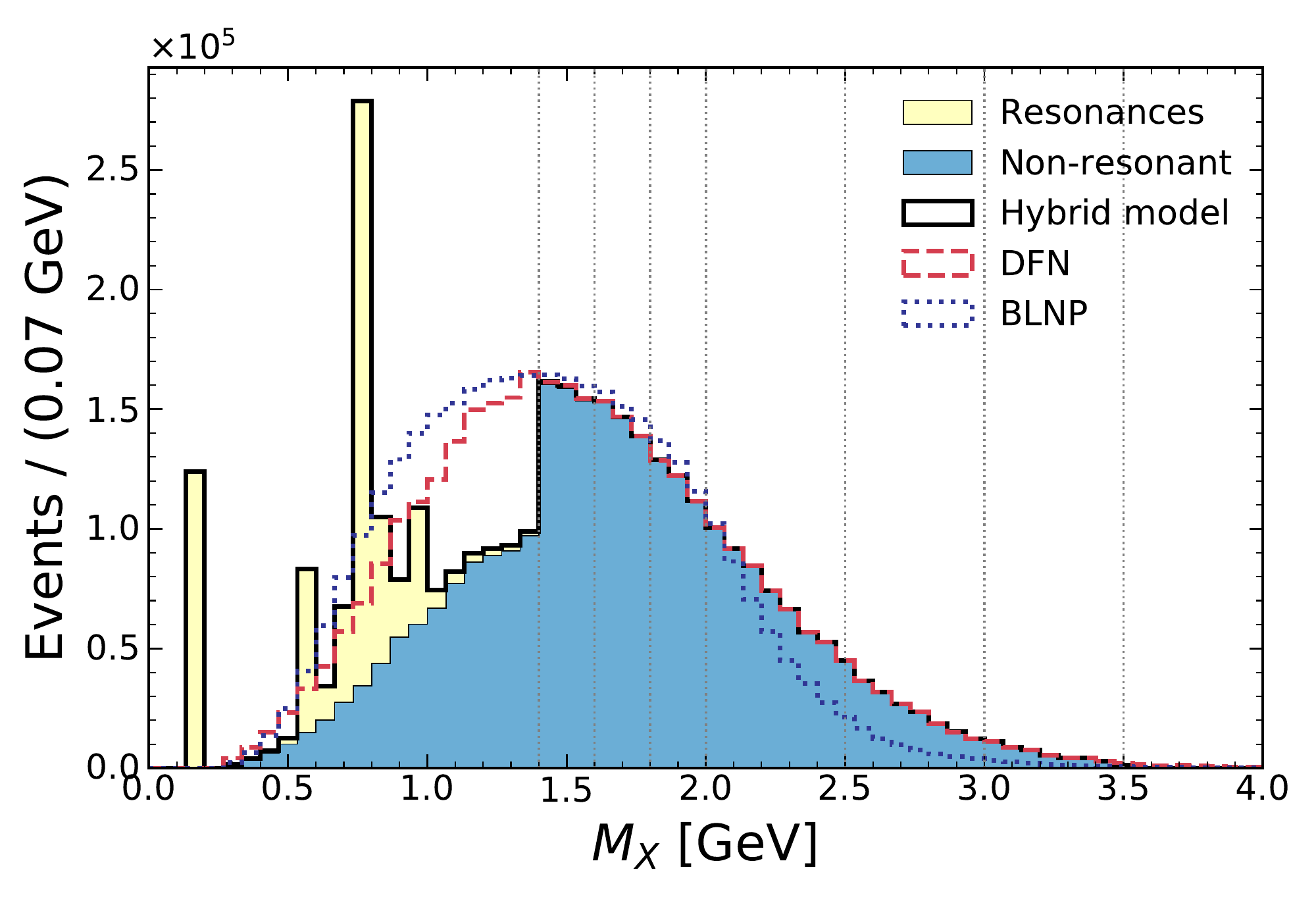}  \\
  \includegraphics[width=0.4\textwidth]{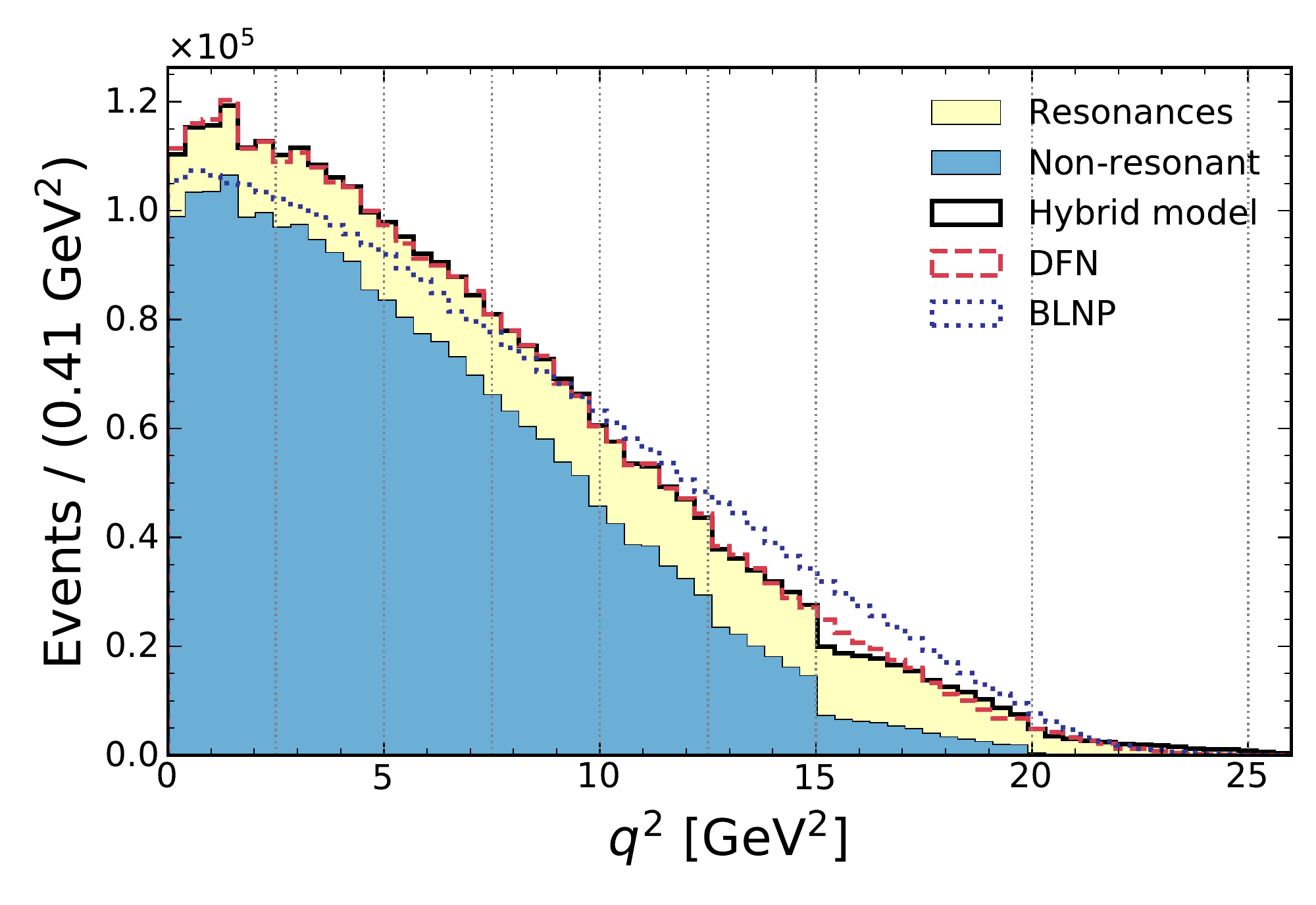} 
  \includegraphics[width=0.4\textwidth]{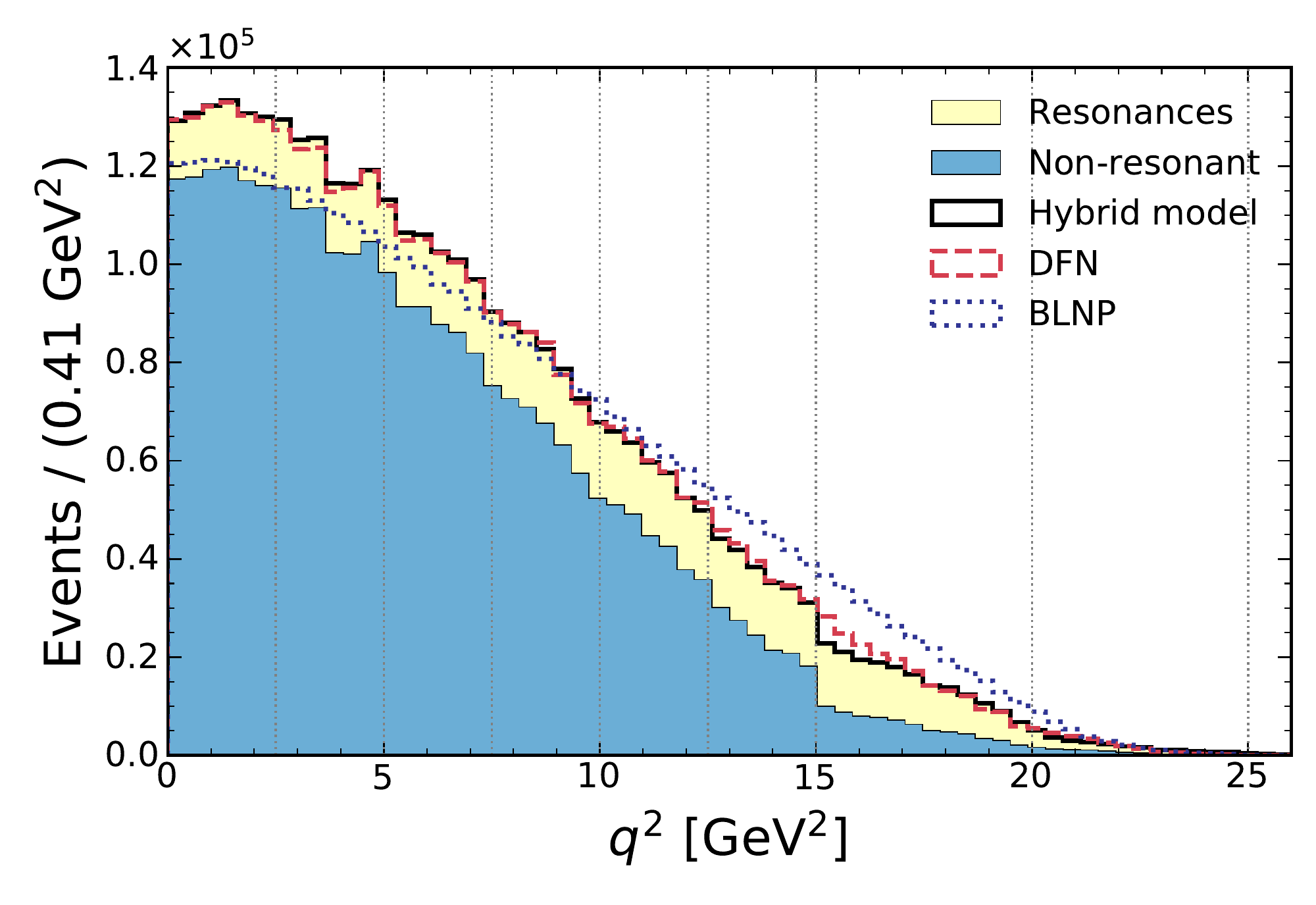} 
\caption{
  The generator level \bulnu distributions $E_\ell^B$, $M_{X}$, and $q^2$ for neutral (left) and charged (right) $B$ mesons are shown. The black histogram shows the merged hybrid model, composed of resonant and non-resonant contributions. For more details on the used models and how the hybrid \bulnu signal sample is constructed, see Section~\ref{sec:data_set_sim_samples}.
 }
\label{fig:hybridvars}
\end{figure}

Figure~\ref{fig:hybridvars} shows the generator level hybrid \bulnu signal sample for $E_\ell^B$, $M_{X}$, and $q^2$ described in Section~\ref{sec:data_set_sim_samples}.

\section*{B. Input variables of $B \to X_c \ell \bar \nu_\ell$ suppression BDT} \label{app:bdtvars}

\begin{figure}[h!]
  \includegraphics[width=0.4\textwidth]{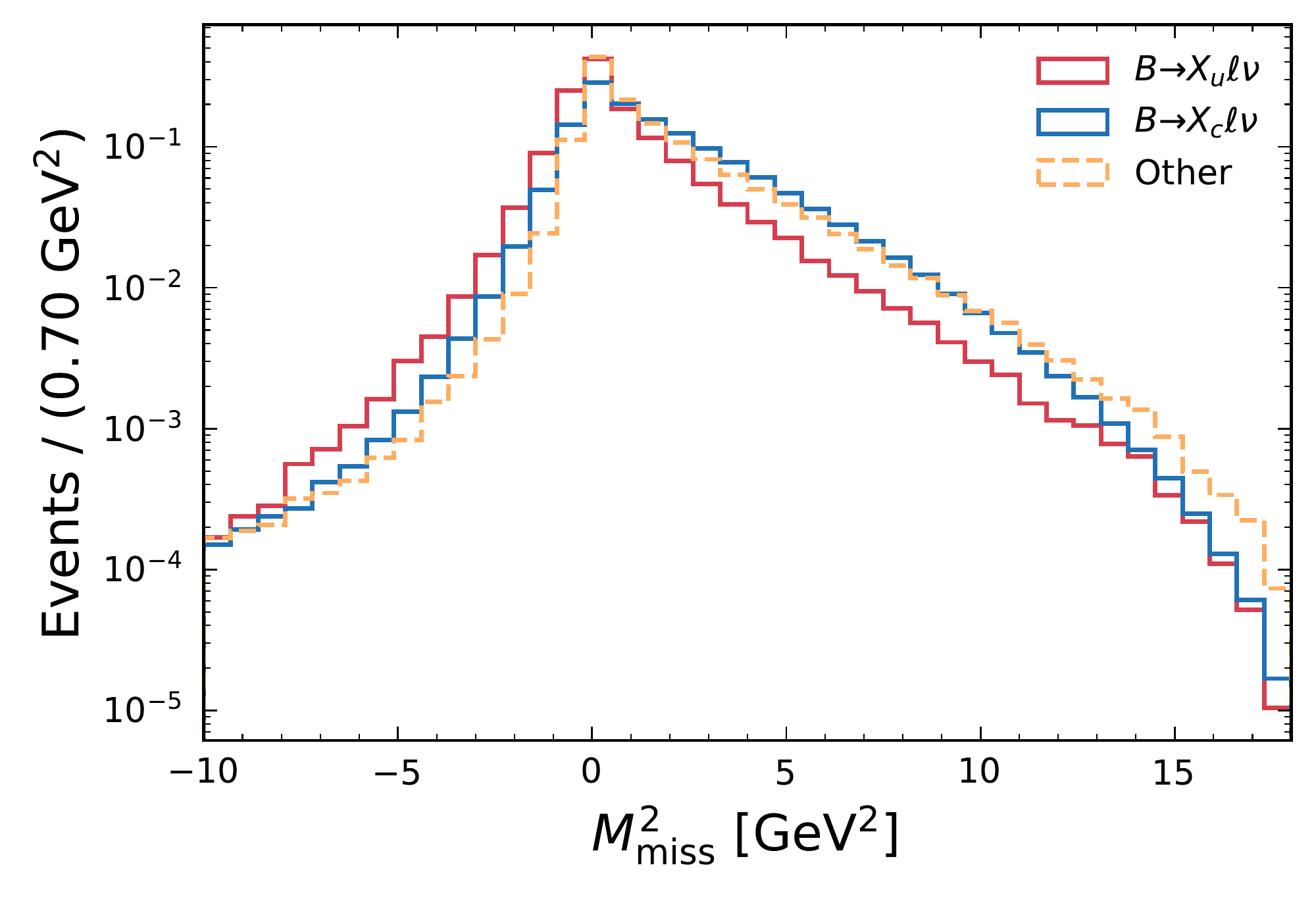} 
  \includegraphics[width=0.4\textwidth]{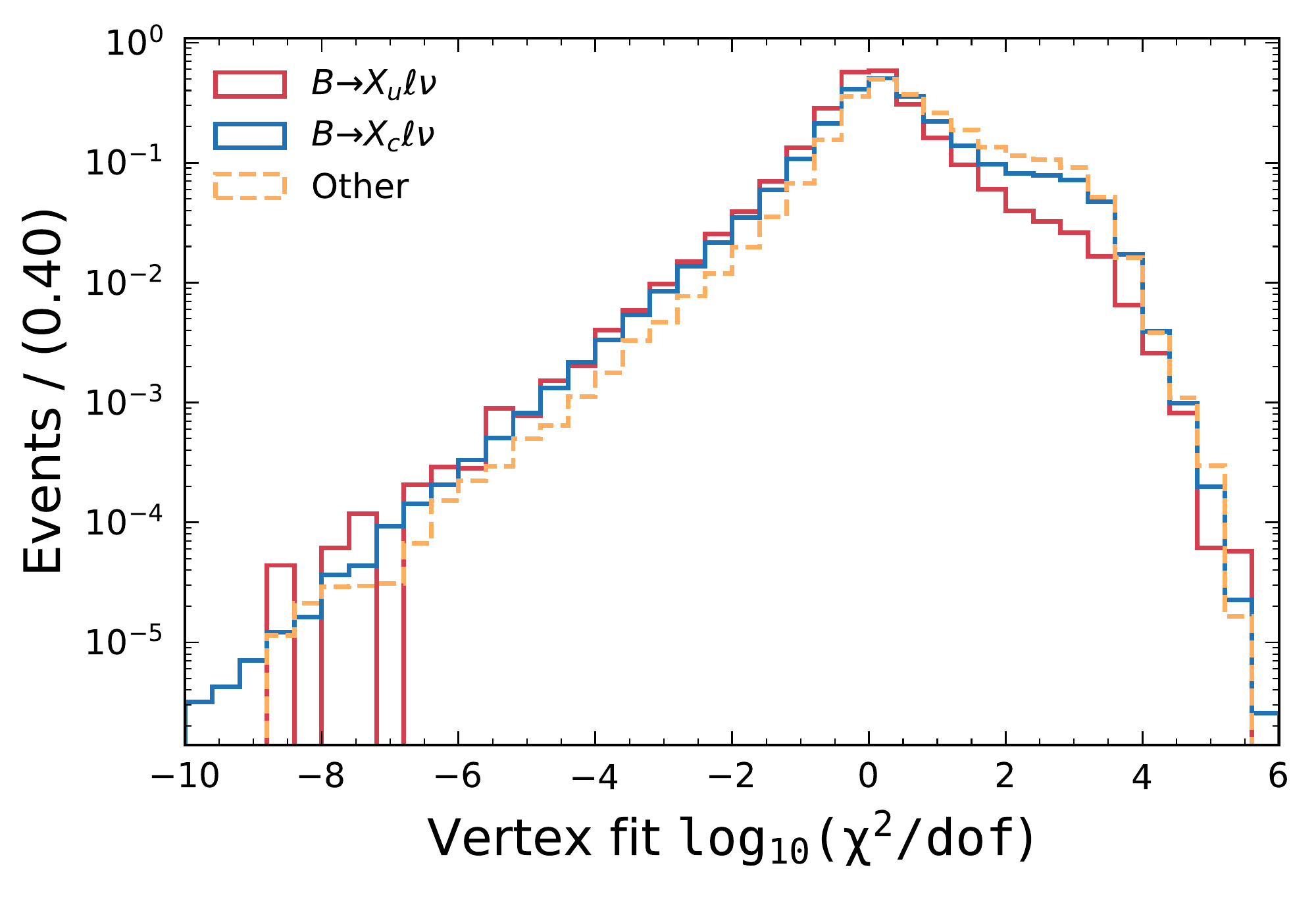}  \\
  \includegraphics[width=0.4\textwidth]{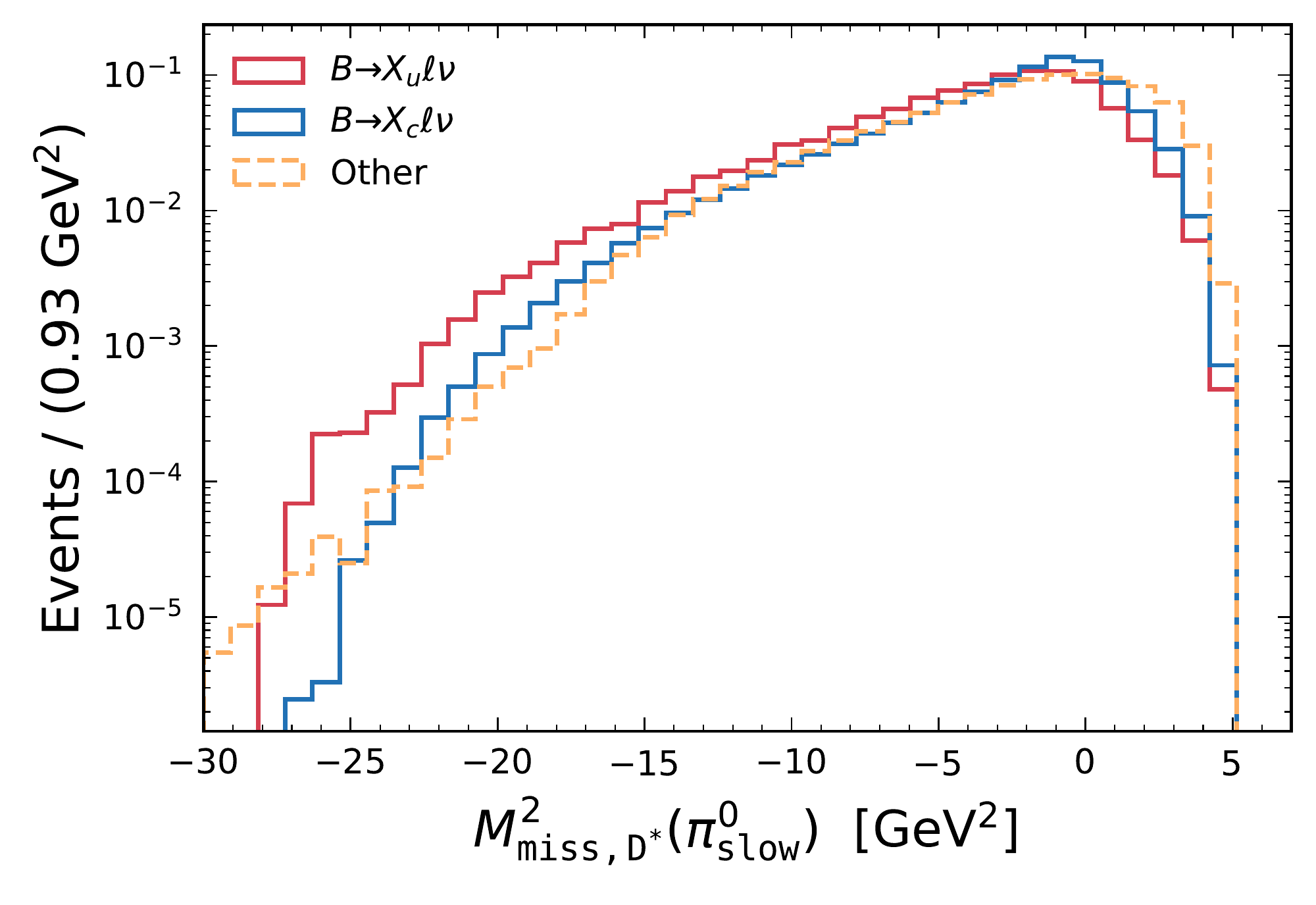}  
  \includegraphics[width=0.4\textwidth]{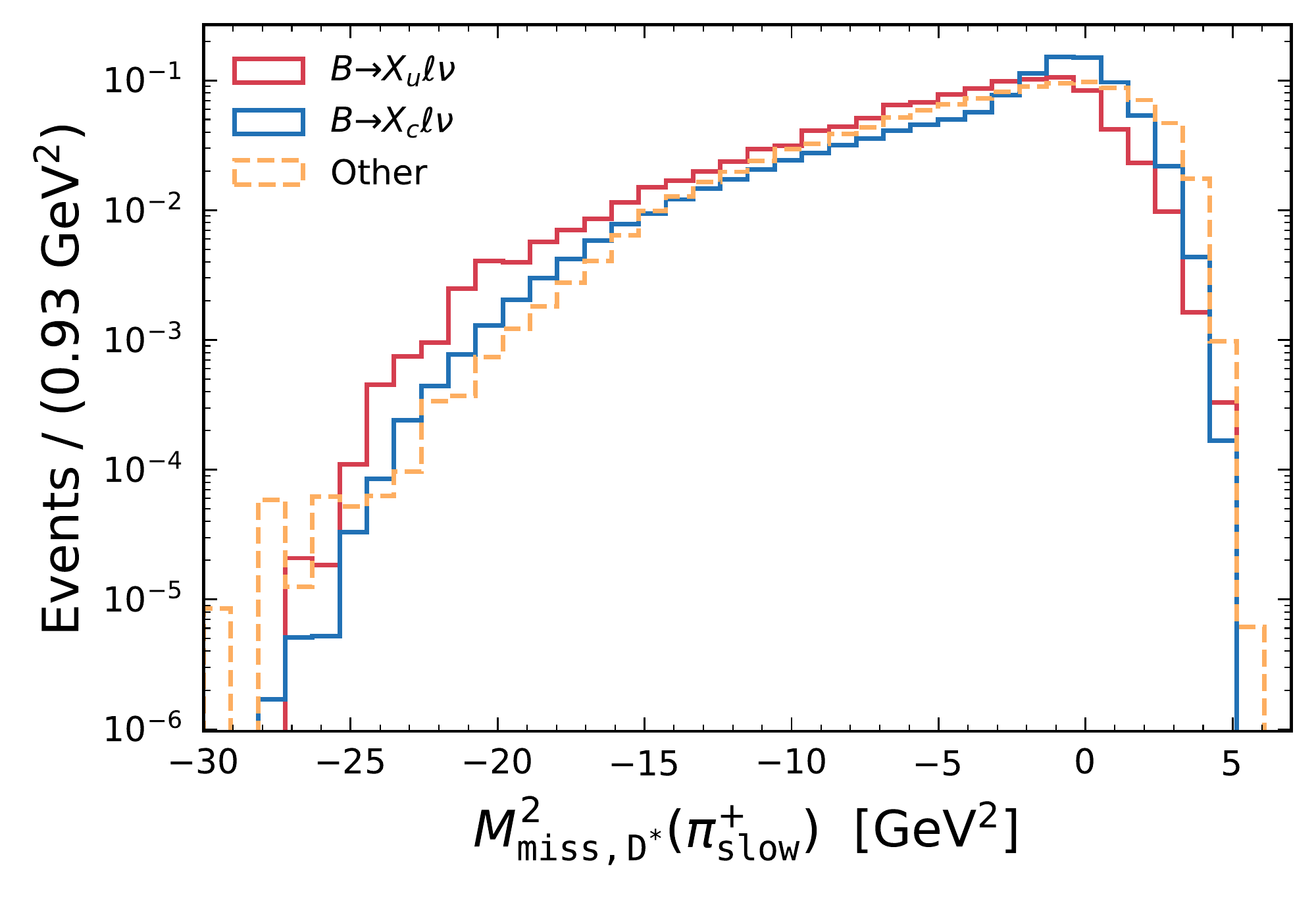} \\
  \includegraphics[width=0.4\textwidth]{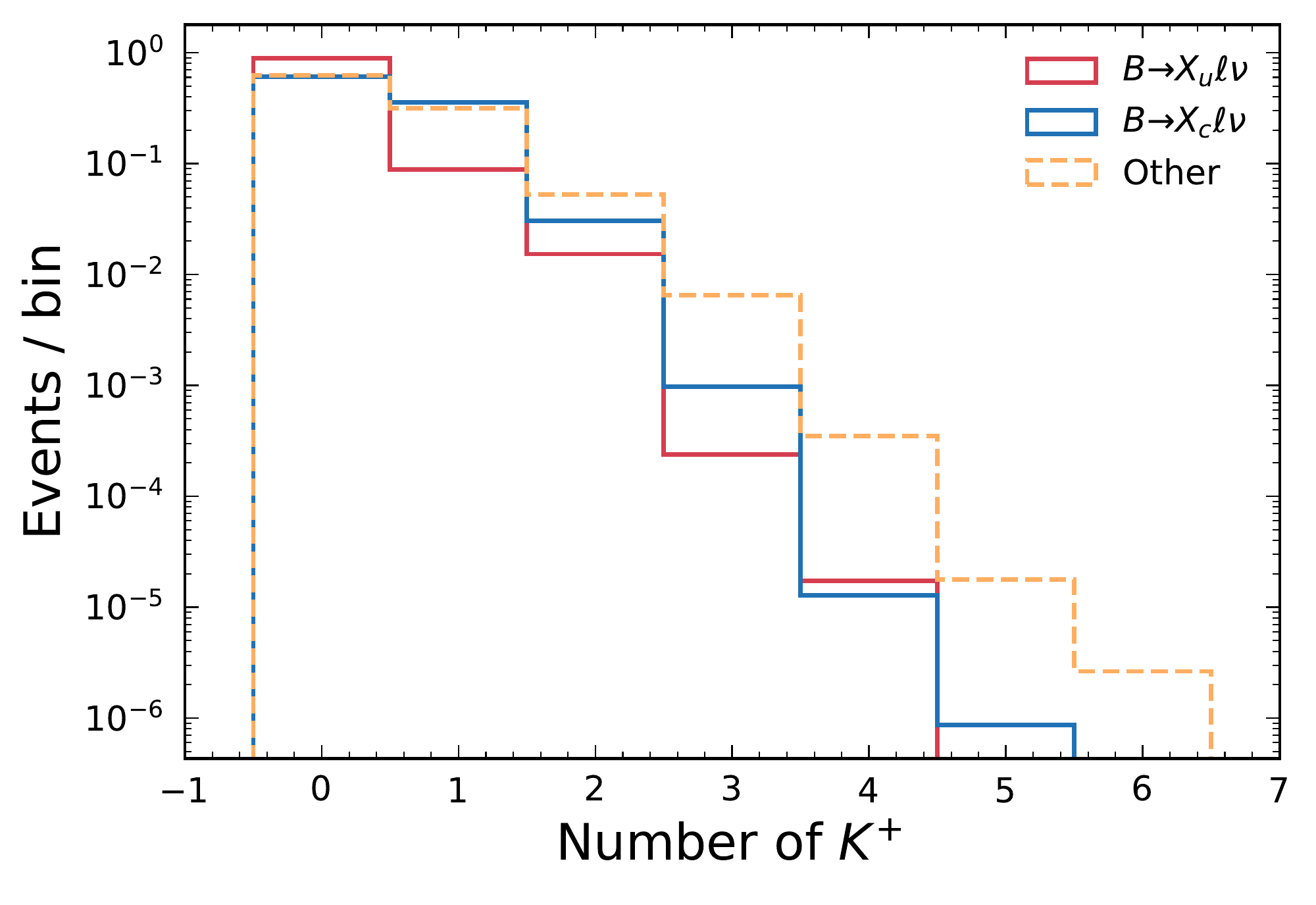} 
  \includegraphics[width=0.4\textwidth]{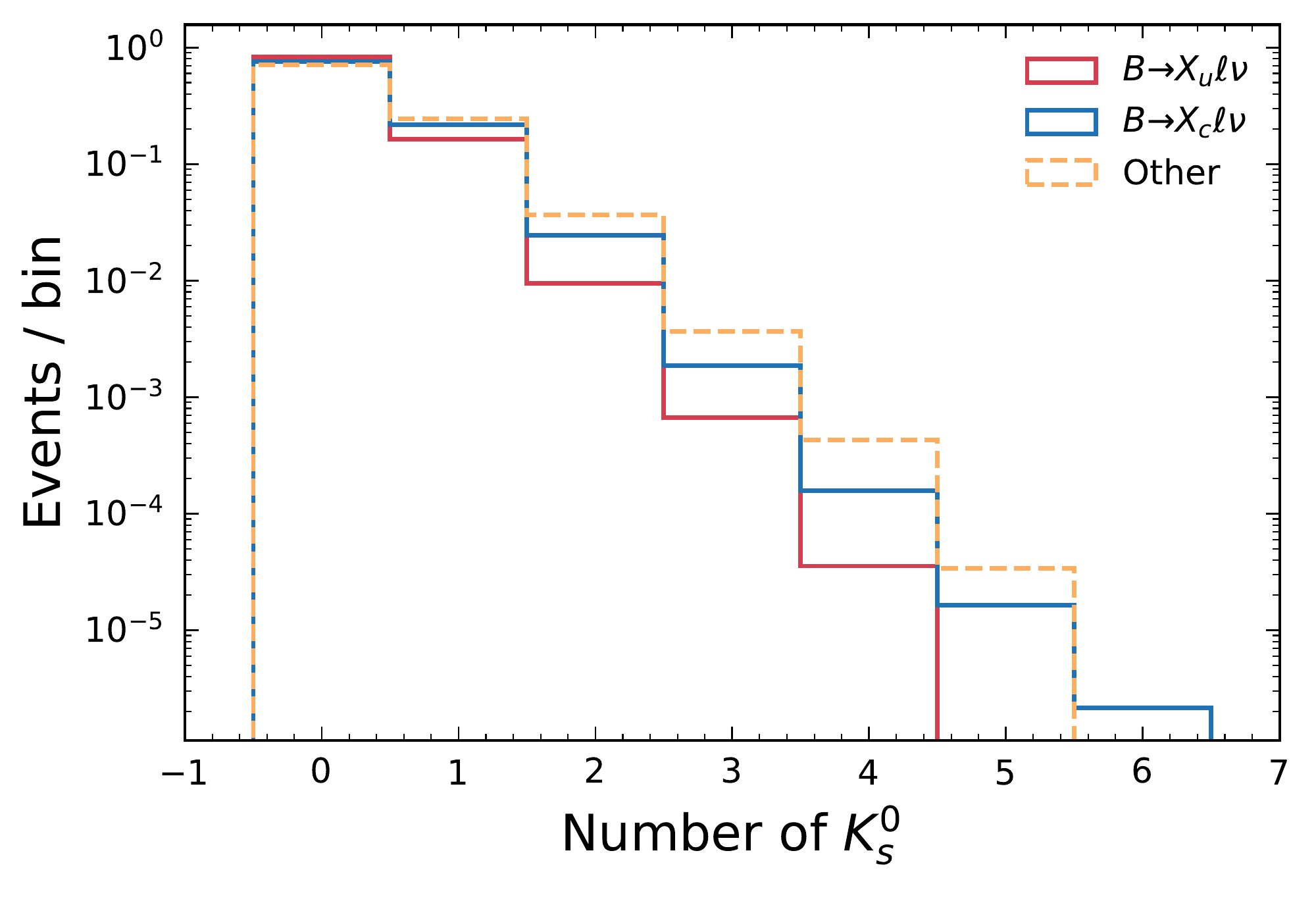}  \\
  \includegraphics[width=0.4\textwidth]{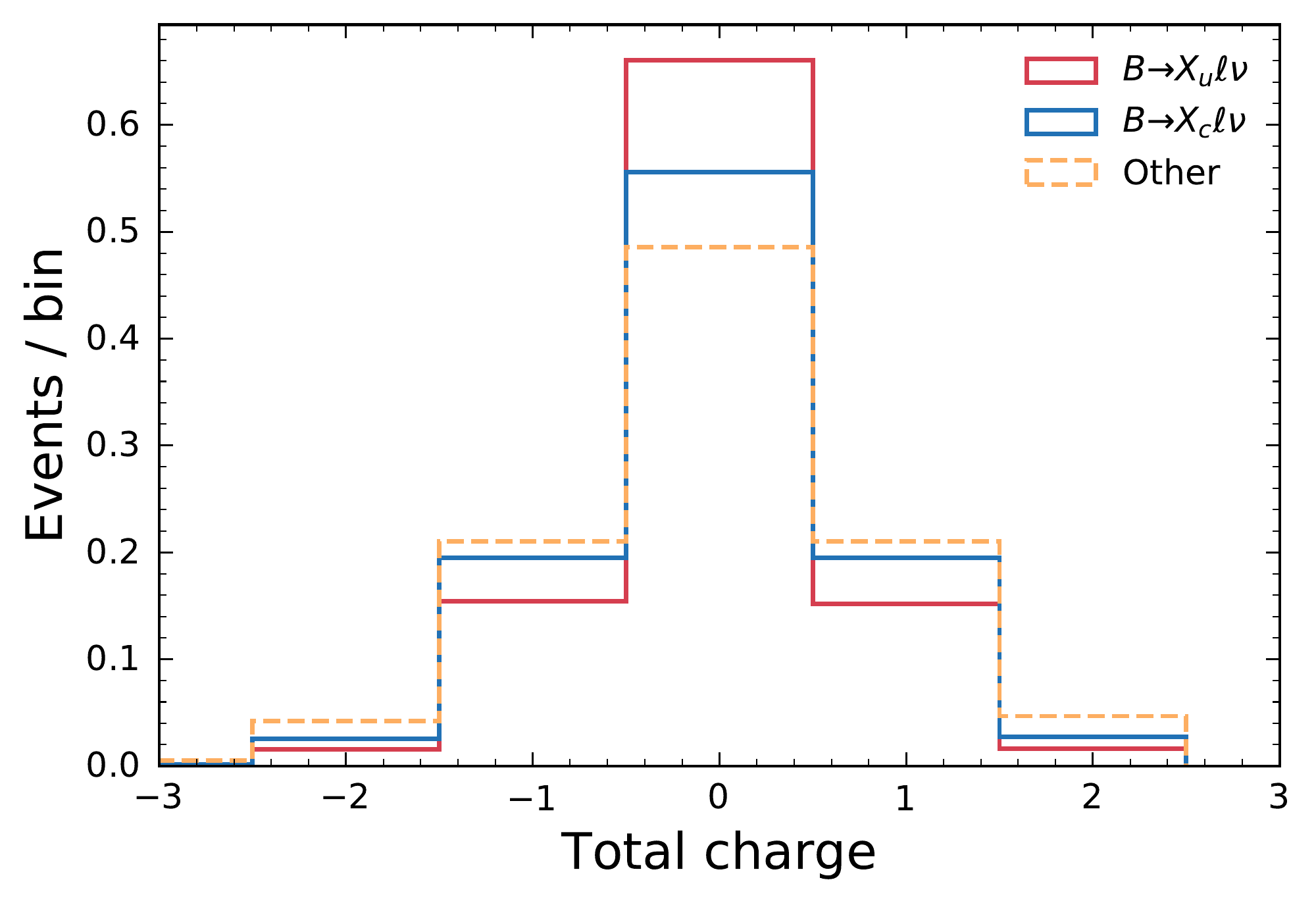}   
\caption{
  The shape of the input variables for the \bclnu background suppression BDT are shown. For details and definitions see Section~\ref{sec:bkg_bdt}.
 }
\label{fig:bdtvars}
\end{figure}

\begin{figure}[h!]
  \includegraphics[width=0.4\textwidth]{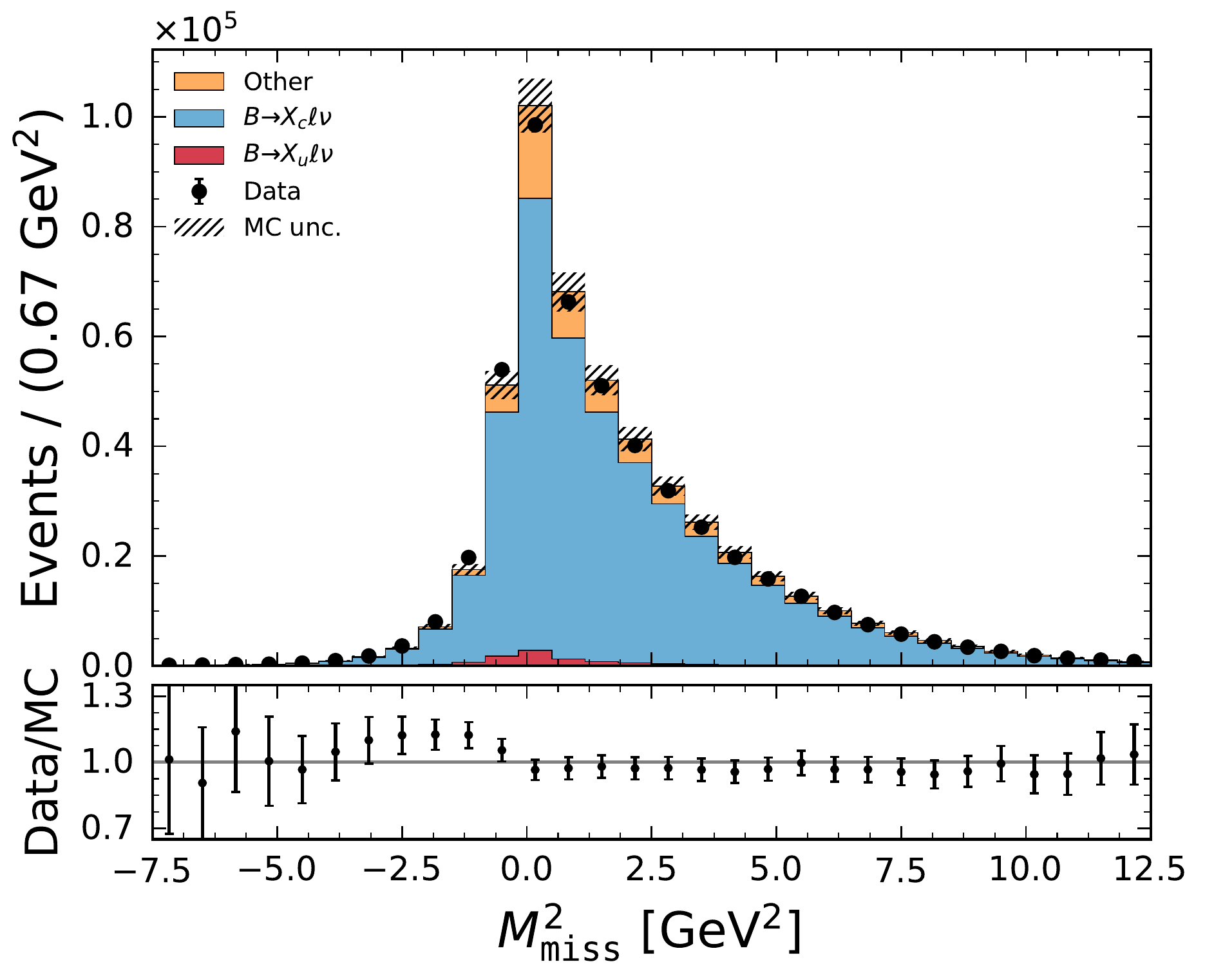} 
  \includegraphics[width=0.4\textwidth]{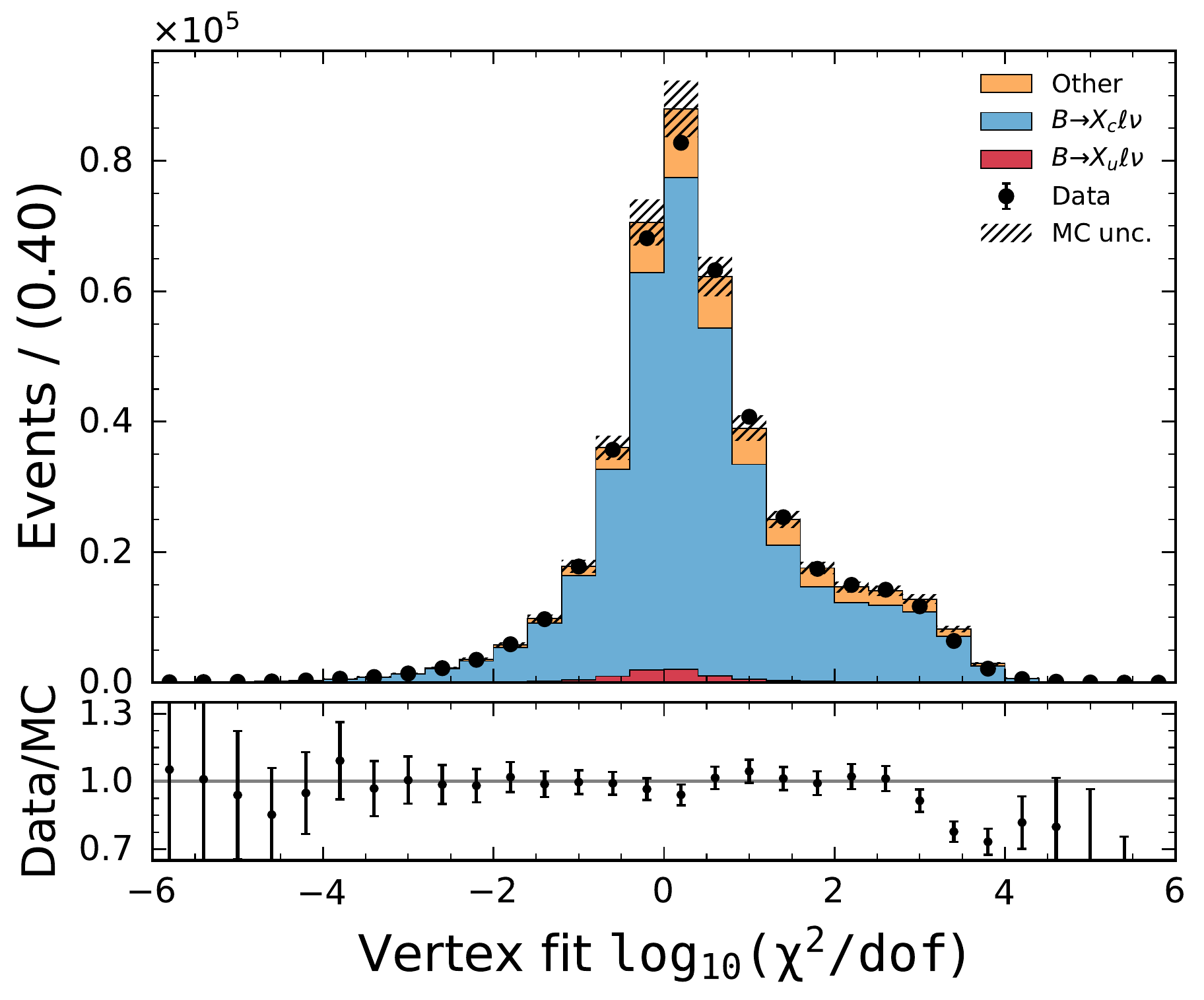}  \\
  \includegraphics[width=0.4\textwidth]{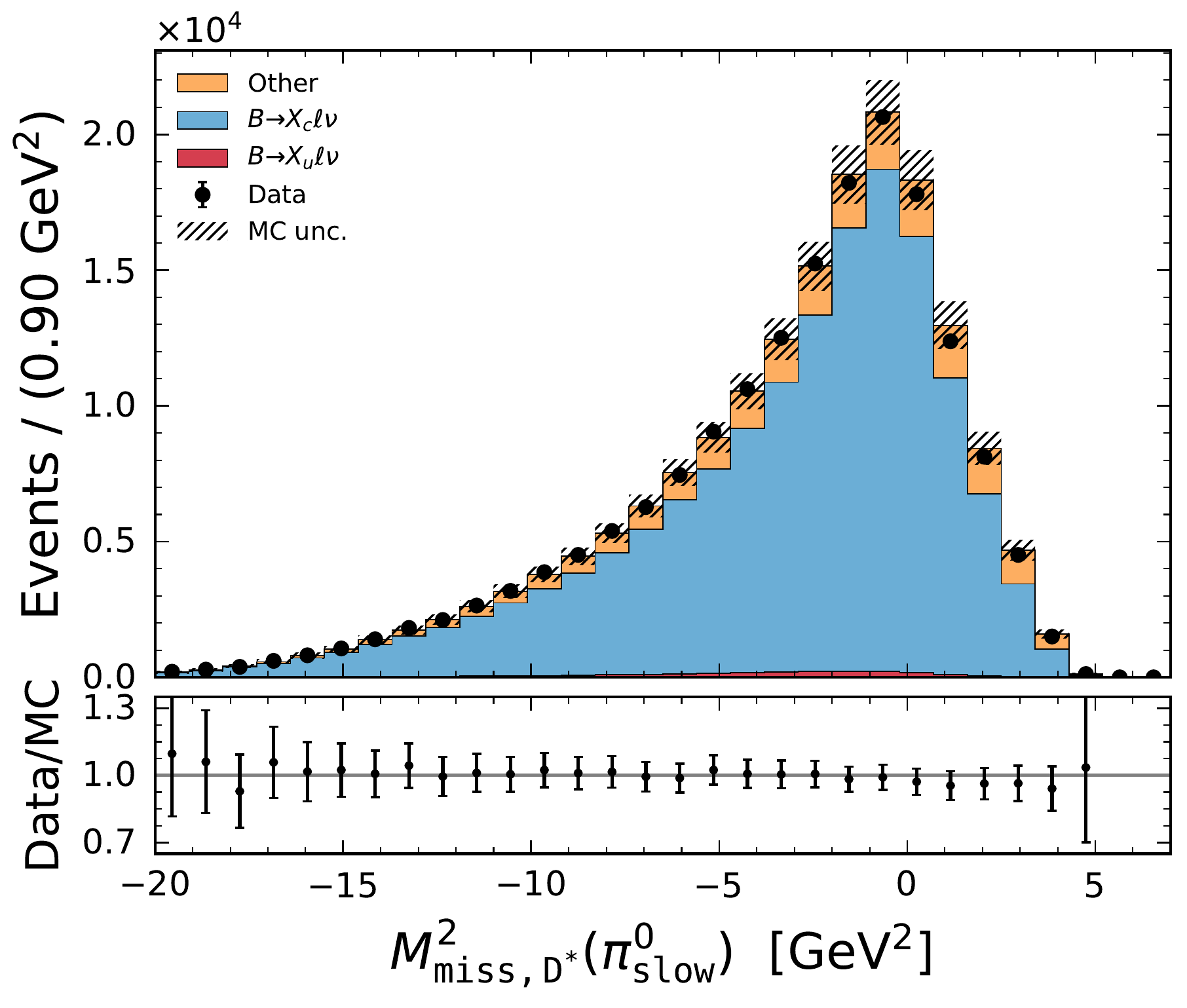}  
  \includegraphics[width=0.4\textwidth]{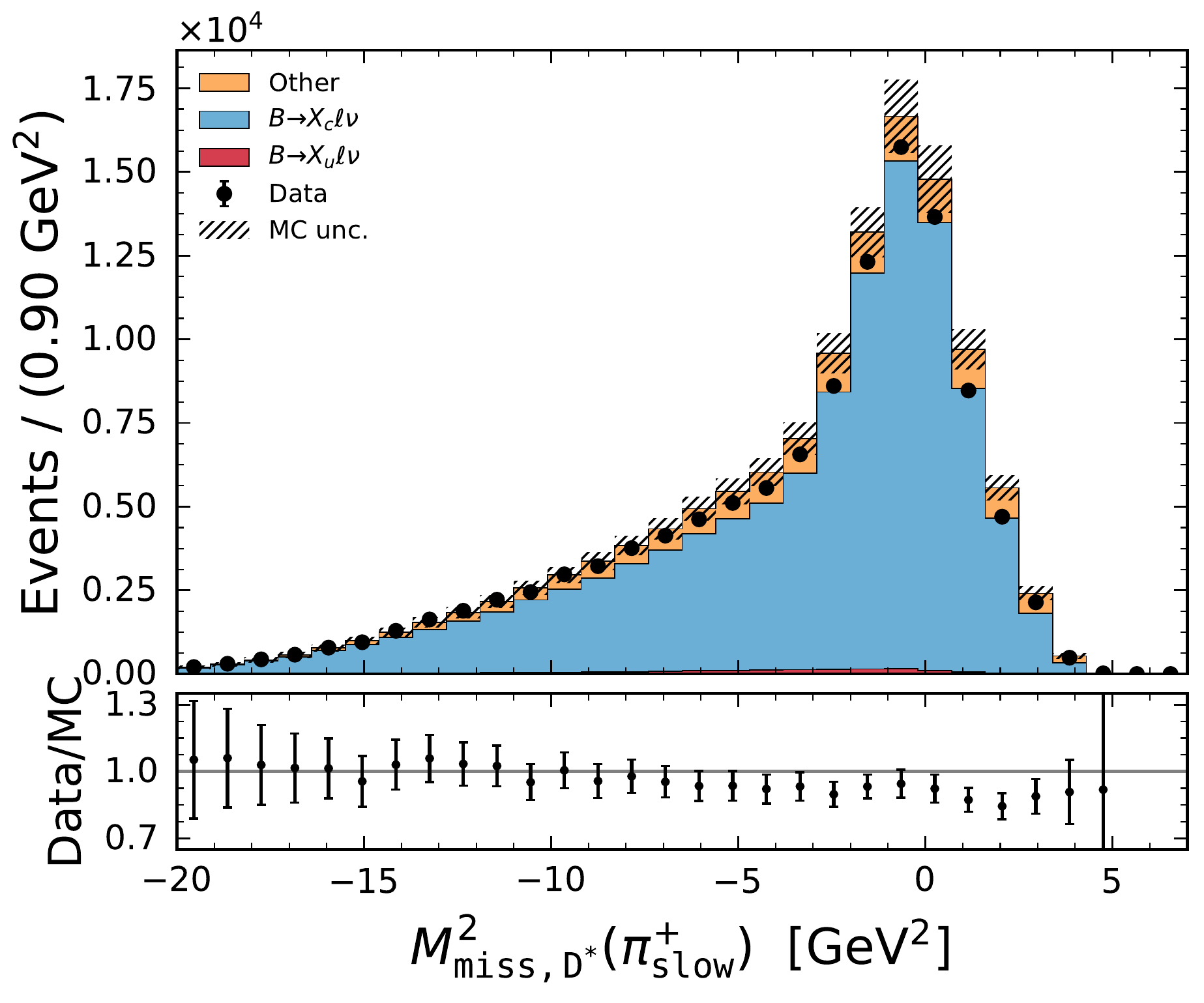} \\
\caption{
  The input variables for the \bclnu background suppression BDT for recorded and simulated events are shown. The uncertainty on the simulated events incorporate the full systematic uncertainties detailed in Section~\ref{sec:syst}.
 }
\label{fig:bdtvars_dataMC}
\end{figure}

\begin{figure}[h!]
  \includegraphics[width=0.4\textwidth]{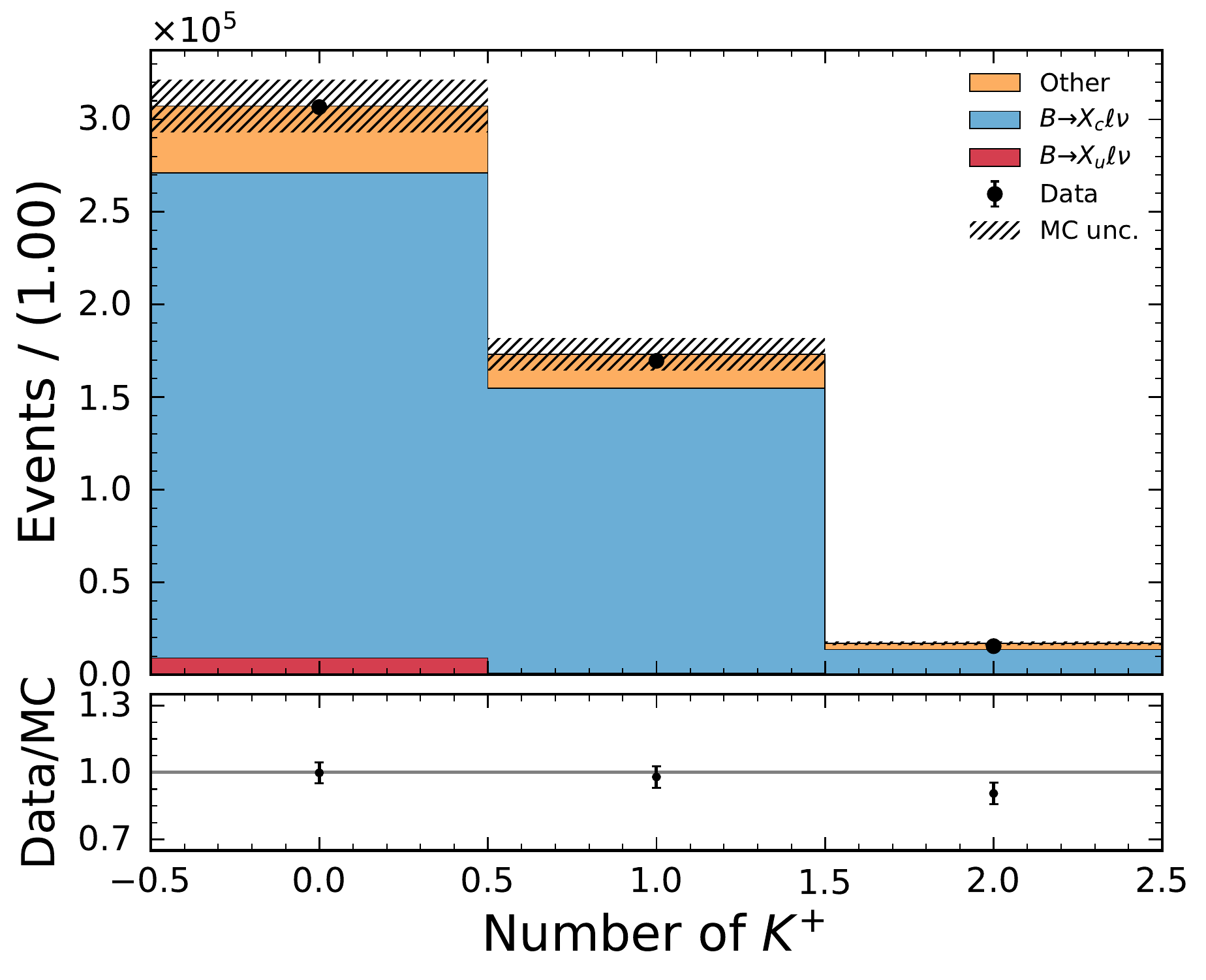} 
  \includegraphics[width=0.4\textwidth]{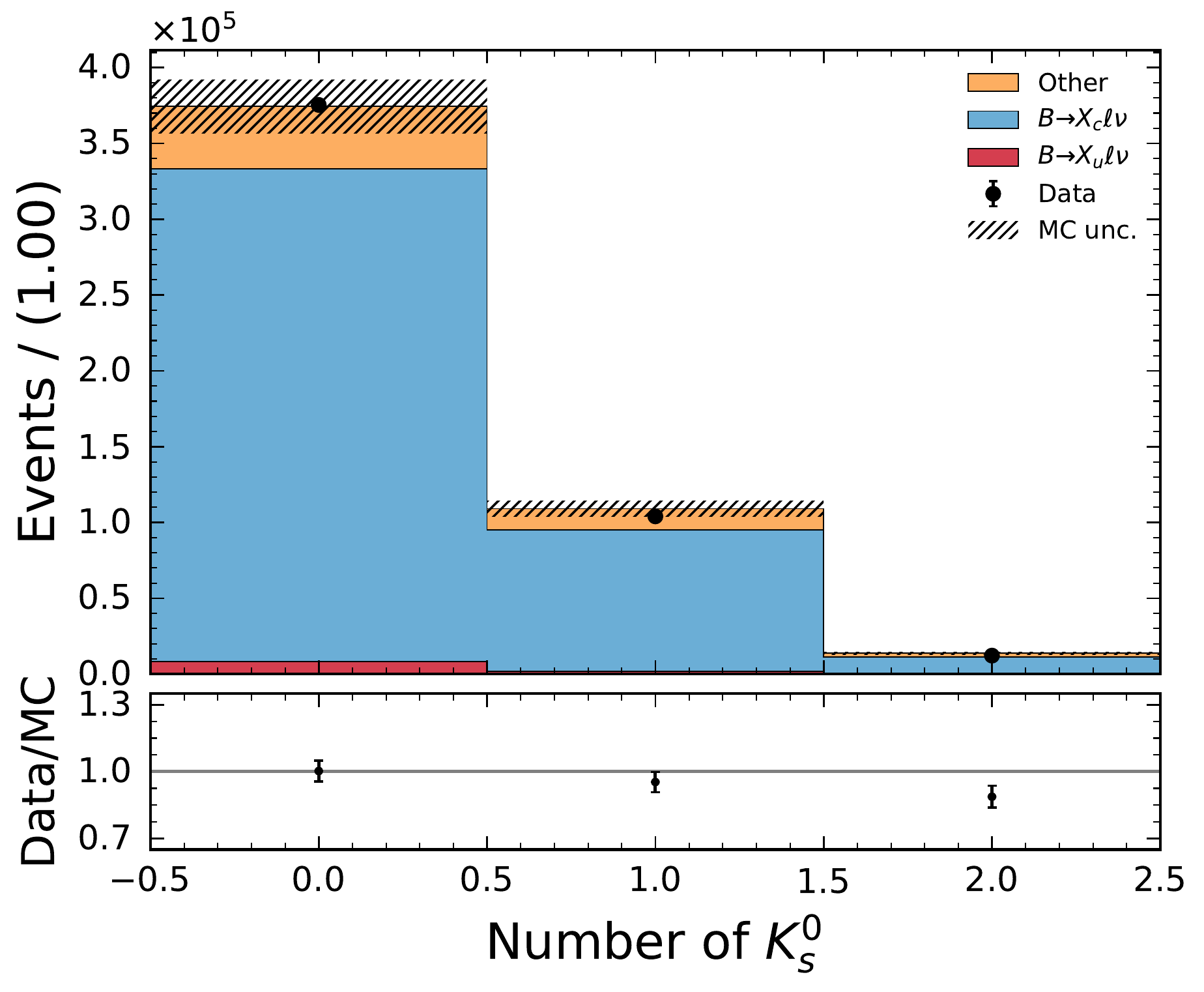}  \\
  \includegraphics[width=0.4\textwidth]{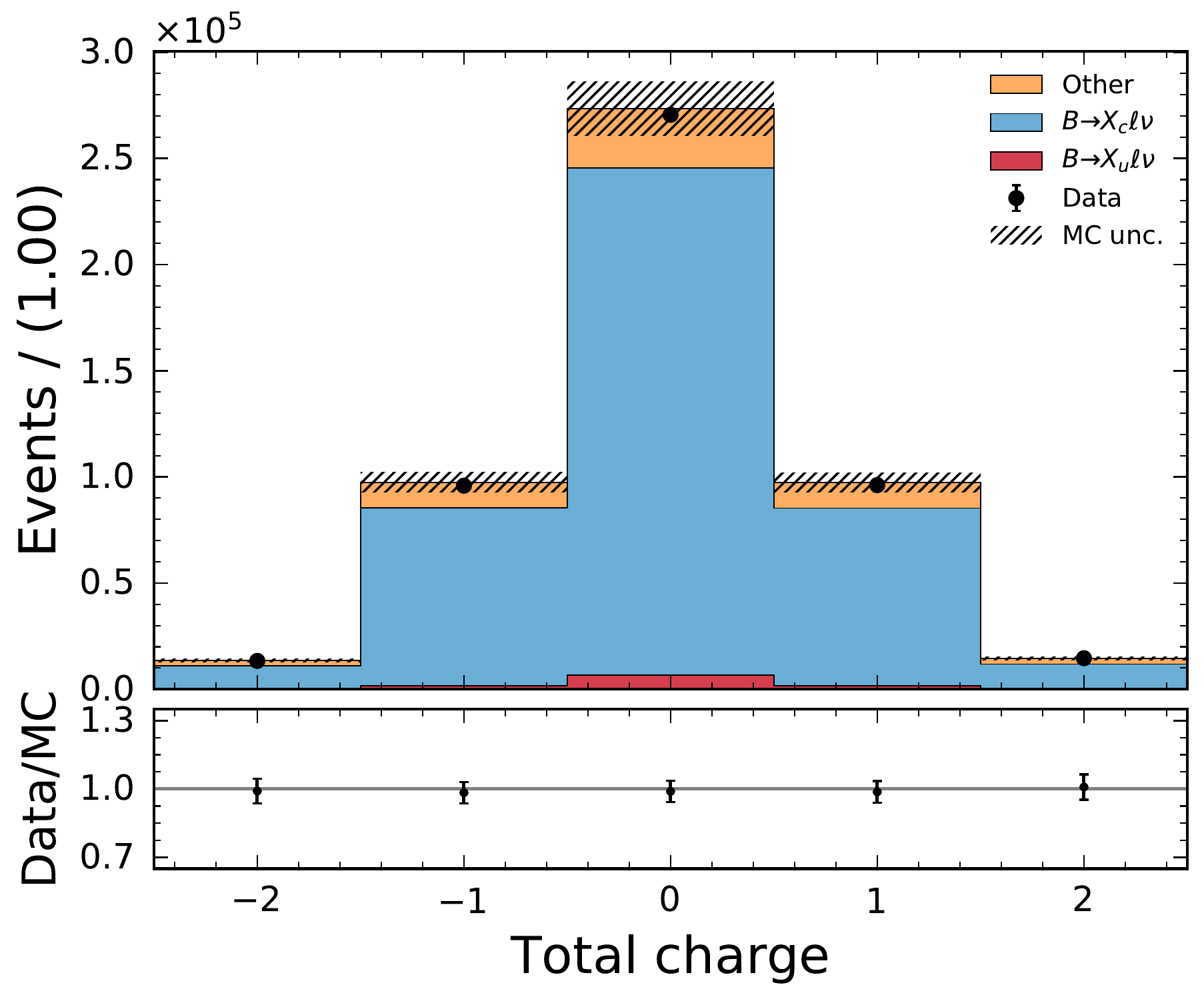}   
\caption{
  The input variables for the \bclnu background suppression BDT for recorded and simulated events are shown. The uncertainty on the simulated events incorporate the full systematic uncertainties detailed in Section~\ref{sec:syst}.
 }
\label{fig:bdtvars_dataMC2}
\end{figure}

\begin{figure}[h!]
  \includegraphics[width=0.4\textwidth]{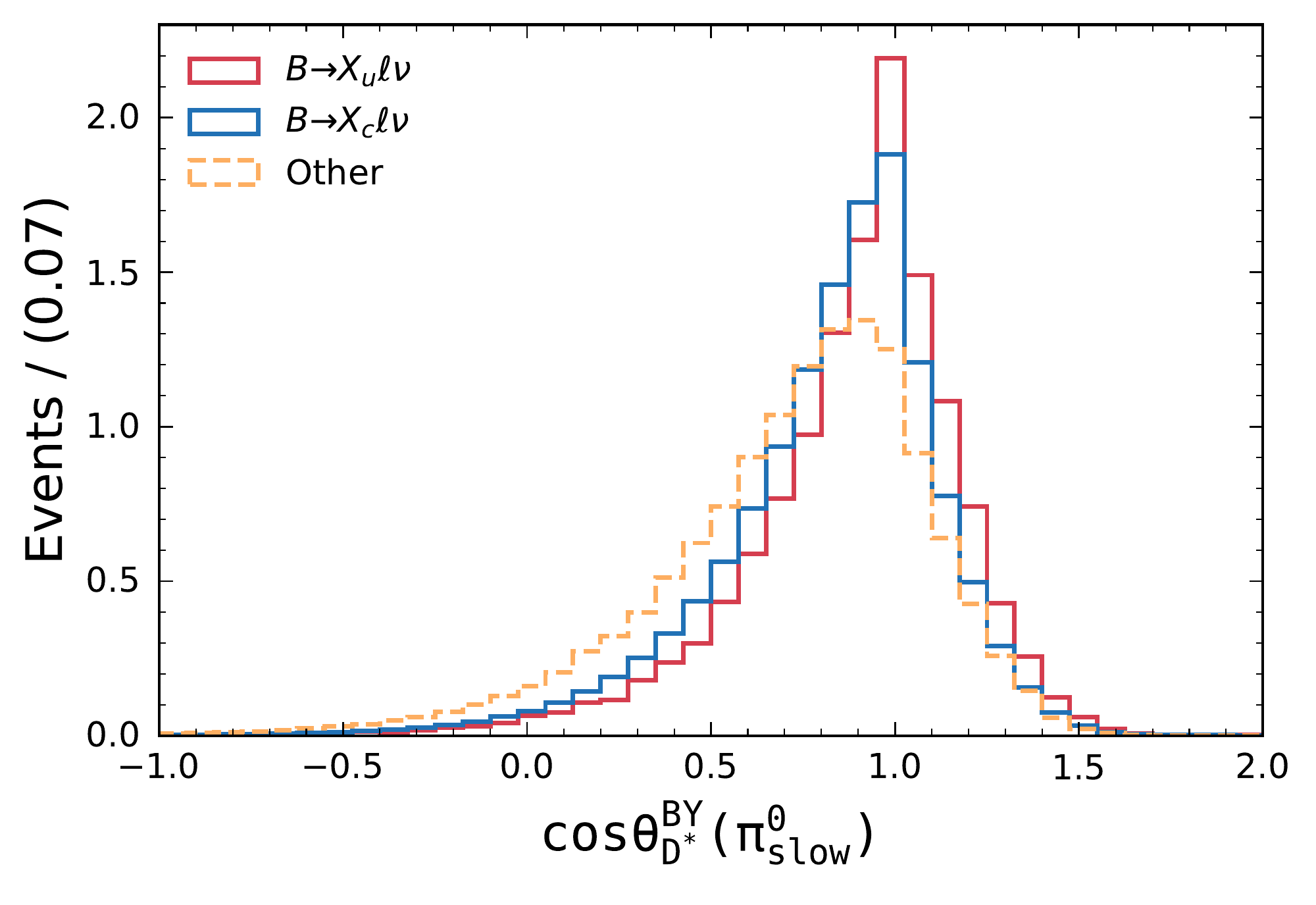} 
  \includegraphics[width=0.4\textwidth]{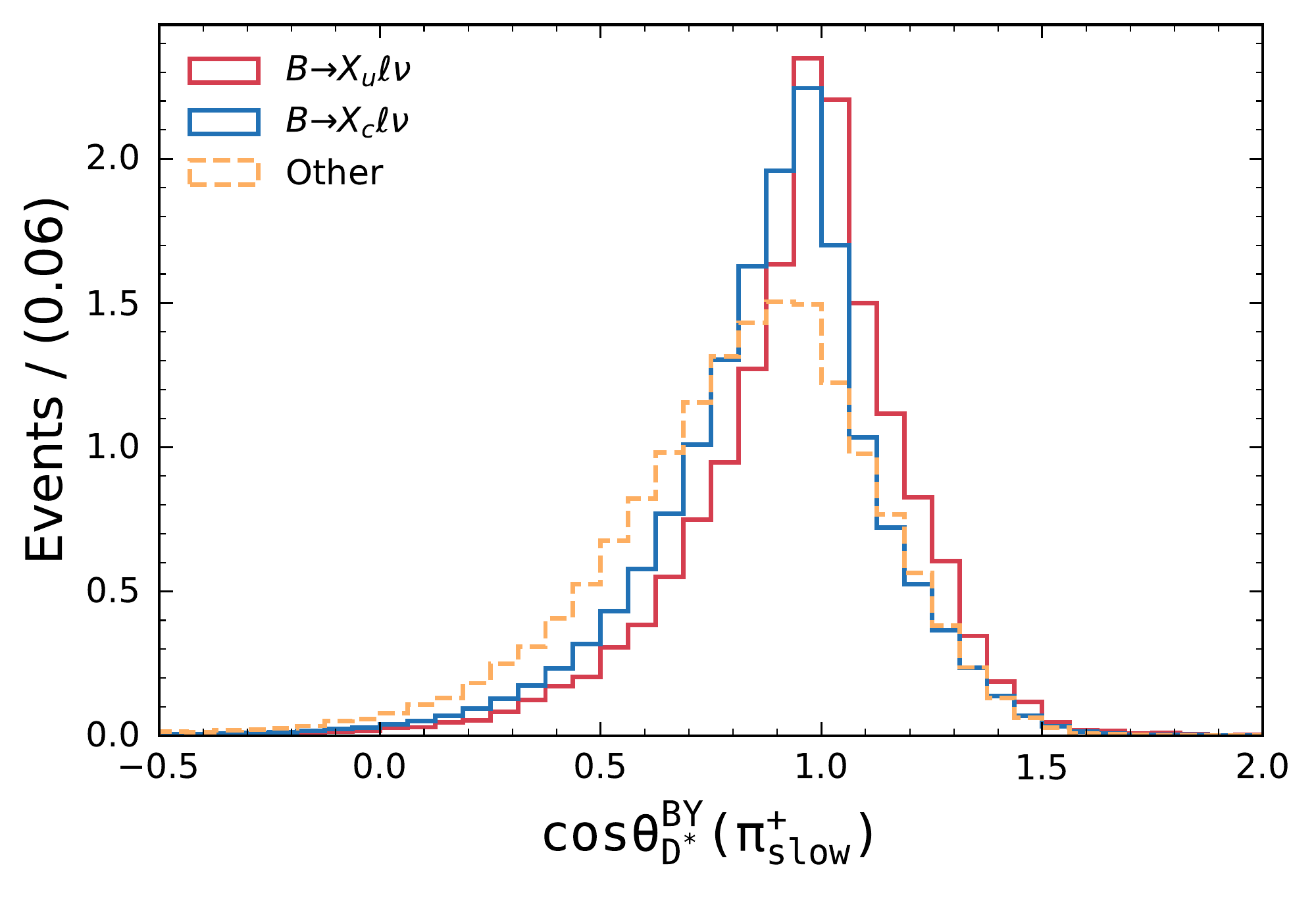}  \\
  \includegraphics[width=0.4\textwidth]{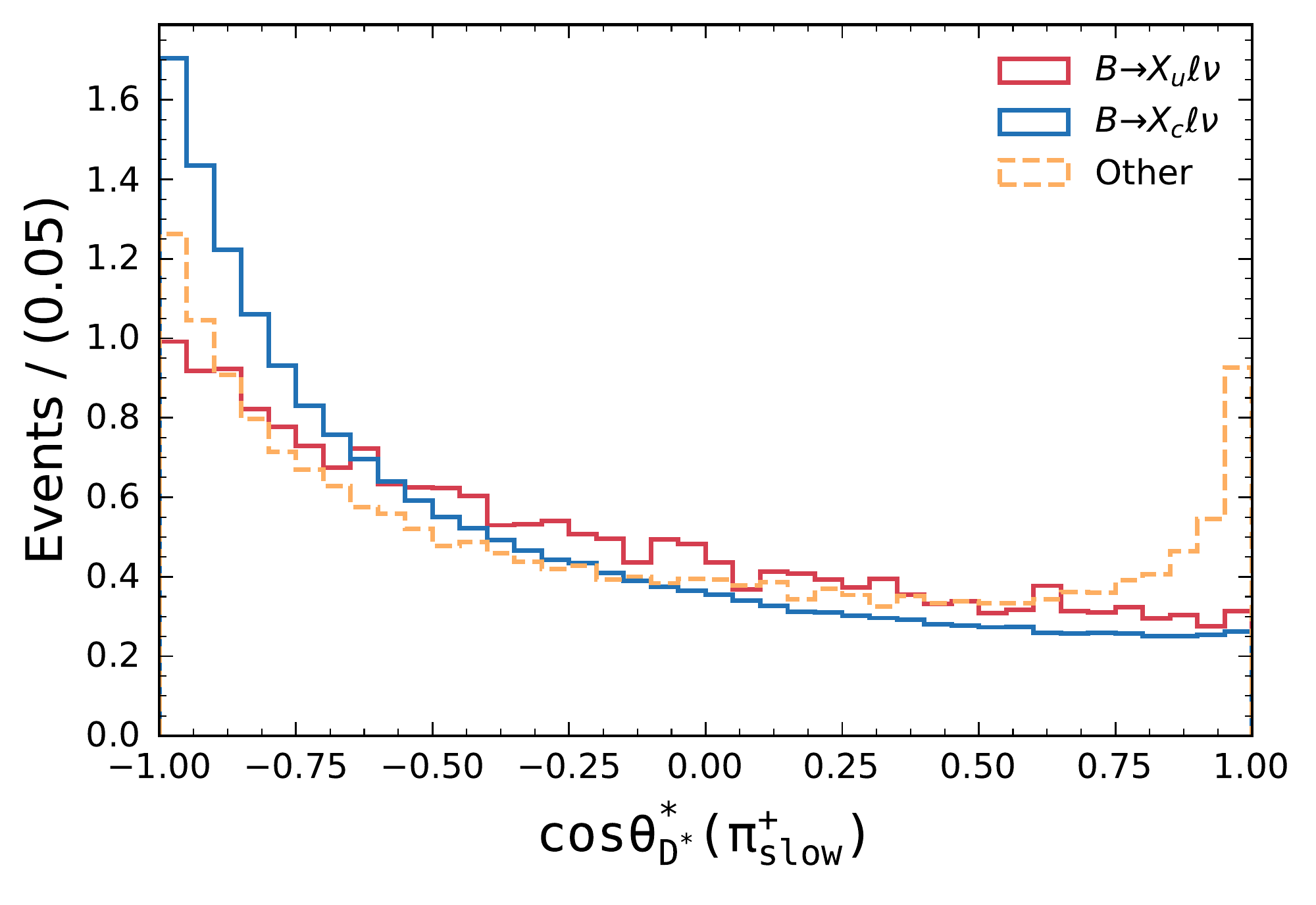}  
  \includegraphics[width=0.4\textwidth]{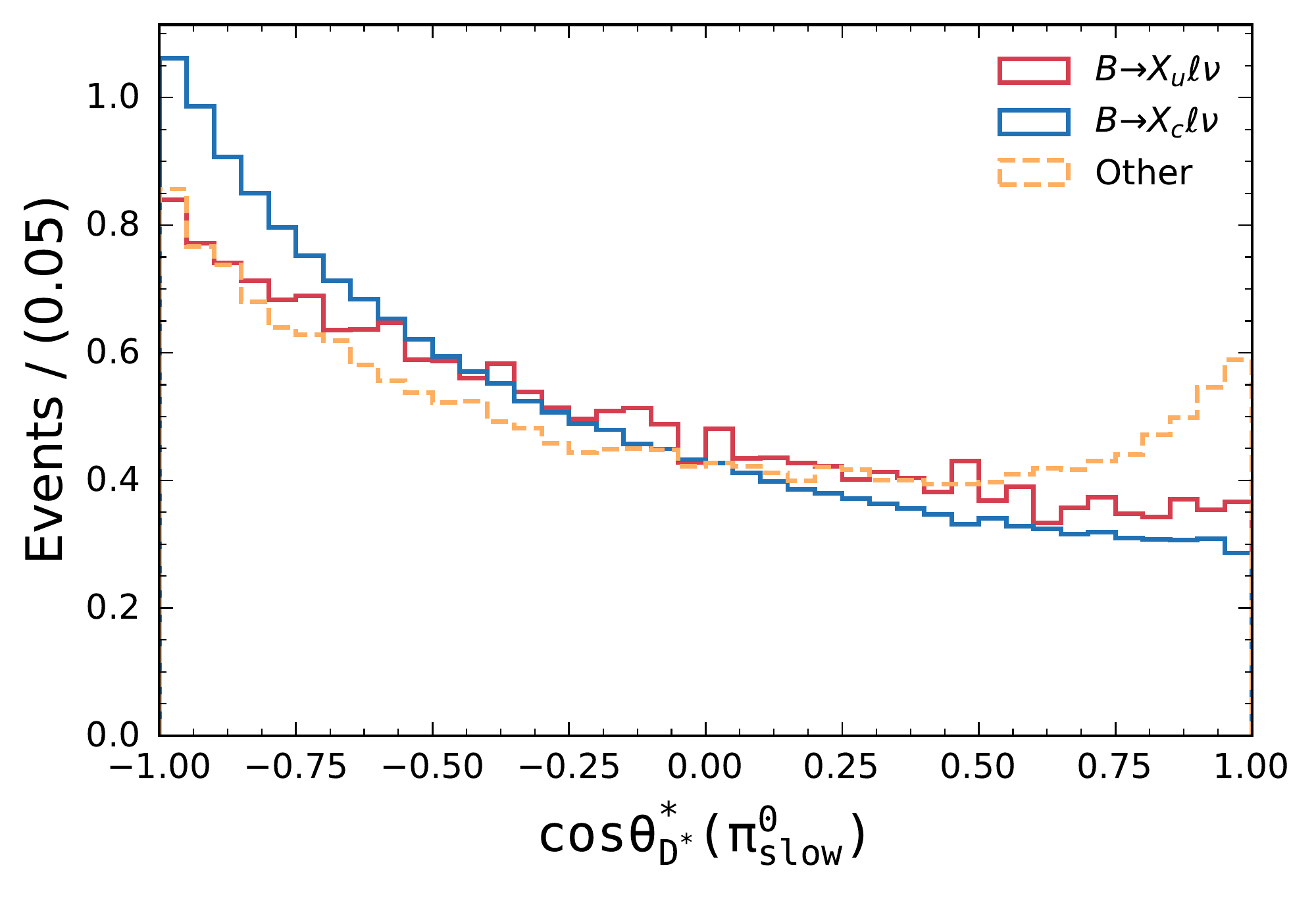} \\
\caption{
  The shape of the input variables for the \bclnu background suppression BDT are shown. For details and definitions see Section~\ref{sec:bkg_bdt}.
 }
\label{fig:bdtvars2}
\end{figure}

\begin{figure}[h!]
  \includegraphics[width=0.4\textwidth]{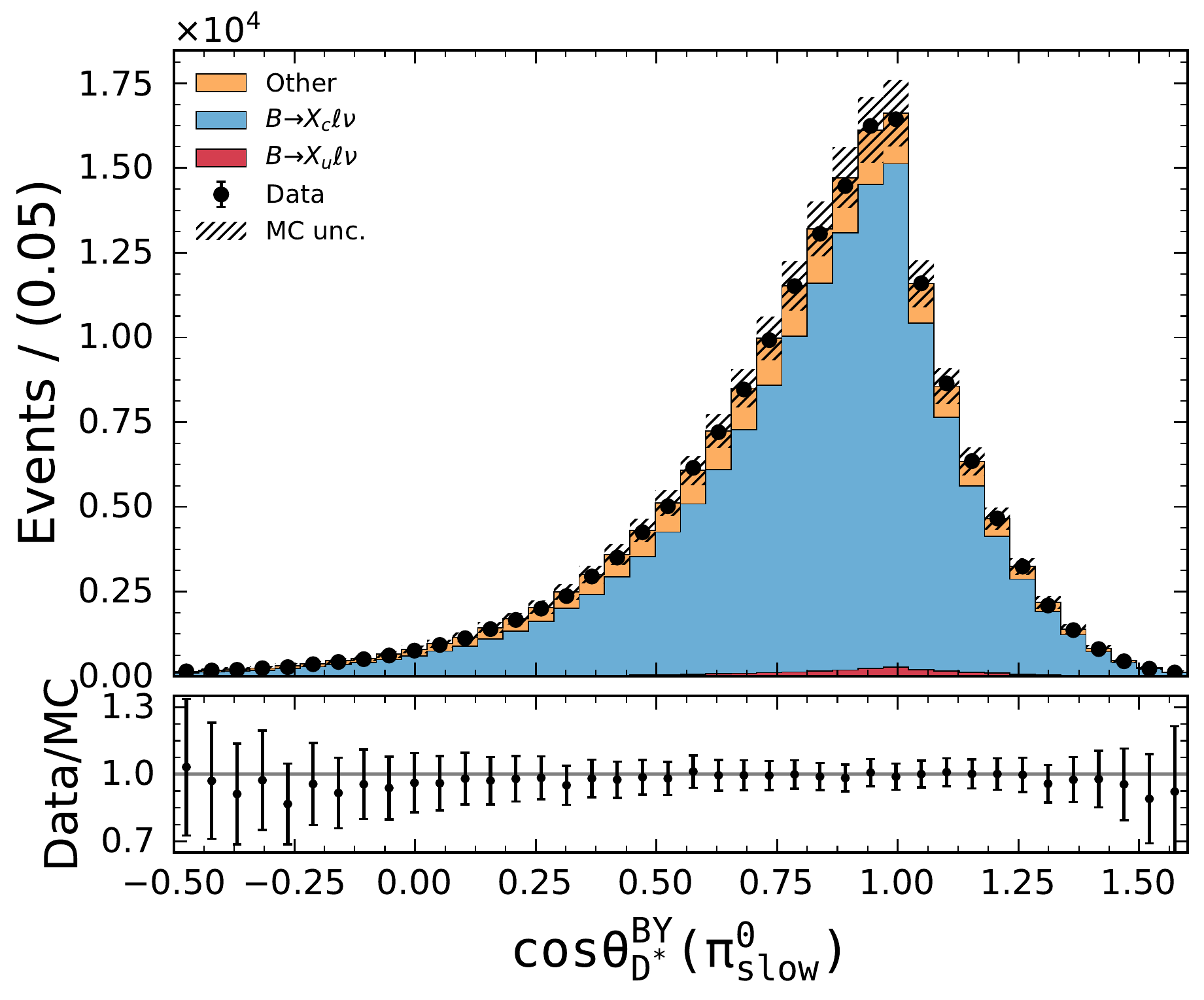} 
  \includegraphics[width=0.4\textwidth]{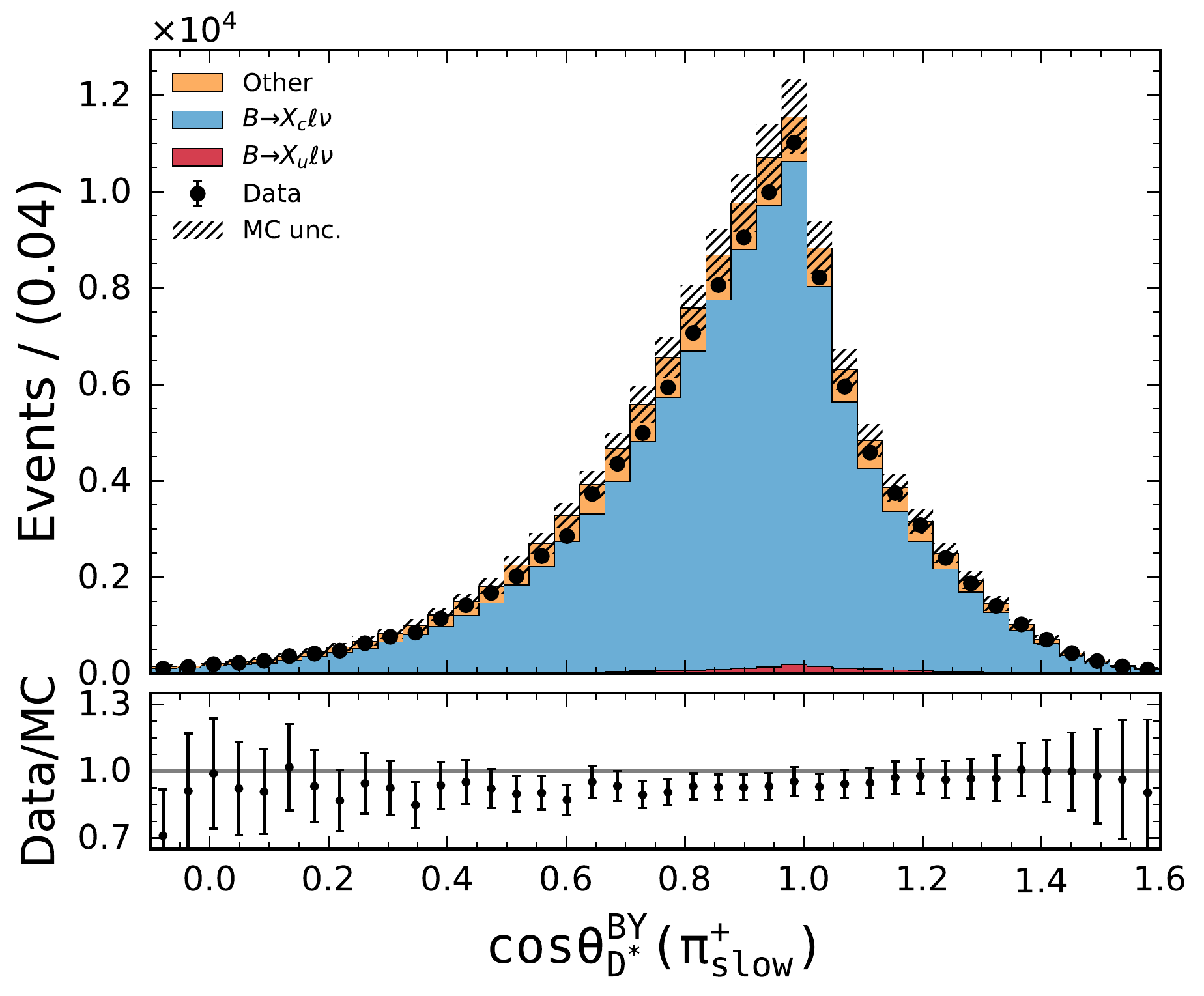}  \\
  \includegraphics[width=0.4\textwidth]{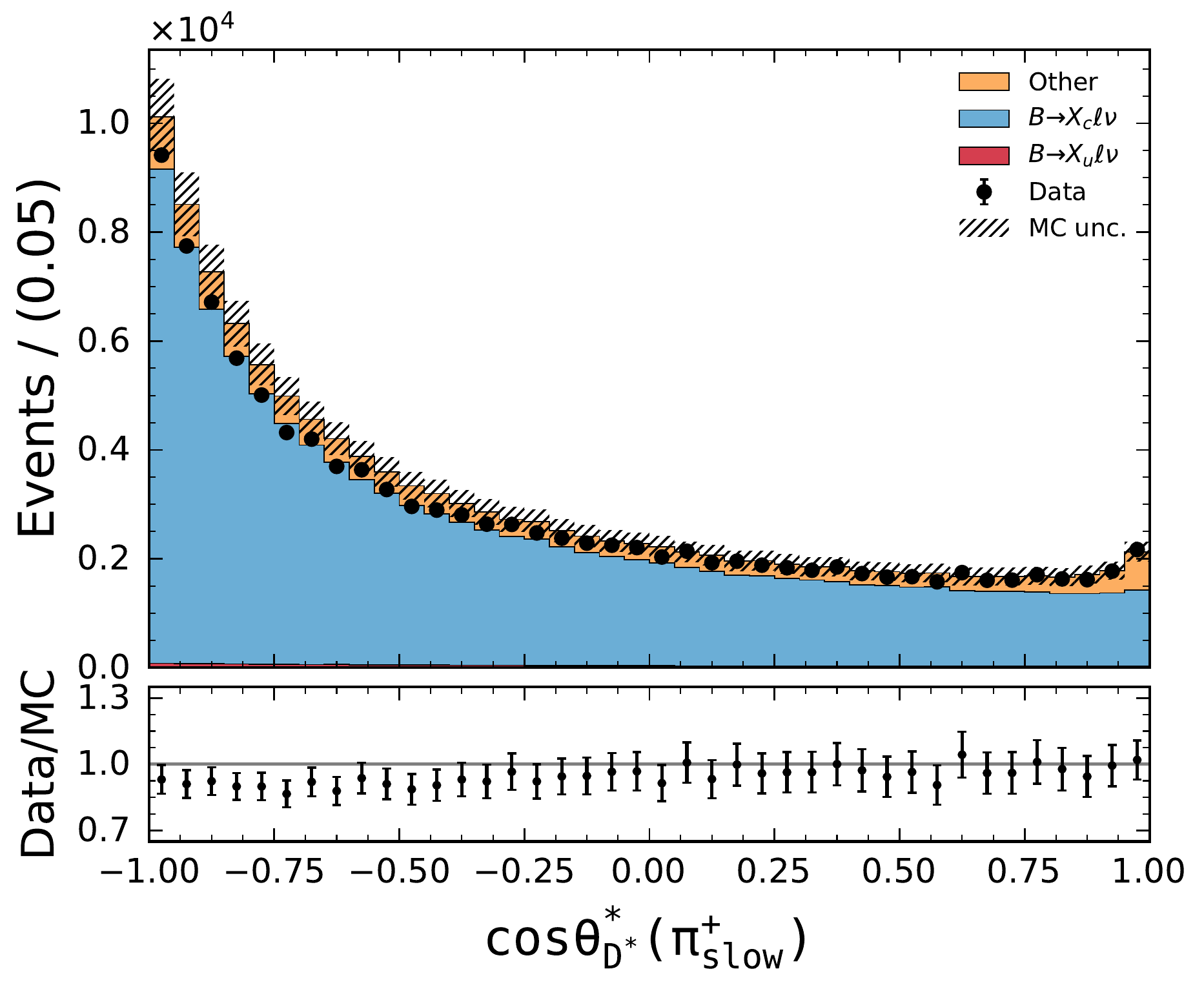}  
  \includegraphics[width=0.4\textwidth]{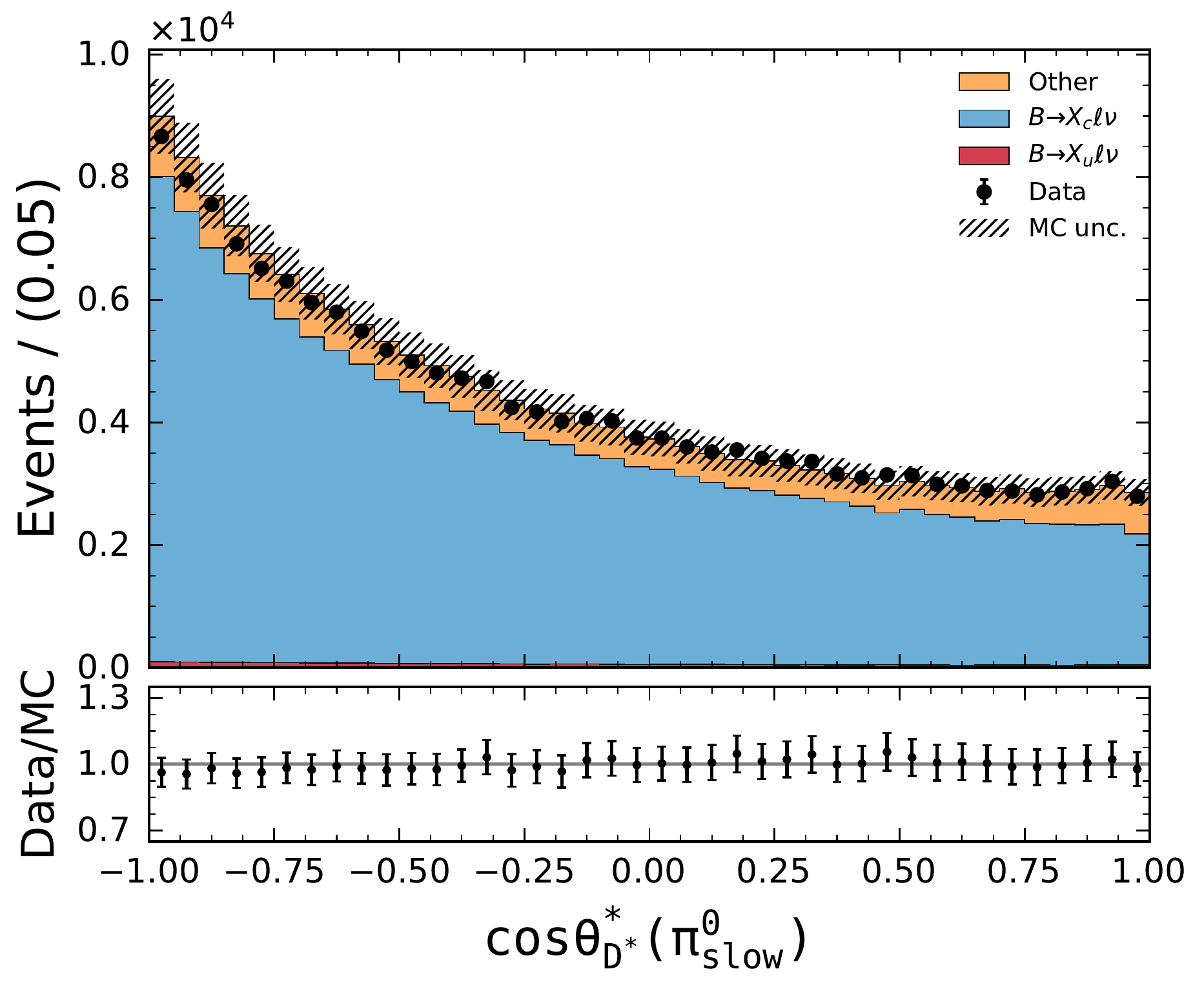} \\
\caption{
  The input variables for the \bclnu background suppression BDT for recorded and simulated events are shown. The uncertainty on the simulated events incorporate the full systematic uncertainties detailed in Section~\ref{sec:syst}.
 }
\label{fig:bdtvars_dataMC3}
\end{figure}

The shapes of the variables used in the \bclnu\ background suppression BDT are shown in Figures~\ref{fig:bdtvars} and \ref{fig:bdtvars2}. The most discriminating variables are $M_{\mathrm{miss}}^2$, the $B_{\mathrm{sig}}$ vertex fit probability, and $M_{\mathrm{miss}, D^*}^2$. Figures~\ref{fig:bdtvars_dataMC}, \ref{fig:bdtvars_dataMC2} and \ref{fig:bdtvars_dataMC3} show the agreement between recorded and simulated events, taking into account the full uncertainties detailed in Section~\ref{sec:syst}. More details about the BDT can be found in Section~\ref{sec:bkg_bdt}. \vspace{4ex}

\clearpage

\section*{C. \bulnu Charged Pion Fragmentation modeling}

\begin{figure}[h!]
  \includegraphics[width=0.4\textwidth]{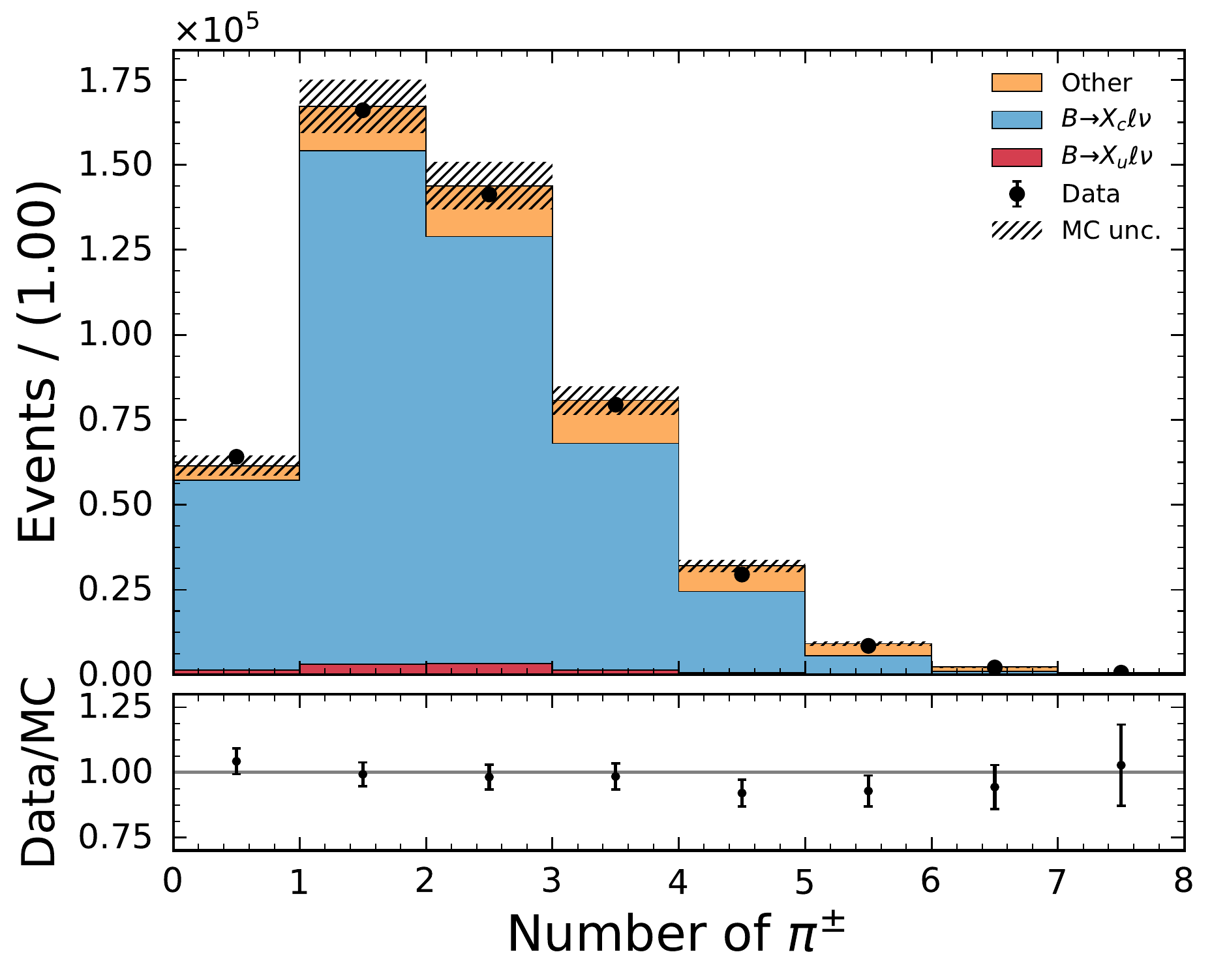}  
  \includegraphics[width=0.4\textwidth]{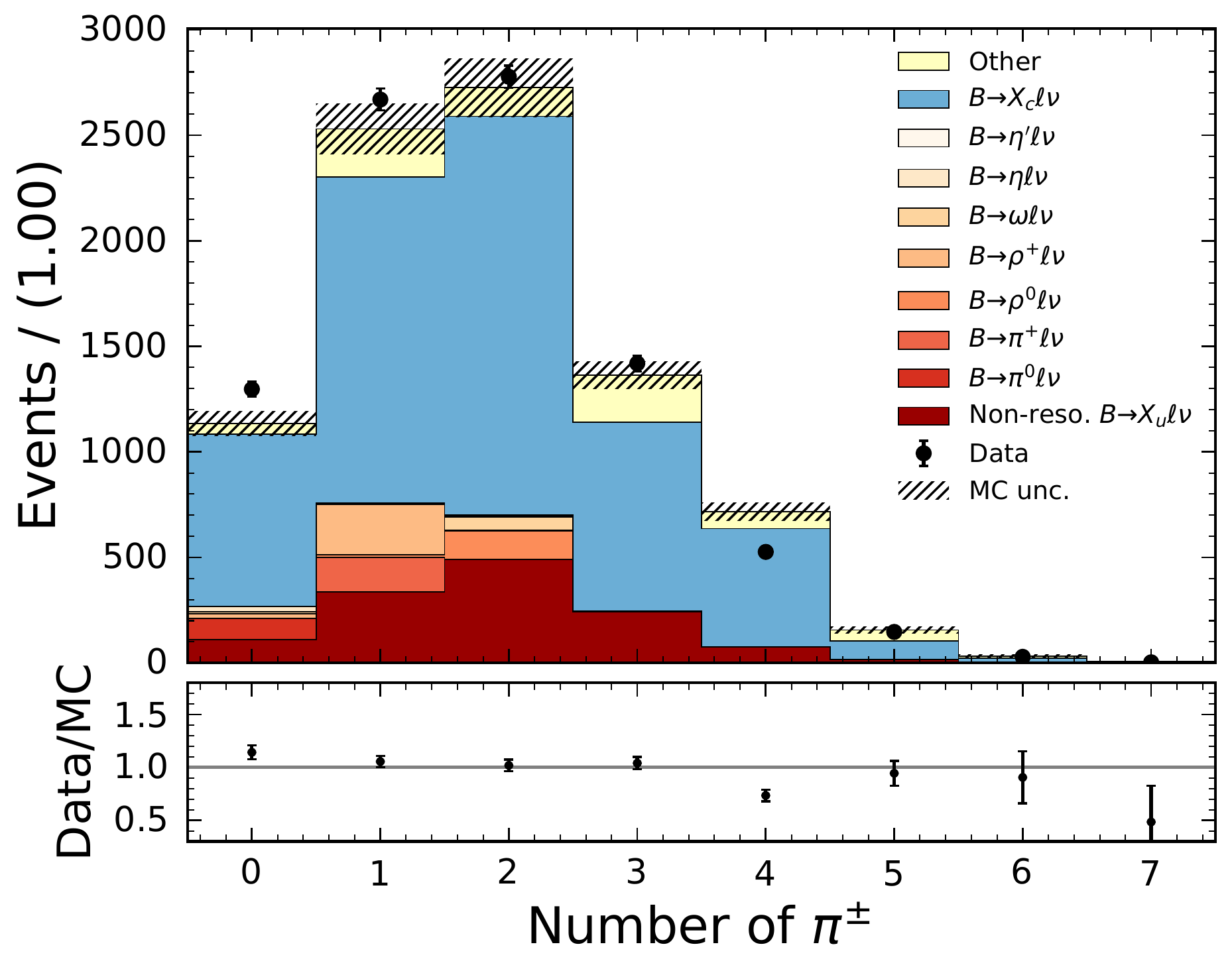} \\
  \includegraphics[width=0.4\textwidth]{figures/npi_pre_scale_wRatio} 
  \includegraphics[width=0.4\textwidth]{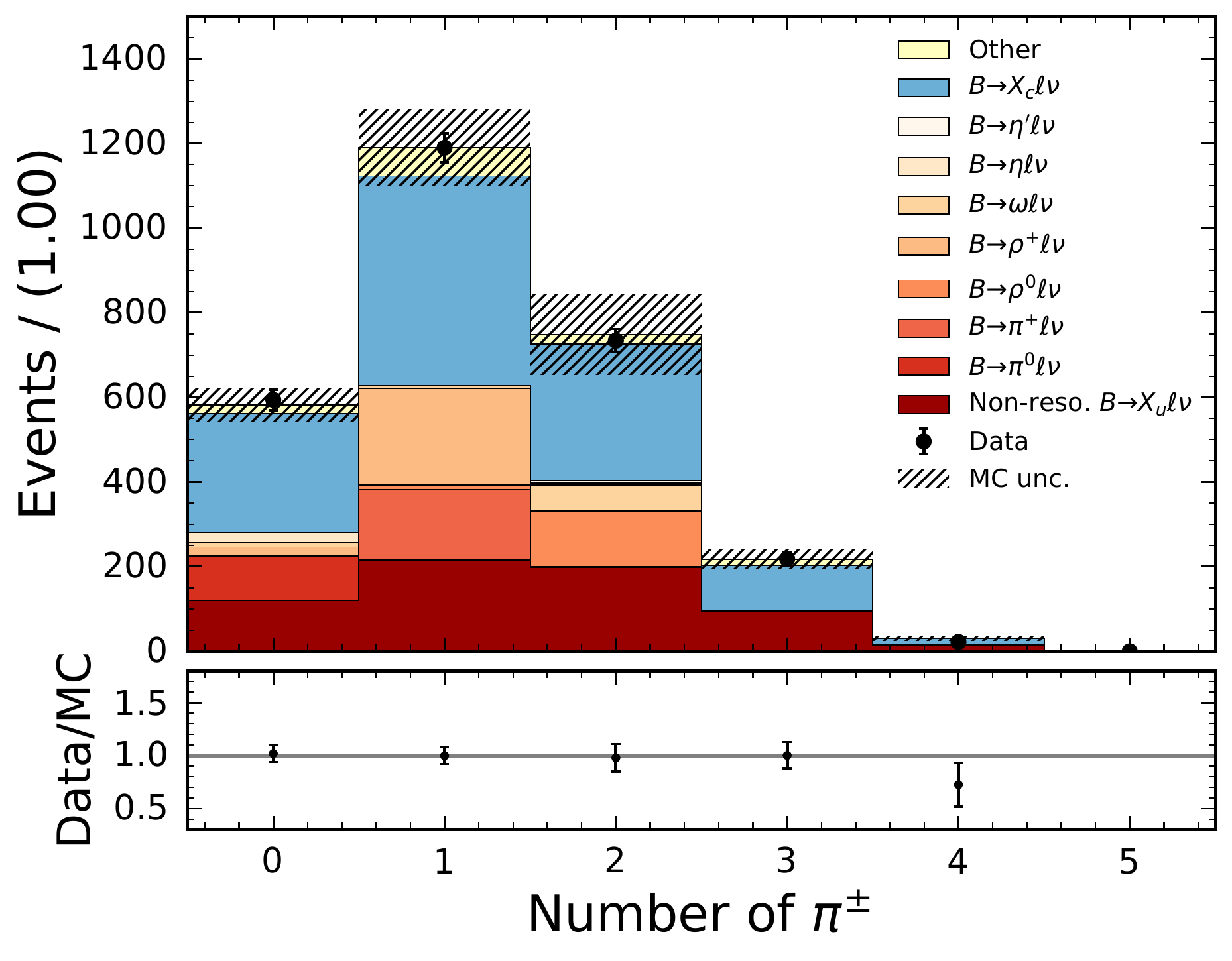} 
\caption{
   The charged pion multiplicity ($n_{\pi^\pm}$) are compared between data and the simulation: (top left) for all events prior the BDT selection; (top right) for all events after the BDT selection; (bottom left): for the signal enriched region of $M_{X} < 1.7$ GeV; (bottom right) for the same region but after rescaling the non-resonant contributions such that the $n_{\pi^\pm}$ fragmentation probability to match the one observed in data.
 }
\label{fig:npicomp}
\end{figure}

Figure~\ref{fig:npicomp} compares the charged pion multiplicity at different stages in the selection. This variable is not used in the signal extraction, but its modeling is tested to make sure that the \bulnu fragmentation probabilities cannot bias the final result. The agreement in the signal enriched region with $M_{X} < 1.7$ GeV after the BDT selection is fair, but shows some deviations. We correct the generator level charged pion multiplicity to match the $n_{\pi^\pm}$ observed in this selection by assigning the non-resonant \bulnu events a correction weight as a function of the true charged pion multiplicity. After this procedure the agreement is perfect and we use the difference in the reconstruction efficiency as an uncertainty on the pion fragmentation on the partial branching fractions and $|V_{ub}|$ (cf. Section~\ref{sec:syst}).

\clearpage

\section*{D. Nuisance Parameter Pulls and Additional Fit Plots} \label{app:NPs}

\begin{figure}[h!]
  \includegraphics[width=0.24\textwidth]{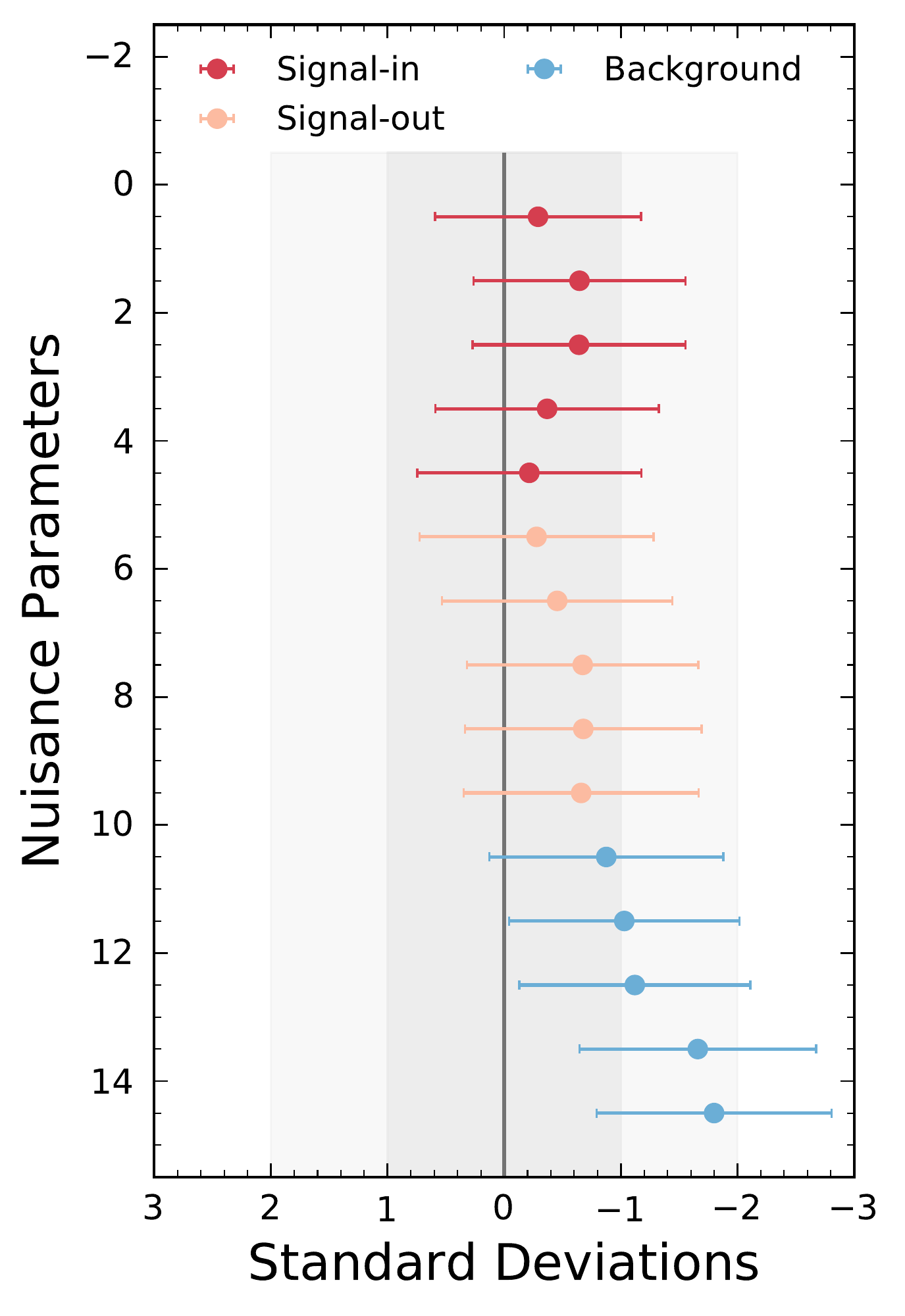} 
  \includegraphics[width=0.24\textwidth]{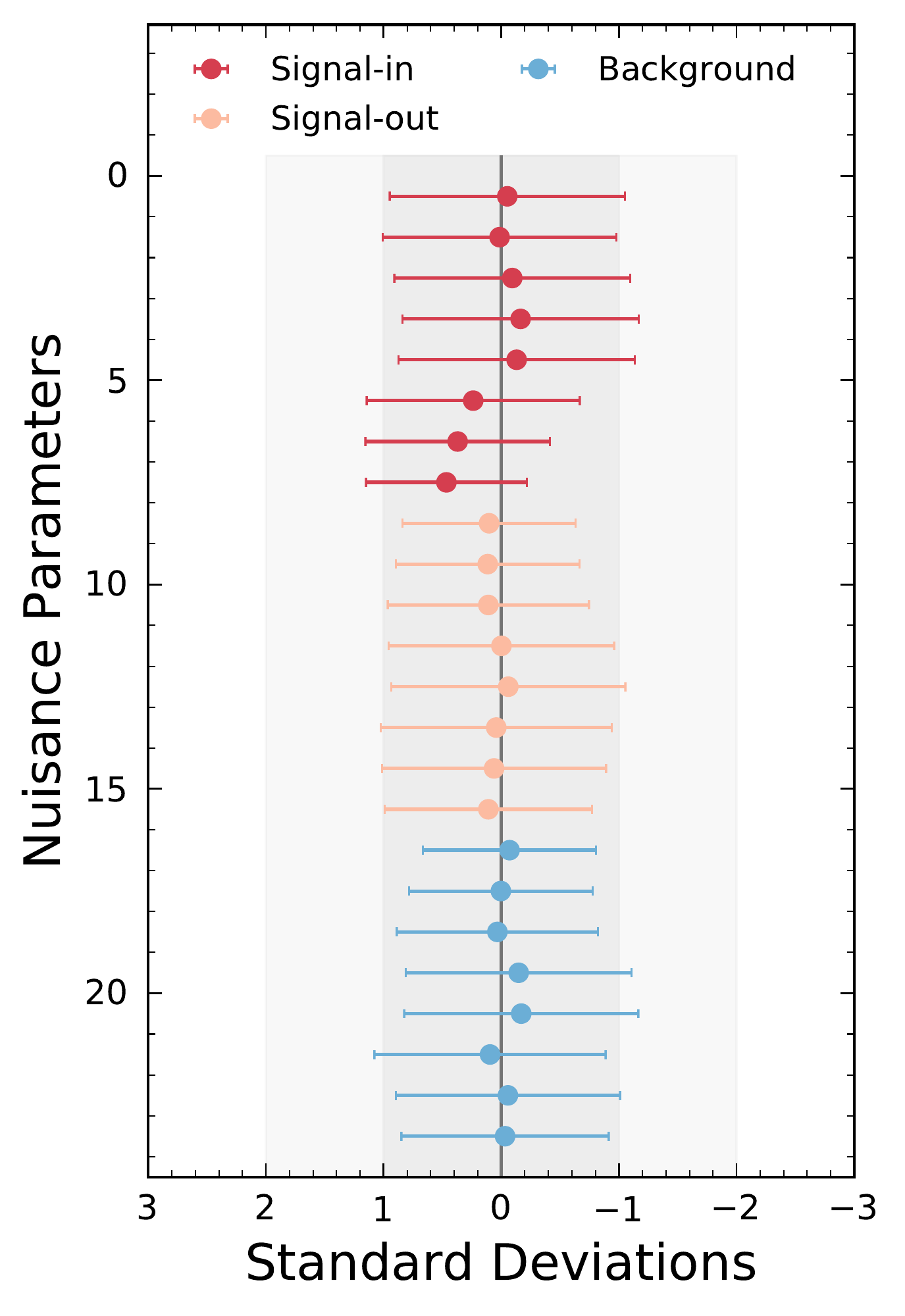}  
  \includegraphics[width=0.24\textwidth]{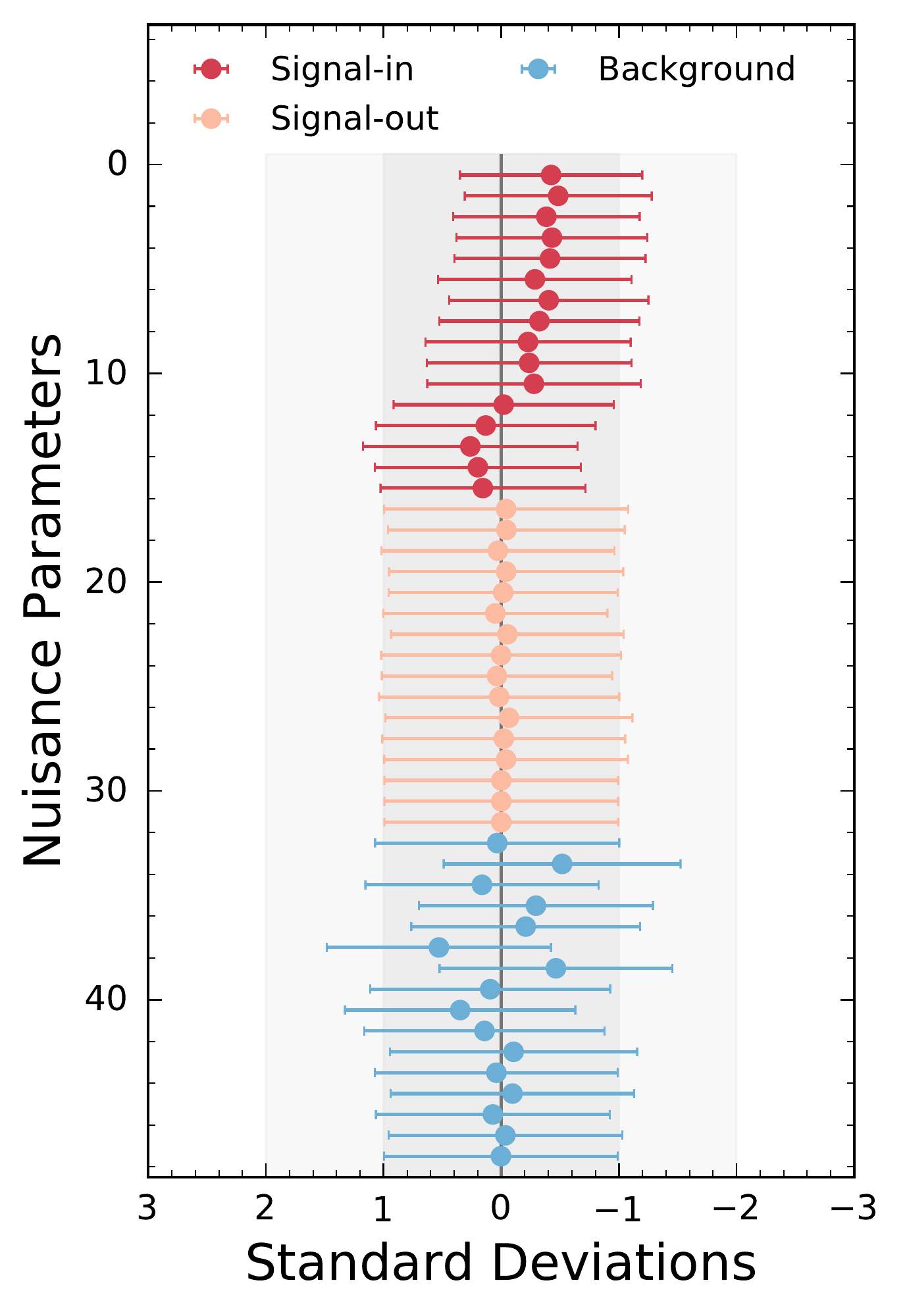} 
  \includegraphics[width=0.24\textwidth]{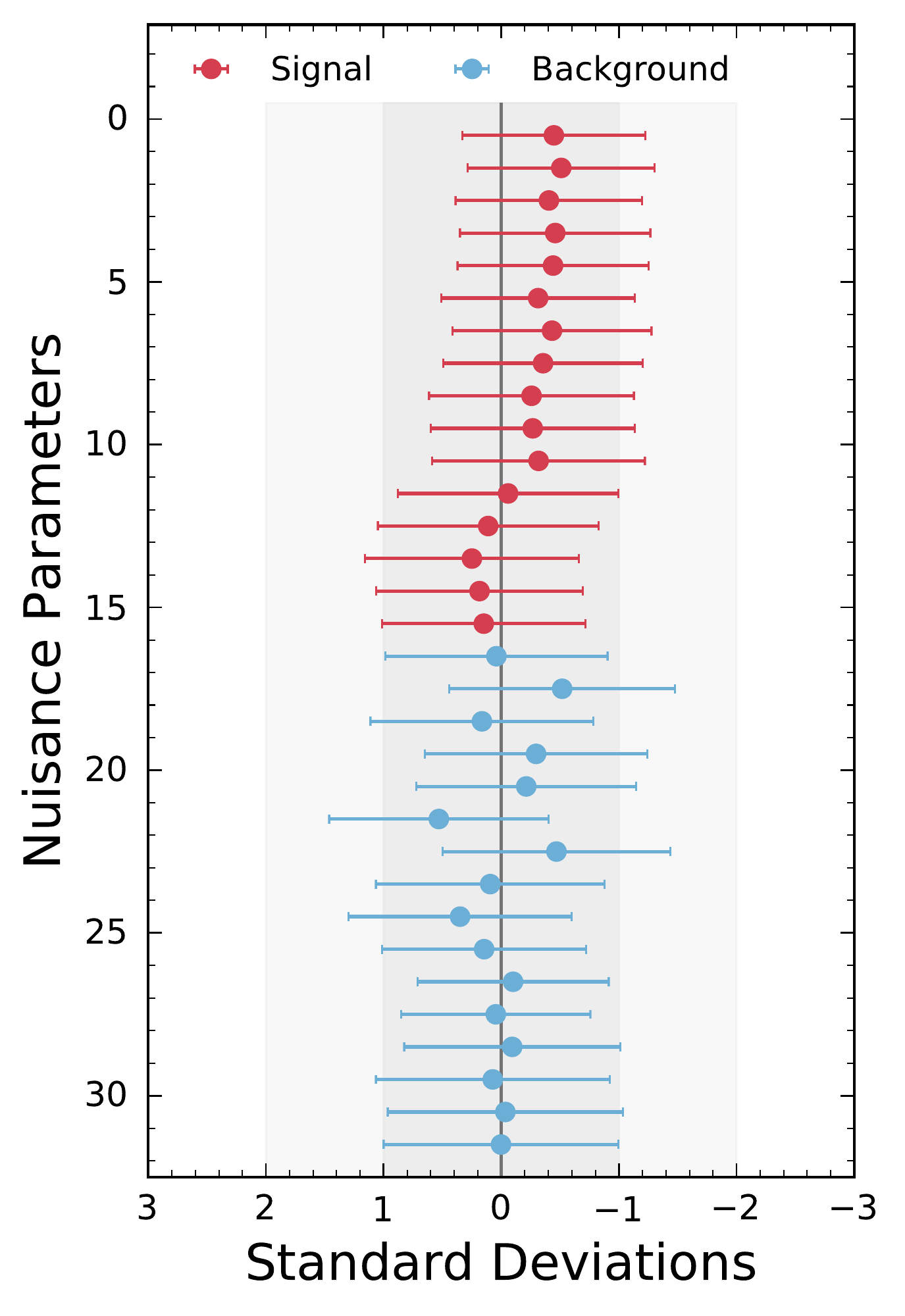} \\
\caption{
  The nuisance parameter pulls on the 1D fits of $M_X$, $q^2$, and $E_\ell^B$ with and without $M_X < 1.7$ GeV events separated out, are shown from left to right.
 }
\label{fig:NP1}
\end{figure}

\begin{figure}[h!]
 \includegraphics[width=0.24\textwidth]{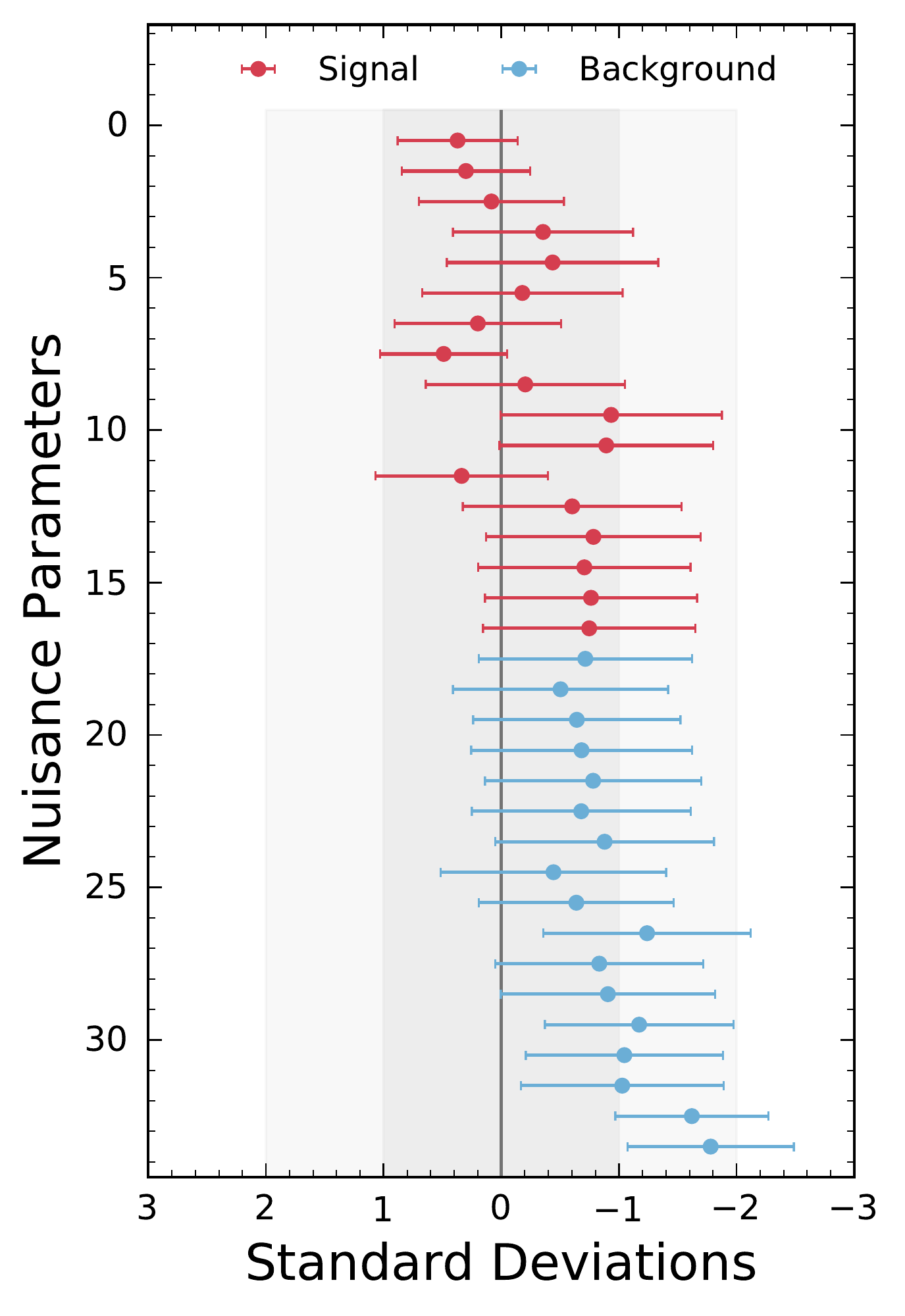} 
\caption{
  The nuisance parameter pulls on the 2D fit of $M_X:q^2$ is shown.
 }
\label{fig:NP2}
\end{figure}

Figures~\ref{fig:NP1} and \ref{fig:NP2} show the nuisance parameter pulls for each fit category $k$ and bin $i$ defined as 
\begin{align}
 \left( \widehat \theta_{ik} -  \theta_{ik}  \right) / \sqrt{ \Sigma_{k,ii}}  \, ,
 \end{align}
 of the partial branching fraction fits, with $\widehat \theta$ ($\theta$) corresponding to the post-fit (pre-fit) value of the nuisance parameter. Note that uncertainties of each pull shows the post-fit error 
 \begin{align}
 \sqrt{ \widehat \Sigma_{k,ii}}
 \end{align}
  normalized to the pre-fit constraint 
  \begin{align}
  \sqrt{ \Sigma_{k,ii}} \, .
  \end{align}

\begin{figure}[h!]
  \includegraphics[width=0.4\textwidth]{figures/post_fit_d_q2_0_differential}
  \includegraphics[width=0.4\textwidth]{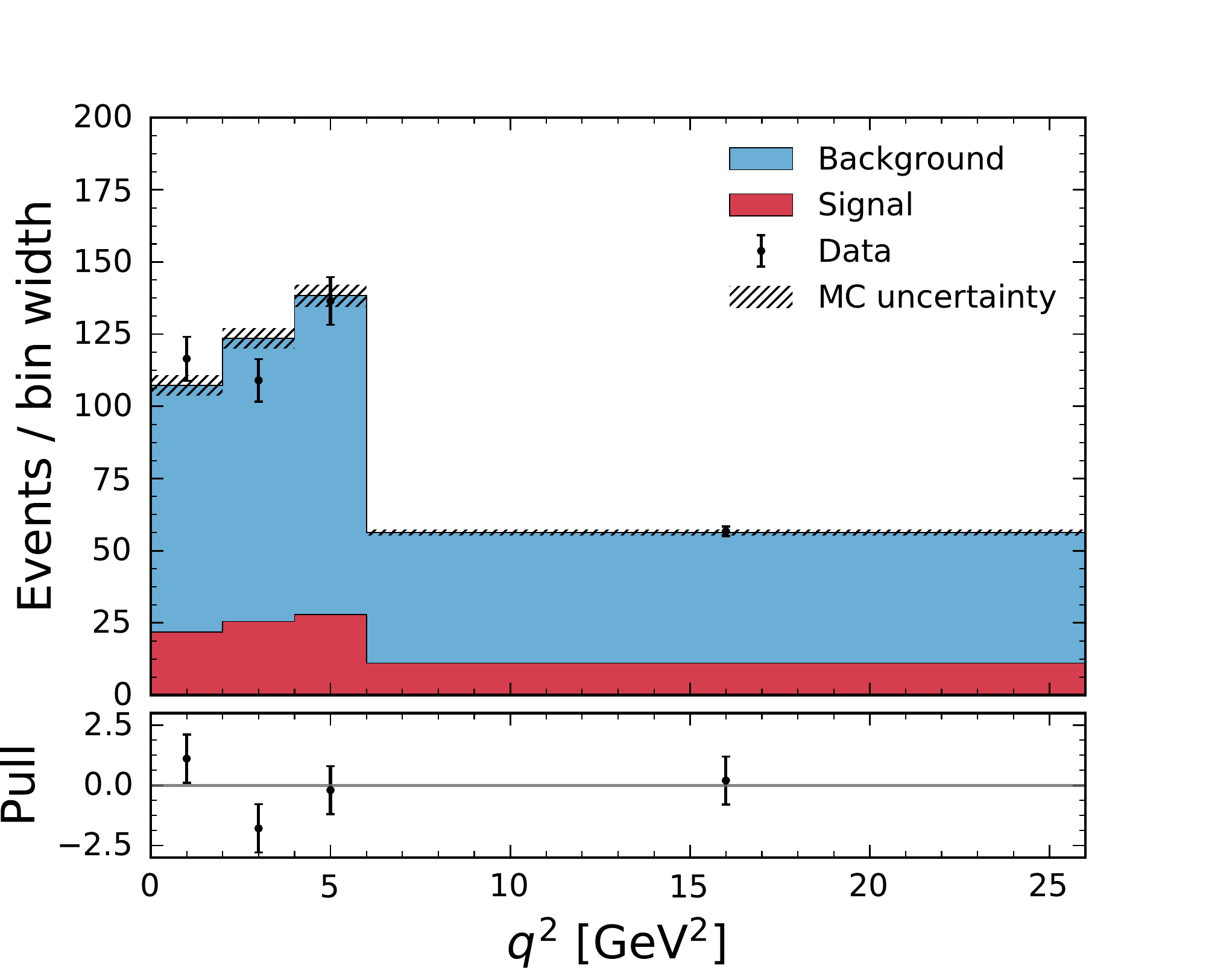} \\
  \includegraphics[width=0.4\textwidth]{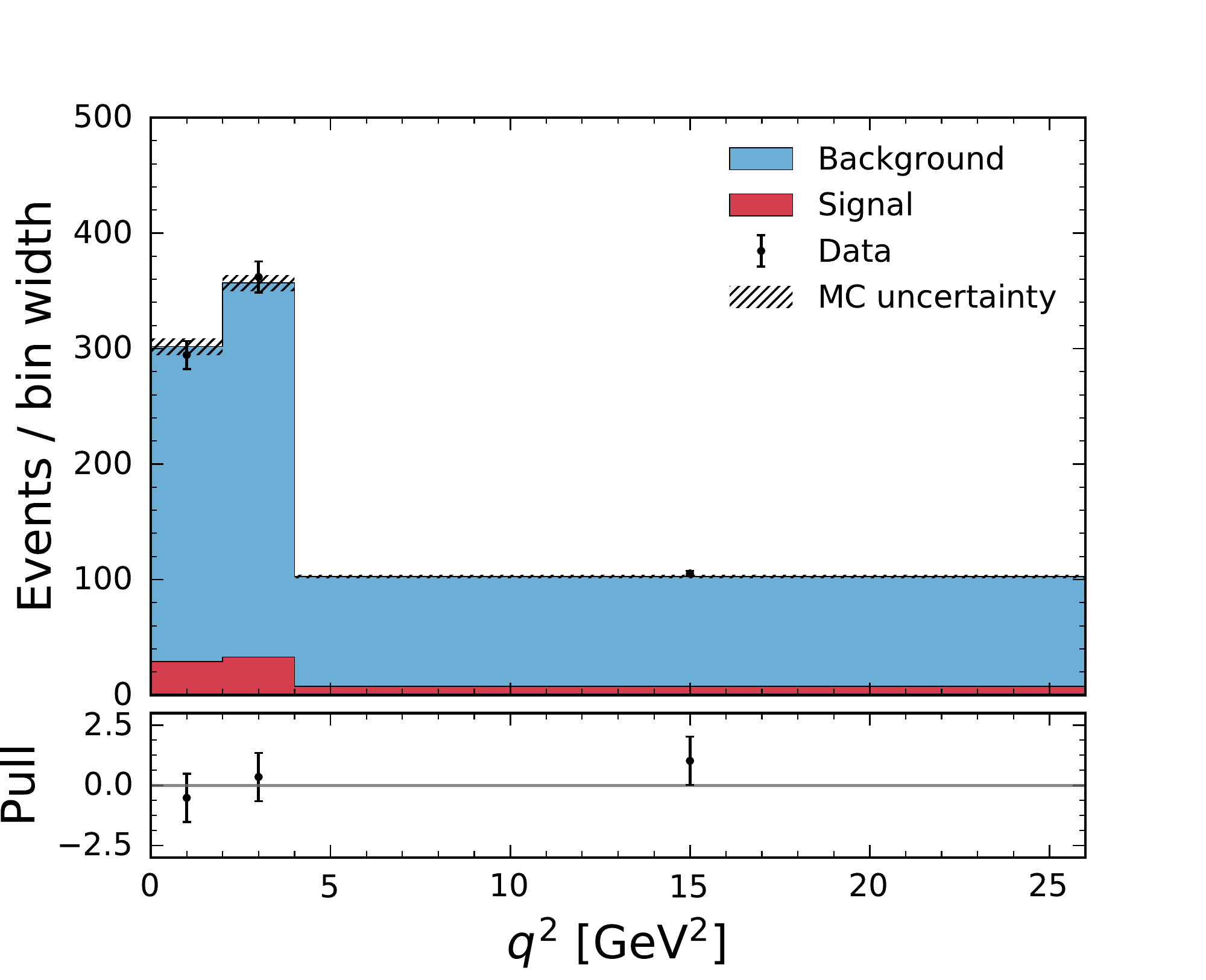} 
  \includegraphics[width=0.4\textwidth]{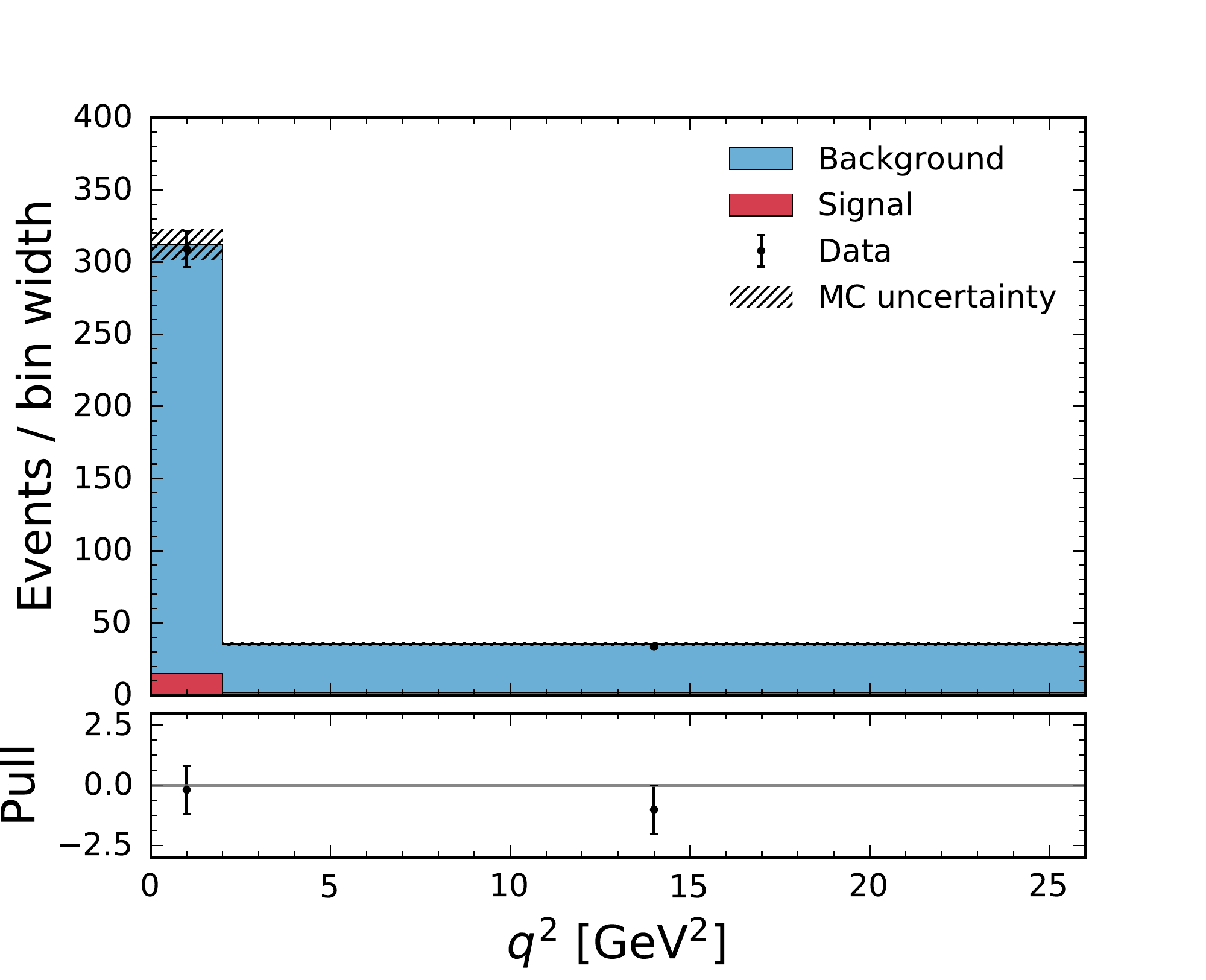}
\caption{
 The post-fit $q^2$ distributions of the two-dimensional fit to $M_X:q^2$ on $M_{X}$ are shown. The panels correspond to: $M_X \in [0, 1.5] \,  \mathrm{GeV}$ (top left), $M_X \in [1.5,1.9] \, \mathrm{GeV}$ (top right), $M_X \in [1.9,2.4]  \, \mathrm{GeV}$ (bottom left)  and $M_X \in [2.4,4] \,  \mathrm{GeV}$ (bottom right). The resulting yields are corrected to correspond to a partial branching fraction with $E_\ell^B > 1 \, \text{GeV}$. 
 }
\label{fig:2Dfit_projections}
\end{figure}

Figure~\ref{fig:2Dfit_projections} shows the post-fit $q^2$ distributions of the two-dimensional fit to $M_X:q^2$ on $M_{X}$. 

\clearpage

\section*{E. Additional Fit Details to the Lepton Flavor Universality and Weak Annihilation Tests}

The fitted yields of the two-dimensional fit to $M_{X}:q^2$ separated in electron and muon candidates, as well as in charged or neutral $B$ mesons are listed in Table~\ref{tab:lfu_fit_details}.

\begin{table*}[]
\caption{
	The fitted yields separated in electron and muon candidates, as well as in charged or neutral $B$ mesons.
}
\renewcommand\arraystretch{1.0}
\centering
\begin{tabular}{lcccc}
\hline
\hline
Decay mode               & $\widehat{\eta}_{\tt{sig}}$      & $\widehat{\eta}_{\tt{bkg}}$           & $10^{3}\left(\epsilon_{\tt{tag}} \cdot \epsilon_{\tt{sel}}\right)$ & $10^{3}\Delta \mathcal{B}$ \\ \hline
$B^{+} \to X_{u}\ell^{+}\nu$   & $914 \pm 56 \pm 64$        & $3667  \pm 77 \pm 63 $                  & $0.30 \pm 0.13$                                                   & $1.65 \pm 0.10 \pm 0.18 $ \\ 
$B^{0} \to X_{u}\ell^{+}\nu$   & $879 \pm 58 \pm 65$        & $3373  \pm 76 \pm 64 $                  & $0.33 \pm 0.11$                                                   & $1.51 \pm 0.10 \pm 0.16 $ \\ 
\hline
$B \to X_{u} e^{+}\nu$         & $870 \pm 56 \pm 59$        & $3311  \pm 75 \pm 60 $                  & $0.31 \pm 0.12$                                                   & $1.57 \pm 0.10 \pm 0.16 $ \\ 
$B \to X_{u} \mu^{+}\nu$       & $936 \pm 58 \pm 71$        & $3716  \pm 78 \pm 71 $                  & $0.32 \pm 0.13$                                                   & $1.62 \pm 0.10 \pm 0.18 $ \\ 
\hline\hline
\end{tabular}
\label{tab:lfu_fit_details}
\end{table*}

\section*{F. BDT Efficiencies}

\begin{figure}[h!]
  \includegraphics[width=0.4\textwidth]{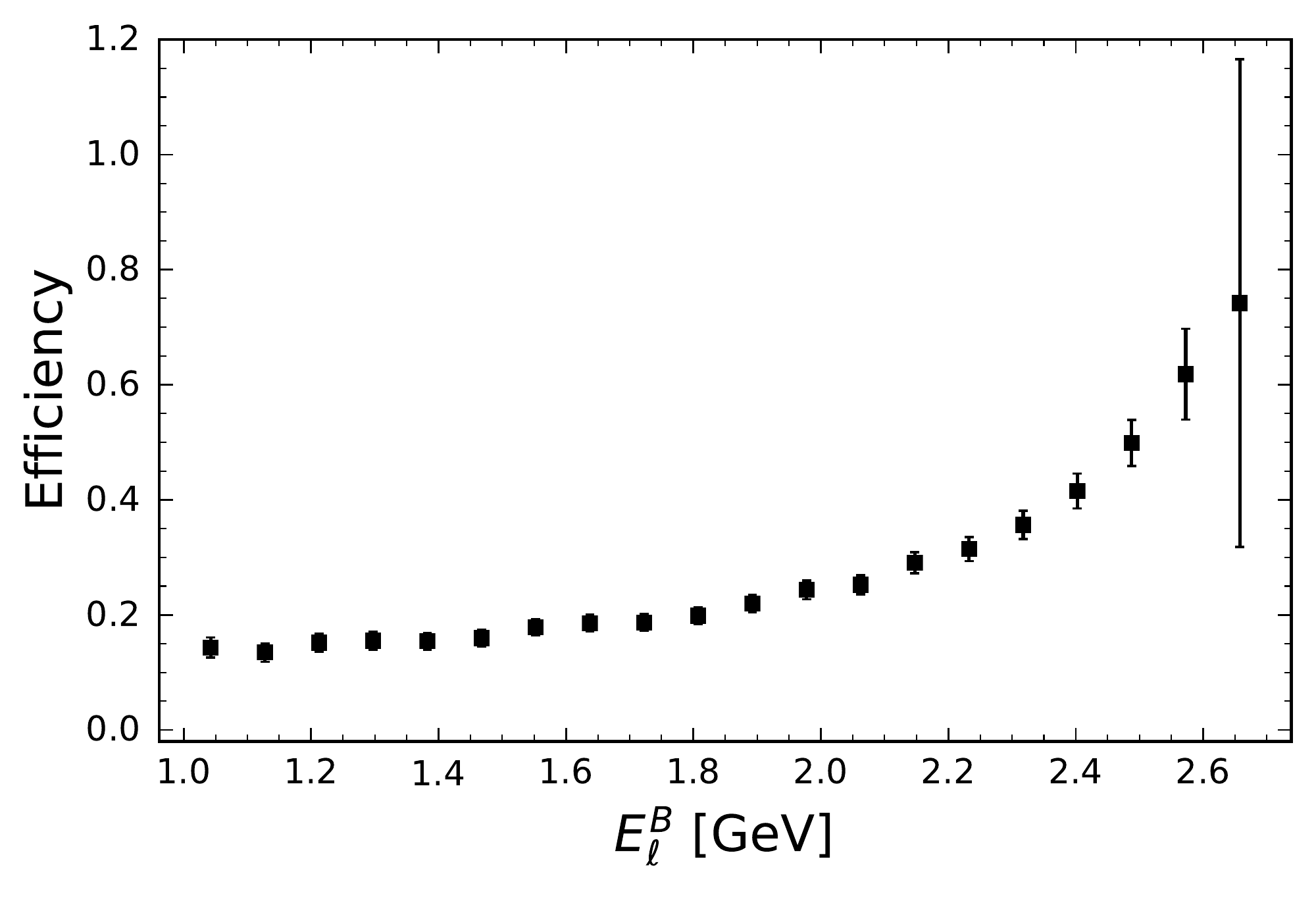}  
  \includegraphics[width=0.4\textwidth]{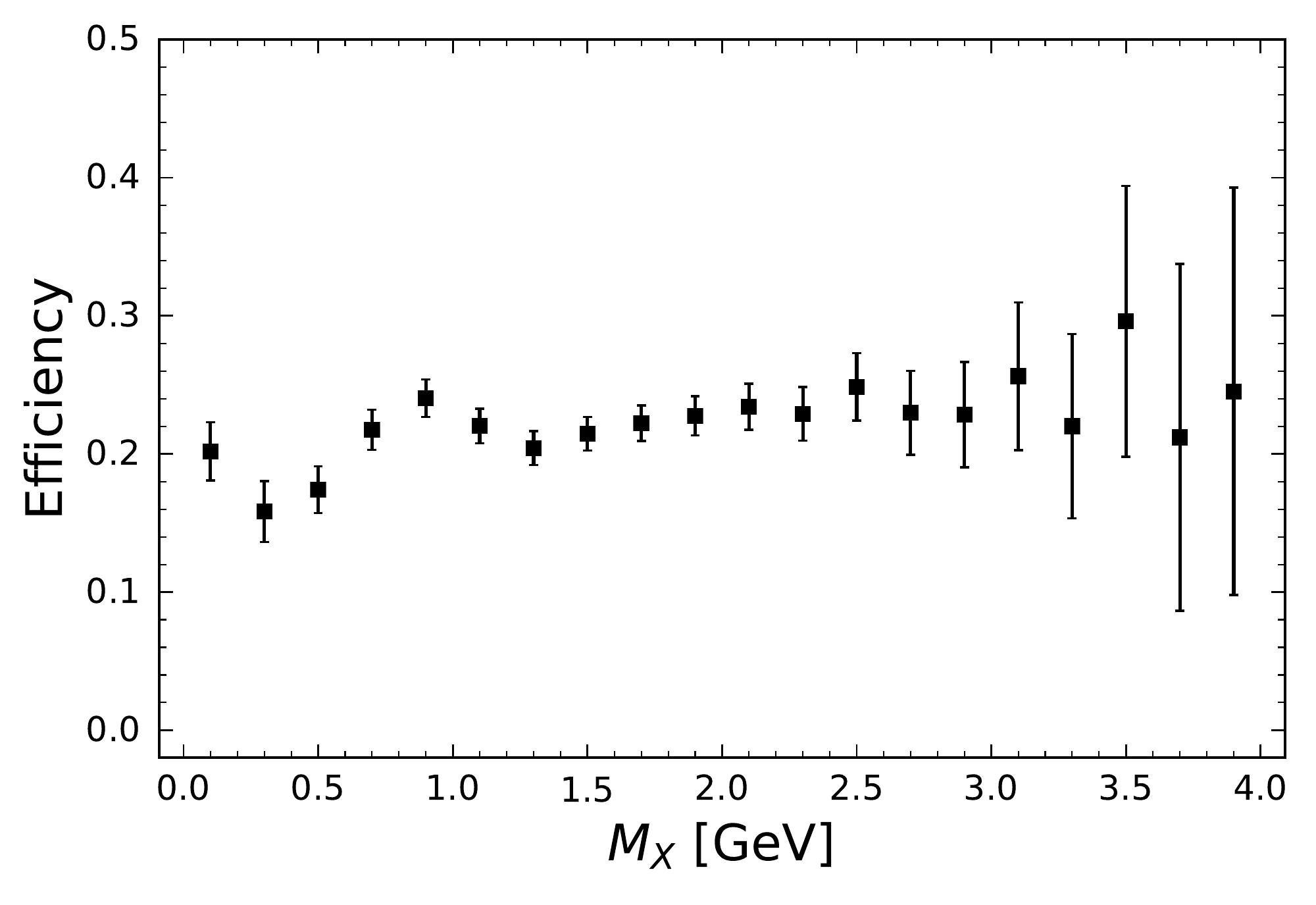}  \\
  \includegraphics[width=0.4\textwidth]{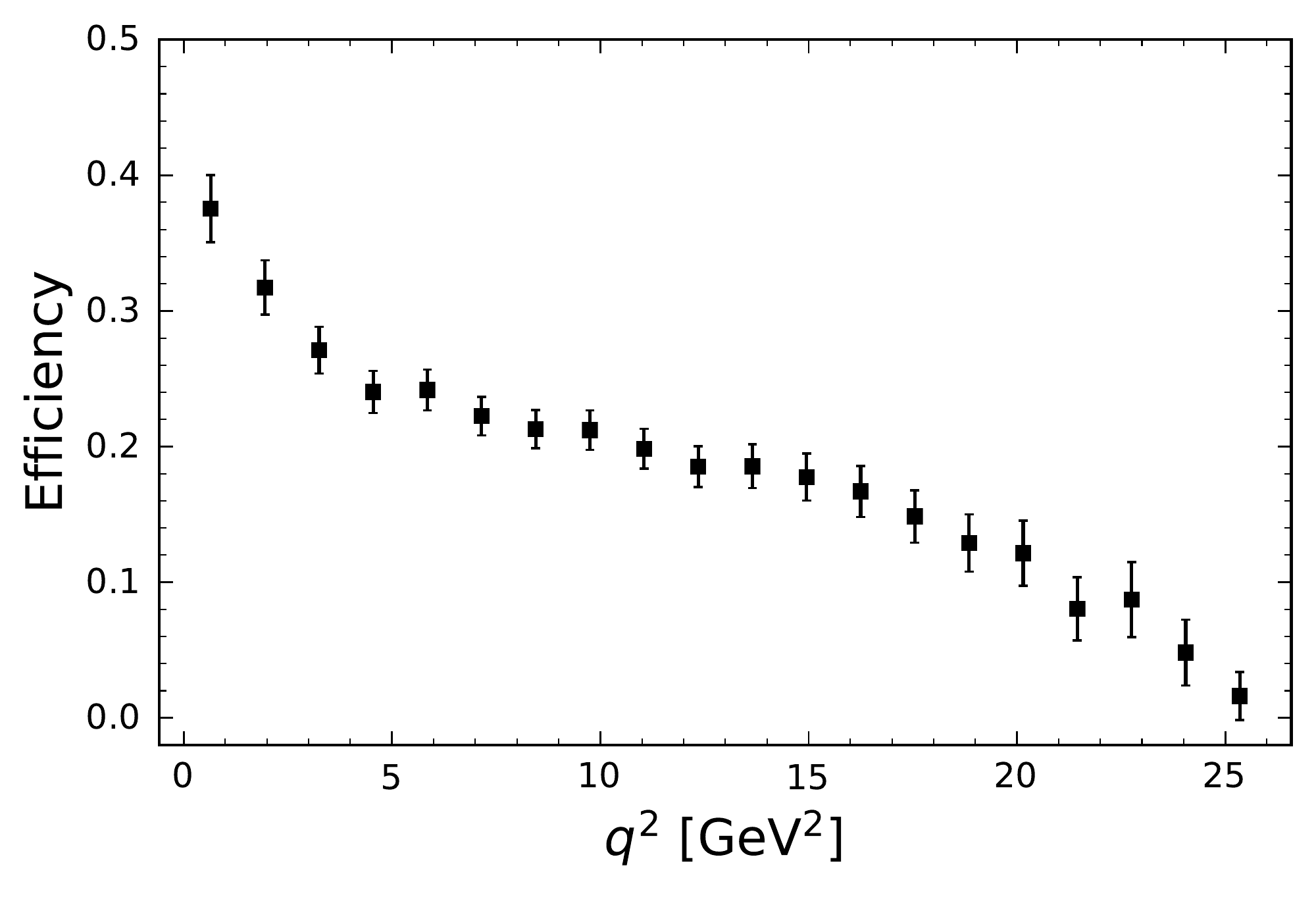}  
  \includegraphics[width=0.4\textwidth]{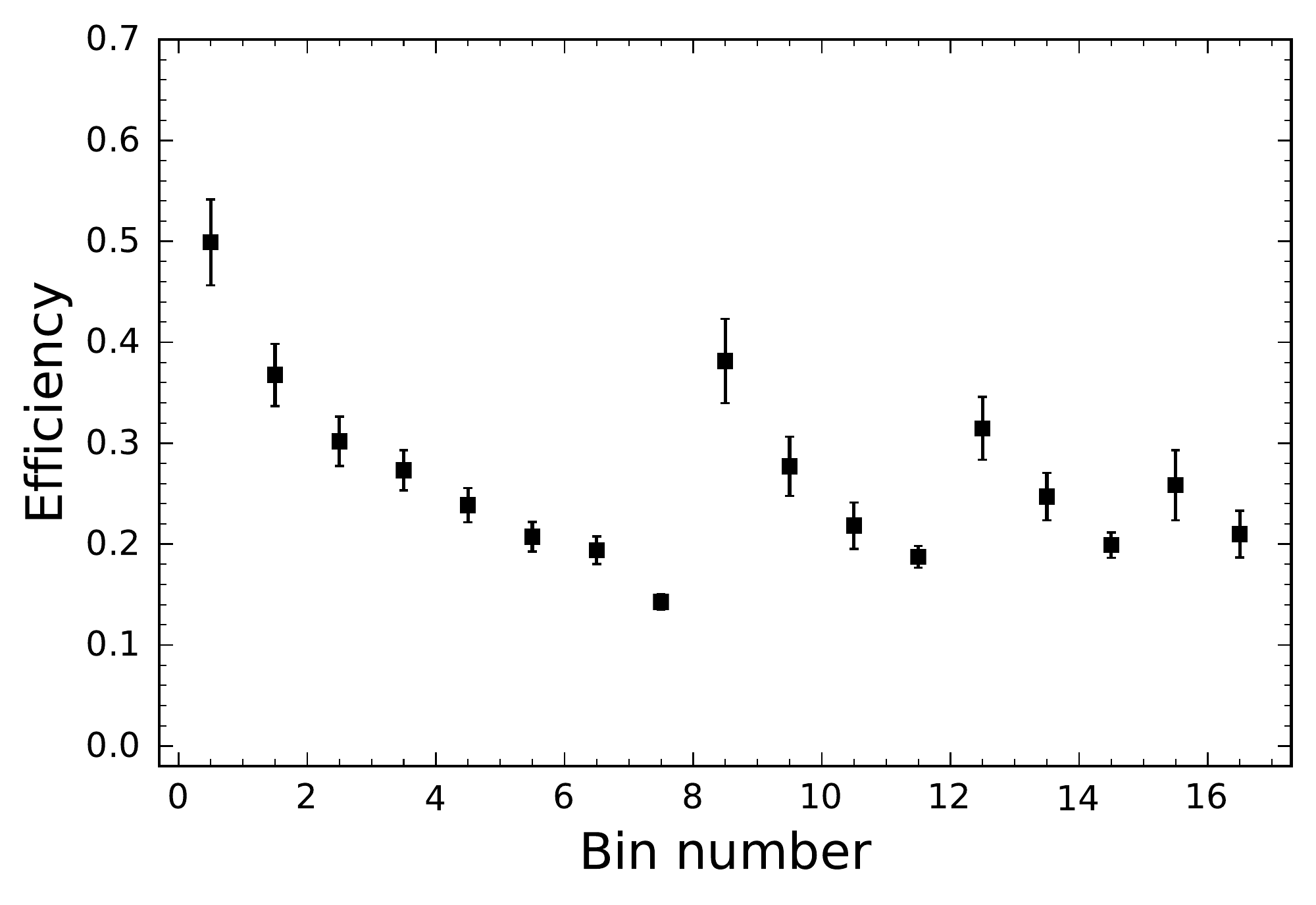} 
\caption{
  The \bulnu efficiency after the BDT selection is shown as a function of the reconstructed kinematic variables ($E_\ell^B$, $M_X$, $q^2$) used in the signal extraction. The bottom right plot shows the efficiencies in the bins of $M_{X}:q^2$ and the binning can be found in the text. The uncertainties are statistical only.
 }
\label{fig:effvars}
\end{figure}

Figure~\ref{fig:effvars} shows the efficiency of the BDT selection as a function of the reconstructed variables $q^2$, $M_{X}$, and the lepton energy $E_\ell^B$ for simulated \bulnu events. Although we avoided using these variables in the boosted decision tree, a residual dependence on the kinematic variables is seen. For instance the efficiency increases with an increase in $E_\ell^B$ and a decrease with respect to high $q^2$. The efficiency on the hadronic mass $M_X$ is relatively flat. This efficiency dependence is linked to the used variables in the BDT. Although we carefully avoided kinematic variables that would allow the BDT to learn these kinematic properties, there are indirect connections: e.g. high $E_\ell^B$ final states have a lower multiplicity as they are dominated by $B \to \pi \ell \bar \nu_\ell$ decays. Further, their corresponding hadronic system carries little momentum and on average such decays retain a better resolution in discriminating variables of the background suppression BDT. A concrete example is $M_{\mathrm{miss}}^2$ (cf. Figure~\ref{fig:bdtvars_dataMC}): high multiplicity \bulnu decays will retain a larger tail in this variable and will be selected with a lower efficiency by the BDT. 

\end{appendix}

\end{document}

%% file: pub569_2.tex
\noaffiliation
\affiliation{Department of Physics, University of the Basque Country UPV/EHU, 48080 Bilbao}
\affiliation{University of Bonn, 53115 Bonn}
\affiliation{Brookhaven National Laboratory, Upton, New York 11973}
\affiliation{Budker Institute of Nuclear Physics SB RAS, Novosibirsk 630090}
\affiliation{Faculty of Mathematics and Physics, Charles University, 121 16 Prague}
\affiliation{Chonnam National University, Gwangju 61186}
\affiliation{University of Cincinnati, Cincinnati, Ohio 45221}
\affiliation{Deutsches Elektronen--Synchrotron, 22607 Hamburg}
\affiliation{Duke University, Durham, North Carolina 27708}
\affiliation{Department of Physics, Fu Jen Catholic University, Taipei 24205}
\affiliation{Key Laboratory of Nuclear Physics and Ion-beam Application (MOE) and Institute of Modern Physics, Fudan University, Shanghai 200443}
\affiliation{Justus-Liebig-Universit\"at Gie\ss{}en, 35392 Gie\ss{}en}
\affiliation{II. Physikalisches Institut, Georg-August-Universit\"at G\"ottingen, 37073 G\"ottingen}
\affiliation{SOKENDAI (The Graduate University for Advanced Studies), Hayama 240-0193}
\affiliation{Gyeongsang National University, Jinju 52828}
\affiliation{Department of Physics and Institute of Natural Sciences, Hanyang University, Seoul 04763}
\affiliation{University of Hawaii, Honolulu, Hawaii 96822}
\affiliation{High Energy Accelerator Research Organization (KEK), Tsukuba 305-0801}
\affiliation{J-PARC Branch, KEK Theory Center, High Energy Accelerator Research Organization (KEK), Tsukuba 305-0801}
\affiliation{Higher School of Economics (HSE), Moscow 101000}
\affiliation{Forschungszentrum J\"{u}lich, 52425 J\"{u}lich}
\affiliation{IKERBASQUE, Basque Foundation for Science, 48013 Bilbao}
\affiliation{Indian Institute of Science Education and Research Mohali, SAS Nagar, 140306}
\affiliation{Indian Institute of Technology Hyderabad, Telangana 502285}
\affiliation{Indian Institute of Technology Madras, Chennai 600036}
\affiliation{Indiana University, Bloomington, Indiana 47408}
\affiliation{Institute of High Energy Physics, Chinese Academy of Sciences, Beijing 100049}
\affiliation{Institute of High Energy Physics, Vienna 1050}
\affiliation{Institute for High Energy Physics, Protvino 142281}
\affiliation{INFN - Sezione di Napoli, 80126 Napoli}
\affiliation{INFN - Sezione di Torino, 10125 Torino}
\affiliation{Advanced Science Research Center, Japan Atomic Energy Agency, Naka 319-1195}
\affiliation{J. Stefan Institute, 1000 Ljubljana}
\affiliation{Institut f\"ur Experimentelle Teilchenphysik, Karlsruher Institut f\"ur Technologie, 76131 Karlsruhe}
\affiliation{Kavli Institute for the Physics and Mathematics of the Universe (WPI), University of Tokyo, Kashiwa 277-8583}
\affiliation{Department of Physics, Faculty of Science, King Abdulaziz University, Jeddah 21589}
\affiliation{Kitasato University, Sagamihara 252-0373}
\affiliation{Korea Institute of Science and Technology Information, Daejeon 34141}
\affiliation{Korea University, Seoul 02841}
\affiliation{Kyoto Sangyo University, Kyoto 603-8555}
\affiliation{Kyungpook National University, Daegu 41566}
\affiliation{Universit\'{e} Paris-Saclay, CNRS/IN2P3, IJCLab, 91405 Orsay}
\affiliation{P.N. Lebedev Physical Institute of the Russian Academy of Sciences, Moscow 119991}
\affiliation{Liaoning Normal University, Dalian 116029}
\affiliation{Faculty of Mathematics and Physics, University of Ljubljana, 1000 Ljubljana}
\affiliation{Ludwig Maximilians University, 80539 Munich}
\affiliation{Malaviya National Institute of Technology Jaipur, Jaipur 302017}
\affiliation{University of Maribor, 2000 Maribor}
\affiliation{Max-Planck-Institut f\"ur Physik, 80805 M\"unchen}
\affiliation{School of Physics, University of Melbourne, Victoria 3010}
\affiliation{University of Mississippi, University, Mississippi 38677}
\affiliation{University of Miyazaki, Miyazaki 889-2192}
\affiliation{Moscow Physical Engineering Institute, Moscow 115409}
\affiliation{Graduate School of Science, Nagoya University, Nagoya 464-8602}
\affiliation{Kobayashi-Maskawa Institute, Nagoya University, Nagoya 464-8602}
\affiliation{Universit\`{a} di Napoli Federico II, 80126 Napoli}
\affiliation{Nara Women's University, Nara 630-8506}
\affiliation{National Central University, Chung-li 32054}
\affiliation{National United University, Miao Li 36003}
\affiliation{Department of Physics, National Taiwan University, Taipei 10617}
\affiliation{H. Niewodniczanski Institute of Nuclear Physics, Krakow 31-342}
\affiliation{Nippon Dental University, Niigata 951-8580}
\affiliation{Niigata University, Niigata 950-2181}
\affiliation{University of Nova Gorica, 5000 Nova Gorica}
\affiliation{Novosibirsk State University, Novosibirsk 630090}
\affiliation{Osaka City University, Osaka 558-8585}
\affiliation{Pacific Northwest National Laboratory, Richland, Washington 99352}
\affiliation{Panjab University, Chandigarh 160014}
\affiliation{Peking University, Beijing 100871}
\affiliation{University of Pittsburgh, Pittsburgh, Pennsylvania 15260}
\affiliation{Punjab Agricultural University, Ludhiana 141004}
\affiliation{Research Center for Nuclear Physics, Osaka University, Osaka 567-0047}
\affiliation{Meson Science Laboratory, Cluster for Pioneering Research, RIKEN, Saitama 351-0198}
\affiliation{Department of Modern Physics and State Key Laboratory of Particle Detection and Electronics, University of Science and Technology of China, Hefei 230026}
\affiliation{Seoul National University, Seoul 08826}
\affiliation{Showa Pharmaceutical University, Tokyo 194-8543}
\affiliation{Soochow University, Suzhou 215006}
\affiliation{Soongsil University, Seoul 06978}
\affiliation{Sungkyunkwan University, Suwon 16419}
\affiliation{School of Physics, University of Sydney, New South Wales 2006}
\affiliation{Department of Physics, Faculty of Science, University of Tabuk, Tabuk 71451}
\affiliation{Tata Institute of Fundamental Research, Mumbai 400005}
\affiliation{Department of Physics, Technische Universit\"at M\"unchen, 85748 Garching}
\affiliation{School of Physics and Astronomy, Tel Aviv University, Tel Aviv 69978}
\affiliation{Toho University, Funabashi 274-8510}
\affiliation{Department of Physics, Tohoku University, Sendai 980-8578}
\affiliation{Earthquake Research Institute, University of Tokyo, Tokyo 113-0032}
\affiliation{Department of Physics, University of Tokyo, Tokyo 113-0033}
\affiliation{Tokyo Institute of Technology, Tokyo 152-8550}
\affiliation{Tokyo Metropolitan University, Tokyo 192-0397}
\affiliation{Virginia Polytechnic Institute and State University, Blacksburg, Virginia 24061}
\affiliation{Wayne State University, Detroit, Michigan 48202}
\affiliation{Yamagata University, Yamagata 990-8560}
\affiliation{Yonsei University, Seoul 03722}
  \author{I.~Adachi}\affiliation{High Energy Accelerator Research Organization (KEK), Tsukuba 305-0801}\affiliation{SOKENDAI (The Graduate University for Advanced Studies), Hayama 240-0193} 
  \author{H.~Aihara}\affiliation{Department of Physics, University of Tokyo, Tokyo 113-0033} 
  \author{S.~Al~Said}\affiliation{Department of Physics, Faculty of Science, University of Tabuk, Tabuk 71451}\affiliation{Department of Physics, Faculty of Science, King Abdulaziz University, Jeddah 21589} 
  \author{D.~M.~Asner}\affiliation{Brookhaven National Laboratory, Upton, New York 11973} 
  \author{H.~Atmacan}\affiliation{University of Cincinnati, Cincinnati, Ohio 45221} 
  \author{T.~Aushev}\affiliation{Higher School of Economics (HSE), Moscow 101000} 
  \author{R.~Ayad}\affiliation{Department of Physics, Faculty of Science, University of Tabuk, Tabuk 71451} 
  \author{V.~Babu}\affiliation{Deutsches Elektronen--Synchrotron, 22607 Hamburg} 
  \author{M.~Bauer}\affiliation{Institut f\"ur Experimentelle Teilchenphysik, Karlsruher Institut f\"ur Technologie, 76131 Karlsruhe} 
  \author{P.~Behera}\affiliation{Indian Institute of Technology Madras, Chennai 600036} 
% ...
\collaboration{The Belle Collaboration}